\documentclass[a4paper,11pt]{article}
\usepackage{jheppub} 
\usepackage{lineno}
\usepackage{dsfont}
\usepackage{braket}

\arxivnumber{2209.04548} 

\title{\boldmath Encoded information of mixed correlations: the views from one dimension higher}

\author{Mahdis Ghodrati}
\affiliation{Asia Pacific Center for Theoretical Physics, Pohang 37673, Republic of Korea}

\emailAdd{mahdis.ghodrati@apctp.org}

\abstract{After reviewing the JT gravity, we discuss the four saddles in the mixed correlation measures of black holes Hawking radiation in the setup of geometric evaporation of  \cite{Verheijden:2021yrb}. By looking from $1d$ higher point of view and partial dimensional reduction, we examine the phase structures and the universalities for these four saddles. We also discuss the behavior of quantum error correction codes for each of these four phases, reaching to consistent results. Then, instead of dimension reduction between Einstein gravity and JT, we try to explore the connections between partition functions and saddles of $3d$ Chern-Simons and $2d$ BF theories, $2d$ Liouville and $2d$ Wess-Zumino-Witten models, and also the dimensionally reduced $1d$ Schwarzian and $1d$  particles on group. We specifically sketch on the connections between these theories in the setup of mixed correlations and island formulation.}

\begin{document}
\maketitle
\flushbottom

\section{Introduction}
Recently, the idea of looking the evolution of black holes and its phase structures, in the setup of quantum gravity, from higher dimensions to lower dimensions has been proved to be very fruitful. This general idea of looking at black hole evaporation from one dimensional higher and specifically the case of $2d$ Jackiw–Teitelboim (JT) gravity from $3d$ Einstein gravity model, has been started in \cite{Almheiri:2019hni}, while the effects of boosted quantum corrections on quantum extremal surface (QESs) and the connections with the Page curve of black holes were first studied in \cite{Almheiri:2019psf}. Related to this idea,  the evolution of the black holes has also been studied by the island/boundary-$\text{CFT}_2$ ($\text{BCFT}_2$) setup in various works which we mention throughout the work here.

We specifically want to extend this method of looking the evaporation of black holes from $3d$ to $2d$ JT gravity, by using mixed correlation measures, and partial, geometric dimension reduction used in \cite{Verheijden:2021yrb}.
After reviewing JT gravity in section \ref{sec:setupJT}, in section \ref{sec:phasesmixed},  we comment in more details on the universalities noted between the phase structures coming from various correlation measures, that we observed in our previous work \cite{Ghodrati:2022kuk}.

 Instead of one angular interval, in our section \ref{sec:calcb}, we consider two interval subsystems, and then using mutual information and the critical bath size, we explore the phase structure of Hawking radiation, for the case before and after the Page time. We find there, many universalities between phases, coming from different correlation measures and models of saddles of Hawking radiation.

On the other hand, there are several formulations for holographic bulk reconstruction such as tensor network, HKLL, quantum recovery channel, or modular flow, where in \cite{Ghodrati:2020vzm}, the connections between some of them have been assessed. The boundary information data which reconstruct the bulk, can in principle specify the phases or saddles as well.  However, the transitions between these saddles in the setup of black holes would be subtle. Viewing this problem from one dimension higher and then using dimension reduction would shed light on several insightful issues, which is the aim of this work. In section \ref{subsec:QECPage}, we apply the inequalities coming from quantum error corrections (QEC), between these four saddles, where we find consistent results for the possible ways to move between these saddles, with the results of QEC.

 \begin{figure}[ht!]
 \centering
  \includegraphics[width=8.5cm] {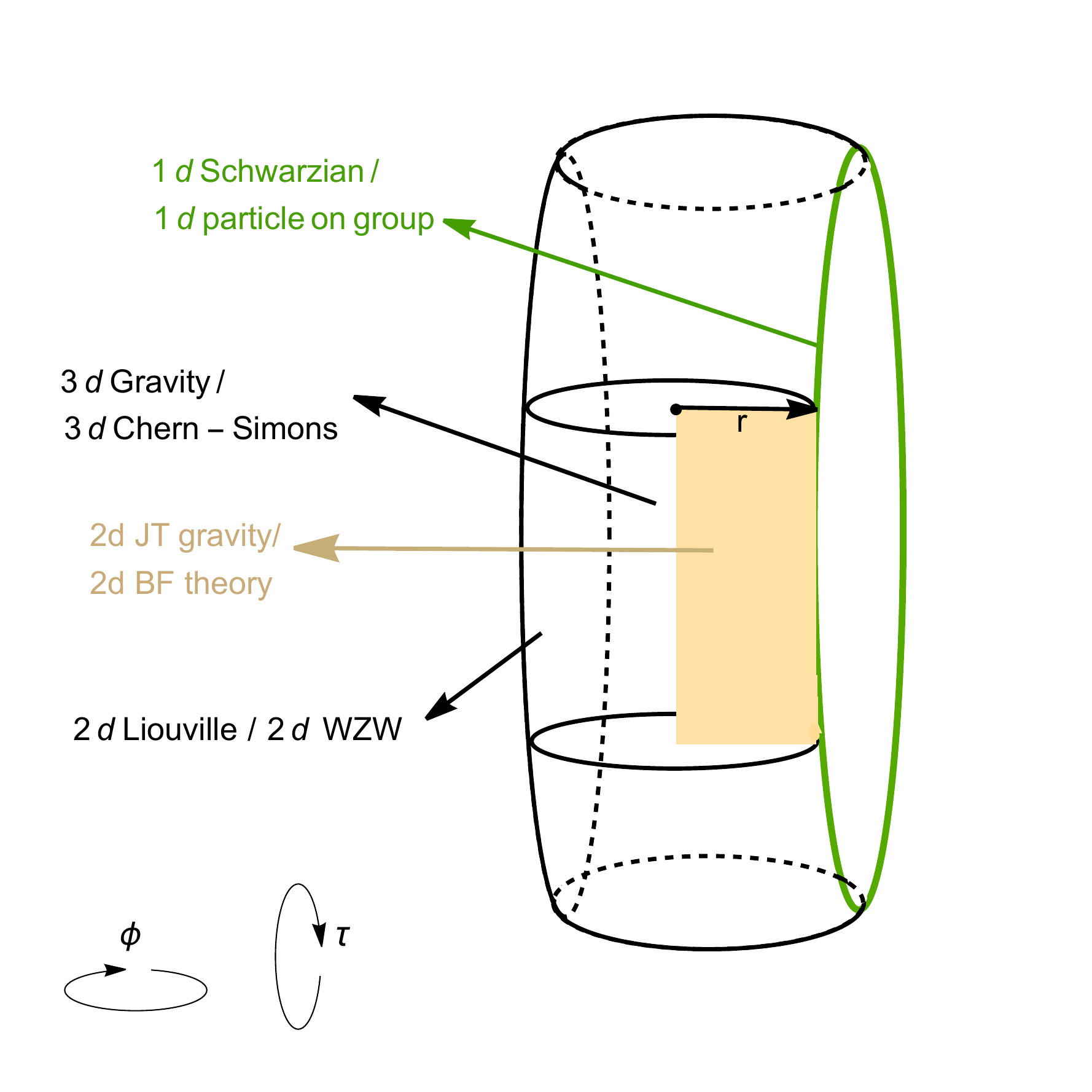}  
  \caption{The connections between various theories in higher dimensions and $1d$ lower dimension case, through holography or dimensional reduction. Inside the torus is the $3d$ Chern-Simons gravity where its boundary is $2d$ Liouville gravity or $2d$ WZW theory in the case of topological theory.  Also, after the dimension reduction, the $3d$ gravity either leads to $2d$ JT gravity or to the $2d$ BF theory. These two theories also have a $1d$ lower dimension boundary theory, namely $1d$ Schwarzian, or $1d$ particle on group theory.}
 \label{fig:dimHoltheories}
\end{figure}

The initial idea of geometric evaporation and dimension reduction in \cite{Verheijden:2021yrb} have been applied between $3d$ Einstein gravity and $2d$ JT model. This idea then can be extended to other interesting and important quantum gravity toy models, such as $3d$ Chern-Simons and $2d$ BF gauged theory, and also for the boundary $2d$ Liouville or $2d$ Wess-Zumino-Witten (WZW) gauged models. Then, one can go even one dimension lower and study the problem for the case of $1d$ Schwarzian or $1d$ particle on group model. The picture in figure \ref{fig:dimHoltheories}, showing the connections between these models would be very illuminative in our study there.  In our section \ref{sec:dimensionsPage},  we then try to sketch how the information is encoded between various dimensions using these models, and the effects of the parameters of partition functions on the behavior of Page curve in these models. Finally, we end up with a short conclusion in section \ref{sec:conclusion}.

\section{The setup}\label{sec:setupJT}
Most of the discussion of black hole evaporation and information loss used the Jackiw-Teitelboim (JT) gravity toy model which could be written as
\begin{gather}
I_{JT} = \frac{1}{16 \pi G} \int d^2 x \sqrt{-g} \left \lbrack \phi_0 R + \phi \left ( R + \frac{2}{L^2} \right ) \right \rbrack\nonumber
+\frac{1}{8\pi G} \int dt \sqrt{-\gamma} \left \lbrack \phi_0 K + \phi \left ( K - \frac{1}{L} \right ) \right \rbrack.
\end{gather}

This gravity model can in fact, be obtained by the spherical reduction of the $4d$ Einstein-Maxwell action,
\begin{gather}
I_{EM} = \frac{1}{16 \pi G_4} \int d^4 x \sqrt{- \hat{g}} \left ( \hat{R} - \frac{1}{4\pi} \hat{F}^2 \right )+ \frac{1}{8 \pi G_4} \int d^3 x \sqrt{\hat{\gamma} } \hat{K},
\end{gather}

Here he equation of motion for the dilaton would be $\phi(r) = \frac{r}{L} \phi_r$.

Generally for simulating the black hole degrees of freedom, an additional term in the from of an end-of-the-world (EOW) brane can be added to the JT action.

 So with the ansatz
\begin{gather}
ds_4^2= g_{\mu \nu} (t,r) dx^\mu dx^\nu + ( \phi_0 + \phi (r,t) ) d \Omega_2^2,
\end{gather}
in works such as \cite{Dong:2021oad}, the action has been chosen as the combination of the JT gravity with an EOW brane which has a tension $\mu$, written in the form of
\begin{gather}
I=I_{JT}+\mu \int_{\text{brane}} dy,
\end{gather}
where the JT action would be written as
\begin{gather}
I_{JT}=-\frac{S_0}{2\pi} \left \lbrack \frac{1}{2} \int_{\mathcal{M}} R + \int_{\partial \mathcal{M}} K \right \rbrack - \left \lbrack \frac{1}{2} \int_{\mathcal{M}} \phi (R+2) + \int_{\partial \mathcal{M}} \phi K \right \rbrack,
\end{gather}
and $S_0$ is the zero temperature entropy of an eternal two-dimensional black hole. In this model the main parameter is the number of orthonormal states or flavors on the EOW brane denoted by $k$ which can actually model the Hawking quanta falling inside the black holes. As $k$ increases, the later regimes of the evaporating black hole could be probed. These states are also entangled with the an auxiliary reference system $R$.

In the work of Verheijden and Verlinde \cite{Verheijden:2021yrb}, which is the base of our work here, however, this EOW brane is replaced with the black hole, as the action for matter $S_{\text{matter}}$, has been added to the JT last action, written in the form of
\begin{gather}
S=\frac{1}{16\pi G} \left \lbrack \int d^2 x \sqrt{-g} \Phi_0 R + \int d^2 x \sqrt{-g} \Phi \left( R + \frac{2}{\ell^2} \right) \right \rbrack + S_{\text{matter}},
\end{gather}
where here $\Phi_0$ is a constant. The first term which is topological, determines the extremal entropy and after adding the appropriate boundary term, it would give the Euler characteristic of the manifold, as it will be shown next in the relation  \ref{eq:eulerJT}.  The last term, $S_\text{matter}$, is some arbitrary matter system which couples to the metric but not to the dilaton.

In the work of \cite{Anderson:2020vwi}, the action for the JT gravity with the EOW brane has been written as
\begin{gather}\label{eq:eulerJT}
I \lbrack \phi, g \rbrack = -S_0 \chi - \frac{1}{4\pi} \Big \lbrack \int_M \sqrt{g} \phi ( R+2) + \int_{\partial M} \sqrt{h} \phi K \Big \rbrack + \phi_r \mu \int_{\text{ETW brane} } ds,
\end{gather}
where the role of Euler character and topology would be more clear, as $\chi$ is the Euler character of the Euclidean spacetime, $S_0$ is the extremal entropy, $\mu$ is the tension of the EOW brane and $\phi$ is the dilaton. The boundary condition for this model similar to Xi Dong's case \cite{Dong:2021oad} is
\begin{gather}
\text{Asymptotic AdS boundary:} \ \ \ \ \ \ \phi= \frac{\phi_r}{\epsilon}, \ \ \ \ \ du^2\equiv \epsilon^2  ds^2|_{bd},\nonumber\\
\text{EOW brane boundary:} \ \ \ \ \ \ \ n^\alpha \partial_\alpha \phi=\mu, \ \ \ \ \ K=0,
\end{gather}

As in Dong's paper \cite{Dong:2021oad}, the boundary would also be the Dirichlet condition on an asymptotic boundary interval, while on the EOW branes, it would be Neumann boundary conditions, as
\begin{gather}
ds^2 \big |_{\partial \mathcal{M}} = \frac{1}{\epsilon^2} d\tau^2, \ \ \ \ \phi= \frac{1}{\epsilon}, \ \ \ \ \epsilon \to 0, \nonumber\\
\partial_n \phi |_{\text{brane}} = \mu, \ \ \ \ \ K=0.
\end{gather}

Note that the nice feature of  \cite{Verheijden:2021yrb} is the presence of black hole instead of EOW, which would let us to study correlations and various mixed entanglement measures in a dynamical setup and study the connections between their behaviors and Page curve, and also information loss paradox of black holes, as we discuss in the next sections.

\section{Phase diagrams from different correlation measures}\label{sec:phasesmixed}

Using the JT action as the toy model and its solutions, one could use several different measures of entanglement and mixed correlation to study black hole Hawking radiation. Then, we aim here to check the results coming from these measures from $1d$ higher point of view.

\begin{figure}[ht!]   
\begin{center}
\hspace*{1.9cm}\includegraphics[width=1.2\textwidth]{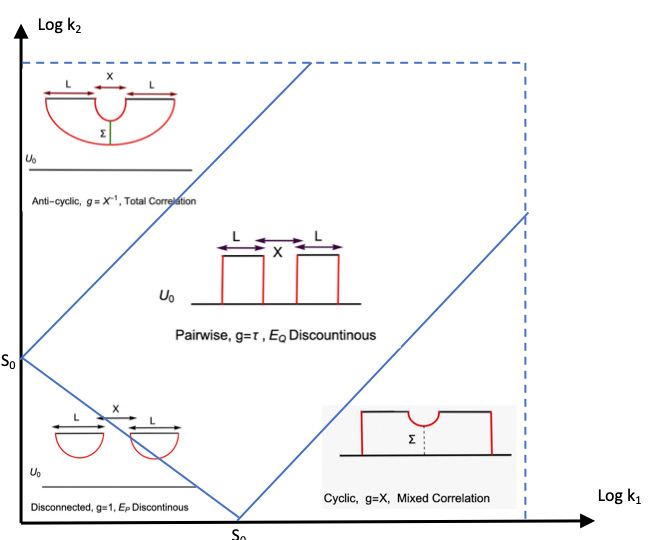}  
\caption{Phase diagram of mixed system using negativity, from \cite{Dong:2021oad}, and comparing with plots of $D_c$ in confining geometries \cite{Ghodrati:2021ozc}.}
\label{fig:fphasesNeg}
\end{center}
\end{figure}

In \cite{Dong:2021oad}, the phase diagram of mixed system using ``entanglement negativity" has been found as shown in figure \ref{fig:fphasesNeg}, which we briefly discussed in \cite{Ghodrati:2022kuk} and commented on its connections with the phase structures of confining models, and here we aim to study them in further details.

 Note that in the calculation of the partition function for deriving the entanglement negativity, four different saddles have been found corresponding to four different types of permutations $g$, where the permutation $g=X$ which corresponds to cyclic phase is related to the mixed correlation phase discussed in  \cite{Agrawal:2021nkw}. The permutation $g=X^{-1}$ which corresponds to anti-cyclic phase is related to the total correlation phase of \cite{Agrawal:2021nkw}. The permutation $g=\tau$ which corresponds to pairwise phase is related to the $E_Q$-discontinuous phase of  \cite{Agrawal:2021nkw}, and the permutation $g=\mathds{1}$ which corresponds to disconnected phase is related to the $E_P$ phase discussed in \cite{Agrawal:2021nkw}. We could connect these saddles with the various phases for two strips in the confining geometries found in \cite{Jain:2020rbb, Ghodrati:2021ozc}.  Note that in figure \ref{fig:fphasesNeg}, $k_2$ and $k_1$ are the number of orthonormal states or flavors in each subsystem.

Then, the R\'{e}nyi generalization of negativity could be written as
\begin{gather}
N_k ( \rho_{AB})= \text{tr} \left ( (\rho_{AB}^{T_B} )^k \right ),
\end{gather}
where $N_k$ is the k-th R\'{e}nyi negativity. This quantity in \cite{Dong:2021clv}  has written as
\begin{gather}
N_k= \text{tr} \left \lbrack  \rho_{AB}^{\otimes k} \big (P_A(X) \otimes P_B(X^{-1})  \big)  \right \rbrack,
\end{gather}
where $X$ is a \textit{k-cycle} and $X^{-1}$ is its inverse, and $P_M(g)$ in general is the representation of a permutation group element $g\in S_k$ on the $k$ copies of some subsystem $M$, and $P_A(X)$ and $P_B(X^{-1})$ are both special cases of it. For the integer $k$, this relation could be written as $N_k= \frac{Z_k}{Z_1^k}$, where $Z_k$ is the boundary partition function on a $k$-fold branched over $M_k^{A,B}$ which can be obtained by gluing $k$ copies of the original boundary spacetime $M_1$ cyclically along $A$ and anti-cyclically along $B$.

Using the holographic duality, $Z_k$ can be calculated as
\begin{gather}
Z_k= e^{-I \lbrack \mathcal{B}_k  \rbrack },
\end{gather}
 where $I \lbrack \mathcal{B}_k \rbrack$ is the on-shell action of proper bulk saddle point solution $\mathcal{B}_k$, where its asymptotic boundary is the $k$-fold cover $M_k^{A,B}$. 
 
 For the case of JT gravity with EOW brane with $k$ orthonormal states or flavors, which are being splitted into two subsystems consisting of $k_1$ and $k_2$ states, such that $k=k_1 k_2$, the pairwise connected geometry would satisfy the relation $k_1 k_2 \gg e^{S_0}$ and $e^{-S_0}  \ll k_1/k_2 \ll e^{S_0}$. These geometries correspond to the set of permutations $\tau$ which are known as non-crossing pairings. An interesting feature of these geometries is that for the even replica index $n$, a pairwise connected geometry is constructed by connecting paired asymptotic boundaries by two-boundary wormholes while for odd $n$, the geometries could be constructed by the similar non-crossing pairings of the boundaries plus a single one-boundary connected component and therefore these geometries would spontaneously break the replica symmetry.

 These geometries are dominant since $k_1$, $k_2$ and $e^{S_0}$ can be put on the most equal footing by maximizing the sum of the three exponents in the relation
 \begin{gather}
 \text{Tr} \lbrack ( \rho_R^{T_2} )^n \rbrack \sim \frac{1}{( k e^{S_0})^n} \sum_{g\in S_n} ( e^{S_0})^{\chi(g)}\ k_1^{\chi(g^{-1} X)} \ k_2^{\chi (g^{-1} X^{-1} )},
 \end{gather}
 and they contribute as
 \begin{gather}
 g=\tau  \to   \left ( e^{S_0} \right)^{\lceil \frac{n}{2} \rceil } k^{\lfloor \frac{n}{2} \rfloor +1},
 \end{gather}
 where $\lceil \frac{n}{2} \rceil $ and $\lfloor \frac{n}{2} \rfloor$ are the ceiling and  floor functions.

 So, the replica saddle points here can then be either symmetric or non-symmetric. In \cite{Dong:2021clv}, it has been shown that, in the phase where $I$ is positive and entanglement wedge is connected, the non-symmetric saddle points are dominant in R\'{e}nyi negativities. Therefore, most of states are in the pairwise geometries which in fact break the replica symmetry there, and that is why in works such as \cite{Dong:2021oad} for the JT gravity with the EOW brane and also in  \cite{Ghodrati:2021ozc} for the supergravity models, most of the space of the parameter regime are covered by the pairwise case.

 In general, a simple relation for logarithmic negativity could be written as
 \begin{gather}
 E_N( \rho_{AB}  ) \simeq \frac{1}{2} I(A:B) + \log \frac{8}{3\pi},
 \end{gather}
which demonstrate the connections between negativity and mutual information as we expected also from the phase structures.

\begin{figure}[ht!]   
\begin{center}
\hspace*{1.1cm} \includegraphics[width=1.2\textwidth]{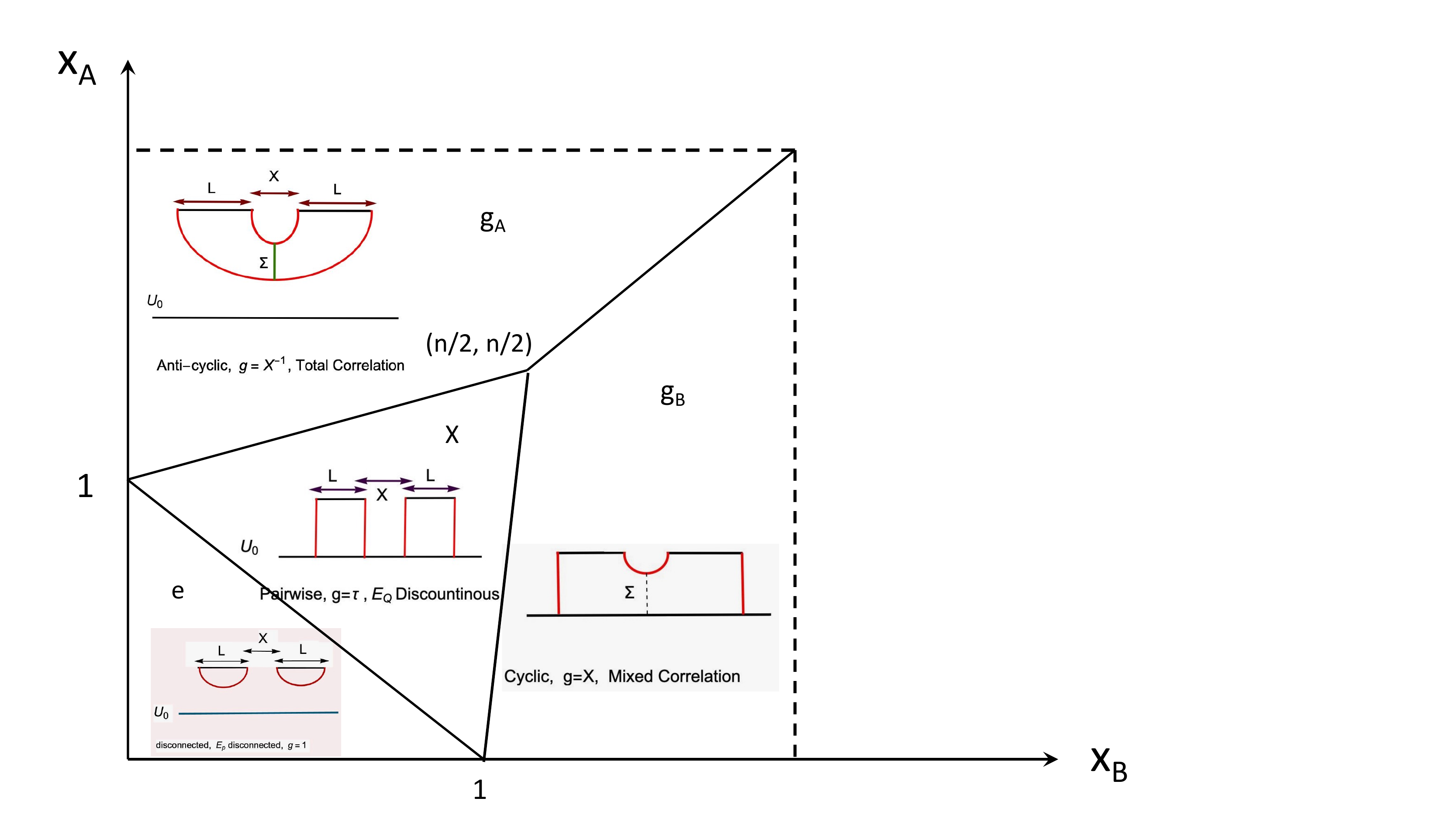}  
\caption{Phase diagram of mixed system using wormholes \cite{Akers:2021pvd}, and comparing with the configurations coming from mutual information and $D_c$ in confining geometries.}
\label{fig:Ackersdiagram}
\end{center}
\end{figure}

\begin{figure}[ht!]   
\begin{center}
\hspace*{1cm}\includegraphics[width=1\textwidth]{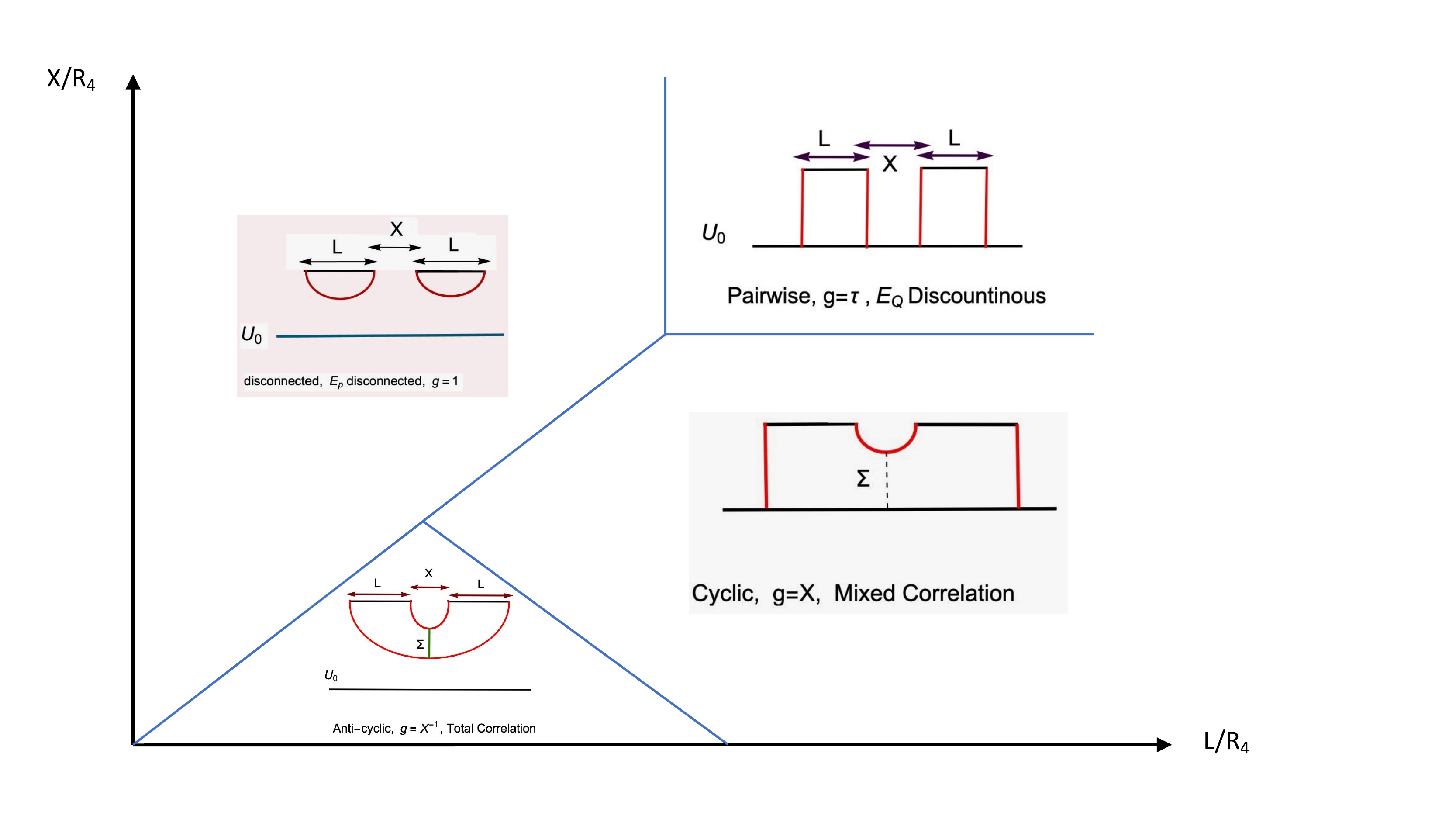}  
\caption{Phase diagram for confining geometries directly coming from mutual information and $D_c$.}
\label{fig:parulphases}
\end{center}
\end{figure}

Then, in \cite{Akers:2021pvd}, the Page curves for ``reflected entropy" have been presented which is shown in figure \ref{fig:Ackersdiagram}, and the connections with the setup of confining geometries are also shown. In that work this quantity has been studied in random tensor network setup. Specifically the problem of the smoothing out of the Page transition has been considered. In that work the phase diagram of the single tripartite tensor for the R\'{e}nyi reflected entropies as the function of the bond dimension have also been presented which is very similar to the phase diagrams found using the negativity in \cite{Dong:2021oad, Jain:2020rbb} and mutual information and critical distance, $D_c$ in \cite{Ghodrati:2021ozc}. For that case, the region where the entanglement wedge is connected would correspond to $X_A+X_B >1$ and $1-X_B < X_A < 1+X_B$, which also corresponds to the pairwise connected saddle for calculating negativity in JT gravity with EOW brane where $k_1 k_2 \gg e^{S_0}$ and $e^{-S_0} \ll k_1/k_2 \ll e^{S_0}$.

There, $X_A$ and $X_B$ are defined as
\begin{gather}
X_A= \frac{\ln \chi_A}{ \ln \chi_C} , \ \ \ \ \ X_B= \frac{\ln \chi_B}{\ln \chi_C},
\end{gather}
where the $\chi_i$ corresponds to the bond dimension or the horizon areas of the wormhole.

The Page phase transition is along the line $X_A+X_B=1$, and there are other phase boundaries along $X_A=1+X_B$ and $X_B=1+X_A$ where the derivative of the mutual information jumps, which in the QCD models would be related to the confinement/deconfinement transitions, and chirality breaking as explained in \cite{Ghodrati:2021ozc}.
The bond dimensions then for the bath or radiation system can be defined and the critical bath size would be connected to this critical bond dimension, which is also related to the critical distance $D_c$ studied in \cite{Ghodrati:2021ozc} for the confining geometries.

The actual phase diagram for confining geometries based on the different scales of the setup, such as the distance between the two strips divided by the AdS scale $X/R_4$, and width of strips $L$ divided by $R_4$, are shown in figure \ref{fig:parulphases}.
Note that in this phase diagram there are actually three scales, the distance between the two equal strips $X$, the width of the strip $L$, and the place of the hard wall in the confining models, $U_0$.  These parameters would correspond to the three constants of $k_1$, $k_2$ and $S_0$ in the phase diagram derived in \cite{Dong:2021oad} and the three bond dimensions or the horizon area of the three-boundary wormhole model of \cite{Akers:2021pvd}, and correspondingly to $S_{A_1}^{\text{eq}}$, $S_A^{\text{(eq)}}$ and $S_B^{\text{(eq)}}$ of \cite{Vardhan:2021npf,Shapourian:2020mkc}. By tuning these parameters in each case the phase space can be probed leading to the phase structures that have compatibilities with each other.

In \cite{Vardhan:2021npf,Shapourian:2020mkc} the entanglement negativity for the infinite and also finite temperatures, using  different methods, have been calculated and the phase structures for the pattern of entanglement have also been constructed. For the infinite temperature they found three phases, no entanglement (NE), maximally entangled phase (ME), and entanglement saturation phase which are shown in figure \ref{fig:shapur}.

In that work for tracking the relative sizes of the subsystem, they used two parameters defined as
\begin{gather}
\lambda:= \frac{S_{A_1}^{\text{(eq)}} }{S_A^{\text{(eq)}} }, \ \ \ \ \ \ c:= \frac{S_A^{\text{(eq)}}}{ S_A^{\text{(eq)}}+S_B^{\text{(eq)}}}.
\end{gather}

 We propose here that the maximally entangled phase (ME) for large values of $c$, for the case of small $\lambda$ and big $\lambda$ of  \cite{Vardhan:2021npf,Shapourian:2020mkc} would be distinct from each other, as the former one corresponds to anti-cyclic phase of \cite{Dong:2021oad} and total correlation phase of \cite{Agrawal:2021nkw}, while the later corresponds to the cyclic phase of \cite{Dong:2021oad} and total correlation phase of \cite{Agrawal:2021nkw}. We also propose that the entanglement saturation phase corresponds to the pairwise phase of \cite{Dong:2021oad} and to the $E_Q$-discontinuous phase of \cite{Agrawal:2021nkw}. In addition, the ``no entanglement" (NE) phase corresponds to the case of $E_P$-discontinuous phase of \cite{Agrawal:2021nkw}.

\begin{figure}[ht!]   
\begin{center}
\hspace*{0cm}\includegraphics[width=1.35\textwidth]{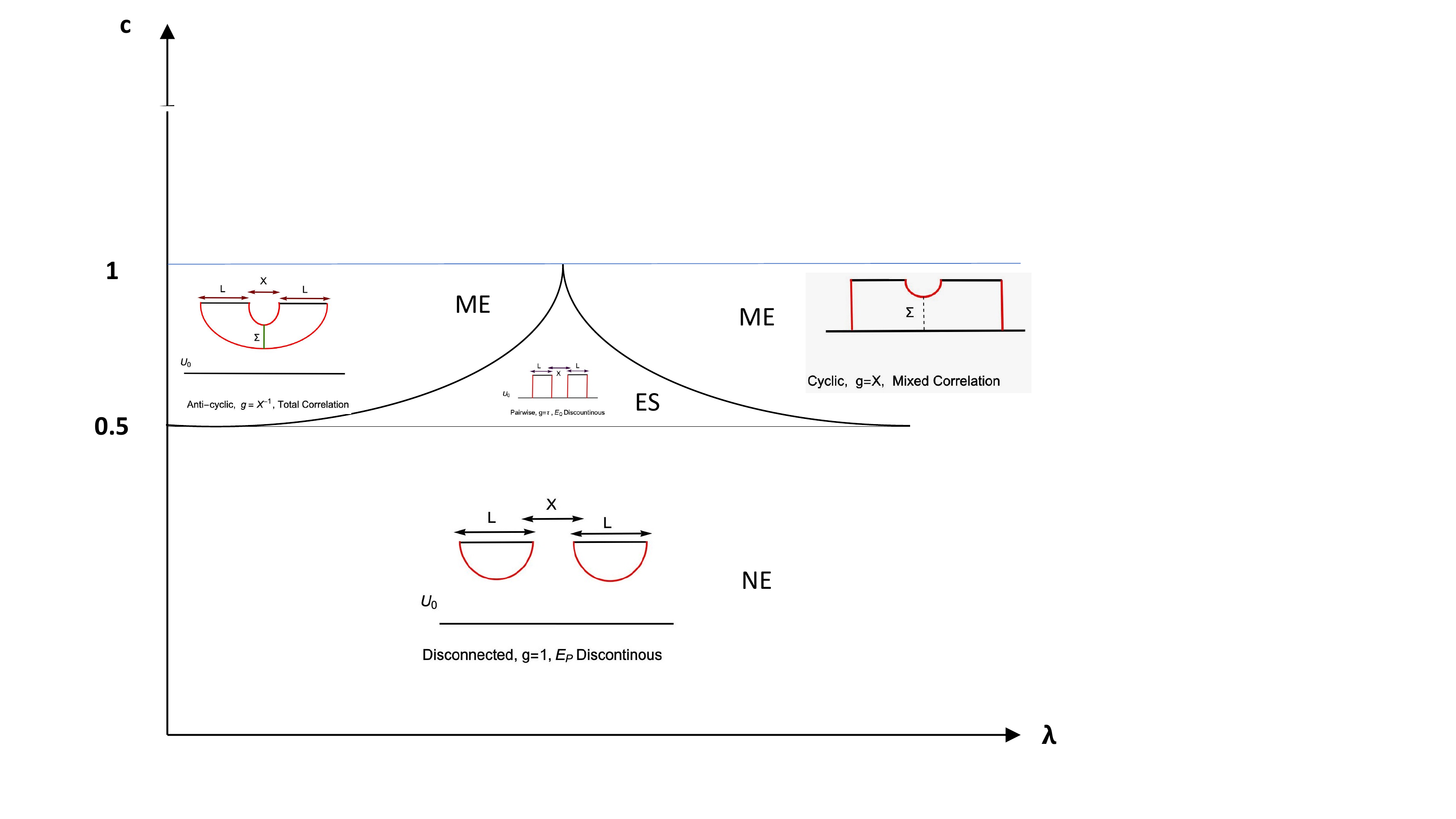}  
\caption{Phase diagram using negativity, found in \cite{Vardhan:2021npf,Shapourian:2020mkc}. }
\label{fig:shapur}
\end{center}
\end{figure}

So the main point is that the similarities and correspondences that is being observed in these phase diagrams point to the universalities in the entanglement structures which shows itself in different measures of mixed correlations and the resulting phase structures.

It should also be noted that the phase transitions between any two of these phases are not sharp, but rather similar to black hole evaporation, in each scenario first islands of other entanglement configurations would appear in each phase region, which then would make that specific phase transition rather smooth, but still the dominant saddle at each phase is the one that we specified in each phase diagram here.

In fact in other works such as \cite{Akers:2021pvd}, it has been shown that for other measures such as $(m,n)$-R\'enyi reflected entropy, there are also some significant non-perturbative effects coming from new other saddles. This fact gives further hints that in each phase with a specific dominant saddle, there are still islands of other entanglement structures and saddles which could get significant non-perturbative effects and then become dominant in different limits, for instance, in the limit of large $m$ or $n$. So, this way, one could find the corresponding  ``Page curve" for any of these specific measures.

One could use other probes of phase structures as well. For instance, in \cite{Jeong:2022zea}, using the entanglement density of strip-subsystems, the entanglement structure of systems where the $U(1)$ or translational symmetry is spontaneously broken is studied. Specifically, there, for classifications and finding the universalities, the area law or the first law of the entanglement entropy have been used. In particular, using the $U(1)$ symmetry breaking pattern, the normal or superconducting phases have been classified. Using one strip with the size of $\ell$, it has been found that, in $(d+1)$ dimensions, the entanglement entropy get a $\ell^d \log \ell$ contribution from the Fermi surface, $\log (\ell^d)$ from the Goldstone bosons and $\ell$-dependent behavior from the topologically ordered degrees of freedom. The system size $\ell$ would be related for instance to $k$ in the study of negativity of black hole radiation.

Another interesting toy model for detecting black hole phases would be moving mirrors \cite{Akal:2022qei}, which can also point to the universalities mentioned before as well. In \cite{Akal:2022qei}, moving mirror models in $2d$ CFT have been studied where based on their late time behaviors, they also found ``four" classes, namely, the type A or timelike mirrors, type B or escaping mirrors, type C or chasing mirrors and type D or terminated mirrors, where each class has its own specific characteristics for the energy stress tensor and entanglement entropy, which essentially can be derived from inspecting the ``endpoints" of these mirrors. Also, the dual of these moving mirrors would be the end of the world branes with different profile structures. We propose here that these ``four" categories of moving mirrors are essentially the four structures of mixed correlation observed in confining backgrounds as in \cite{Jain:2020rbb, Ghodrati:2021ozc} or the structures of replica wormholes observed by studying the logarithmic negativity in \cite{Dong:2021oad}. Even in the dynamical setups of \cite{Haehl:2022uop}, for the localized shockwaves inside the black hole, the four regimes have been observed and their dual quantum circuits model have been constructed. Now the the exact correspondence between each categories (and also the subcategories) of these models can be detected. 

Based on the symmetry structures and boundary conditions, we expect that the type A or the timelike mirror corresponds to the disconnected, $g=1$ or $E_P$ discontinuous regime shown in figure \ref{fig:fphasesNeg}. Then, the type B or the escaping mirror would correspond to the Anti-cyclic, $g=X^{-1}$ case with the total correlation regime. Type C or the chasing mirror would correspond to Cyclic, $g=X$ with Mixed correlation regime and finally type D or terminated mirror would correspond to the Pairwise, $g=\tau$ or $E_Q$ discontinuous regime. The correspondences between the subcategories then similar to what have been found in \cite{Vardhan:2021npf,Shapourian:2020mkc} could also be associated to those found in \cite{Akal:2022qei}.

By tracing the similarities between these models and categories, it could be noticed that the case of black hole radiation where subregions are correlated would be related to type B or escaping mirrors and therefore mostly have the anti-cyclic $g=X^{-1}$ or the "Total correlation" saddle. For the case where particles get created and then evaporated, the process could be modeled by the kink mirrors which is related to a subcategory of type A or timelike mirrors which the essentially is the two disconnected wedges, with $g=1$ and $E_P$ discontinuous. 

When the mirrors move very fast, i.e, $p'(u_{\text{end}}) =\infty$ and $p(u_{\text{end}} )= \infty$, creating the chasing mirror category, where the degrees of freedom get accumulated,  a connected wedge would be constructed, and the cyclic mixed correlation saddle with $g=X$ would come up. Lastly, the terminated mirror where the mirror trajectory would terminated at a particular point in the bulk spacetime, the null points would appear which is related to projection/preparation of direct product state, creating the Pairwise case with two disconnected bulk wedges reaching to the end point of bulk with $p(u_{\text{end}} ) =v_{\text{end}}$ and $p'(u_{\text{end} } ) = \infty/0$.

Another evidence for all of the correspondences between the saddles we mentioned above come from the analysis of the energy flux.  As found in \cite{Akal:2022qei}, for type A and B, the energy flux would be finite and these two correspond to cases where the bulk wedge does not reach to the end wall, and therefore they are related to the disconnected and anti-cyclic saddles correspondingly.

It worths to mention here that, in \cite{Agrawal:2021nkw}, then the optimized correlation measures in two-dimensional thermal states which are dual to spacetimes containing black holes have been studied. These measures were EoP, Q-correlation, R-correlation and squashed entanglement, and they probed the parameter space for the phase diagrams. They proposed that the ``Q-correlation" would have the richest behavior, so it would be interesting to use this measure in the setup of \cite{Verheijden:2021yrb}, to analyze the phase diagram of $2d$ spacetime from the perspective of $3d$ BTZ.

\section{Entanglement structure before and after the Page time: the view from $1d$ higher}\label{sec:calcb}

In this section the entanglement structures of radiation and bath in the two-dimensional model of Verlinde and Verheijden (VV) \cite{Verheijden:2021yrb} and for the cases before and after the Page time are examined.

\begin{figure}[ht!]   
\begin{center}
\hspace*{0cm}\includegraphics[width=0.36\textwidth]{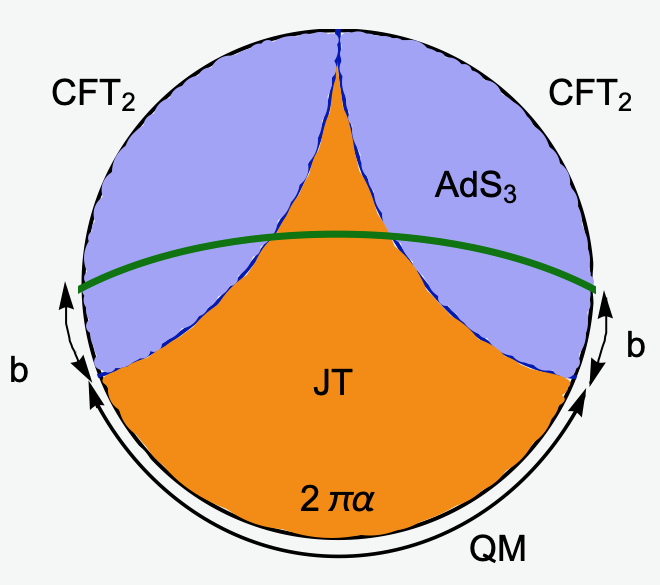}  
\caption{The definition of the parameter $\alpha$ and $b$ in the equation \ref{eq:Sbefore} and \ref{eq:Safter}. }
\label{fig:shapur}
\end{center}
\end{figure}

For the setup of VV, from the length of the geodesic, the entanglement entropy before the Page time has been found as
\begin{gather}\label{eq:Sbefore}
S=\frac{1}{4G}\left ( 2 \log \sinh \frac{\pi}{\beta} ( 2 \pi \ell ( 1 - \alpha ) - 2b ) \right ),
\end{gather}
and then after the Page time, the geodesic would ``jump'' and so the entropy would become
\begin{gather}\label{eq:Safter}
S=\frac{1}{4G} \left ( 2 \log \sinh \frac{\pi} {\beta} ( 2\pi \ell \alpha + 2b ) \right ).
\end{gather}

The parameter $\alpha$ determines the partial dimension reduction in the angular $\varphi$- direction where $\alpha \in (0, 1\rbrack$ defining the angle $2\pi \alpha$ which covers the JT part and the rest defines the bath part.

Now instead of one interval, which corresponds to one angle $2 \pi \alpha$ in the relations \ref{eq:Sbefore} or \ref{eq:Safter}, we take two intervals with angles $\mu$ and with distance $\nu$ among them. Totally, we have the angular distance $2\pi\alpha= 2\mu+\nu$ for the whole angular interval. Then, we can use the mutual information relation as
\begin{gather}\label{eq:MI}
I( A:B) = S(\rho_A) + S (\rho_B) - S ( \rho_{AB}),
\end{gather}
which for our case would be $S_A=S_B= S |_\mu$, and $S_{AB}= S(2\mu+\nu) + S(\nu)$. So the results for these two cases would be the dynamical extension of the work of \cite{Shapourian:2020mkc} to the setup of \cite{Verheijden:2021yrb}.

Using our diagrams, we will show that before the Page time, the critical size of the bath $b_c$ where the phase transition happens is much bigger than both the system size $\mu$ and $\nu$, and also their sum $\mu+\nu$, which corresponds to the case where $q= \frac{L_B}{L_A}  \gg 1$ in \cite{Shapourian:2020mkc}. After the Page time, the critical size of bath would be very small, and therefore corresponds to the case of  $q= \frac{L_B}{L_A}  \ll 1$ studied in \cite{Shapourian:2020mkc}, where $L_B$ is the size of JT which here is $2 \pi \alpha$ and $L_A=1- 2\pi \alpha$.

\subsection{Before the Page time}
From the relation for the mutual information \ref{eq:MI} and entanglement entropies \ref{eq:Sbefore} and \ref{eq:Safter}, for the case before the Page time, we find

\begin{gather}
\sinh^2 \frac{\pi}{\beta} \Big ( 2\pi \ell (1 -\mu) -2b_c \Big) = \sinh \frac{\pi}{\beta} \Big ( 2 \pi \ell ( 1 - \nu ) - 2b_c \Big ) \sinh \frac{\pi} {\beta} \Big ( 2 \pi \ell ( 1 - 2 \mu - \nu ) - 2b_c \Big ),
\end{gather}
which has four solutions as
\begin{flalign}\label{eq:bcbefore}
b_c(1) & \to   \frac{\beta}{4\pi} \left( \log (2)-4 i \pi  c_1   -2 \log \left(-\sqrt{\frac{e^{-\frac{4 \pi ^2 l}{\beta }} \left(\xi  e^{\frac{2 \pi ^2 l (4 \mu +\nu )}{\beta }}+2 e^{\frac{4 \pi ^2 l \mu }{\beta }}-e^{\frac{8 \pi ^2 l \mu }{\beta }}-1\right)}{1-e^{-\frac{4 \pi ^2 l \nu }{\beta }}}}\right)\right),   \\
b_c(2) & \to  \frac{\beta}{4\pi} \left (  \log (2)-4 i \pi  c_1  -\log \left(\frac{e^{-\frac{4 \pi ^2 l}{\beta }} \left(\xi  e^{\frac{2 \pi ^2 l (4 \mu +\nu )}{\beta }}+2 e^{\frac{4 \pi ^2 l \mu }{\beta }}-e^{\frac{8 \pi ^2 l \mu }{\beta }}-1\right)}{1-e^{-\frac{4 \pi ^2 l \nu }{\beta }}}\right) \right ), \\
b_c(3)  & \to \frac{\beta}{4\pi} \left ( \log (2)-4 i \pi  c_1  -2 \log \left(-\sqrt{\frac{e^{-\frac{4 \pi ^2 l}{\beta }} \left(\xi  \left(-e^{\frac{2 \pi ^2 l (4 \mu +\nu )}{\beta }}\right)+2 e^{\frac{4 \pi ^2 l \mu }{\beta }}-e^{\frac{8 \pi ^2 l \mu }{\beta }}-1\right)}{1-e^{-\frac{4 \pi ^2 l \nu }{\beta }}}}\right)  \right ), \\
b_c(4) & \to \frac{\beta}{4\pi} \left (  \log (2)-4 i \pi  c_1 -\log \left(\frac{e^{-\frac{4 \pi ^2 l}{\beta }} \left(\xi  \left(-e^{\frac{2 \pi ^2 l (4 \mu +\nu )}{\beta }}\right)+2 e^{\frac{4 \pi ^2 l \mu }{\beta }}-e^{\frac{8 \pi ^2 l \mu }{\beta }}-1\right)}{1-e^{-\frac{4 \pi ^2 l \nu }{\beta }}}\right) \right ),
\end{flalign}
where in the above relation $\xi$ is
\begin{flalign}
\xi & =\sqrt{e^{-\frac{8 \pi ^2 l (2 \mu +\nu )}{\beta }} \left(-4 e^{\frac{4 \pi ^2 l (\mu +\nu )}{\beta }}+4 e^{\frac{8 \pi ^2 l (\mu +\nu )}{\beta }}-2 e^{\frac{4 \pi ^2 l (2 \mu +\nu )}{\beta }}-4 e^{\frac{4 \pi ^2 l (3 \mu +\nu )}{\beta }}+e^{\frac{4 \pi ^2 l (4 \mu +\nu )}{\beta }}+4 e^{\frac{8 \pi ^2 l \mu }{\beta }}+e^{\frac{4 \pi ^2 l \nu }{\beta }}\right)}.
\end{flalign}

\begin{figure}[ht!]
 \centering
  \includegraphics[width=5 cm] {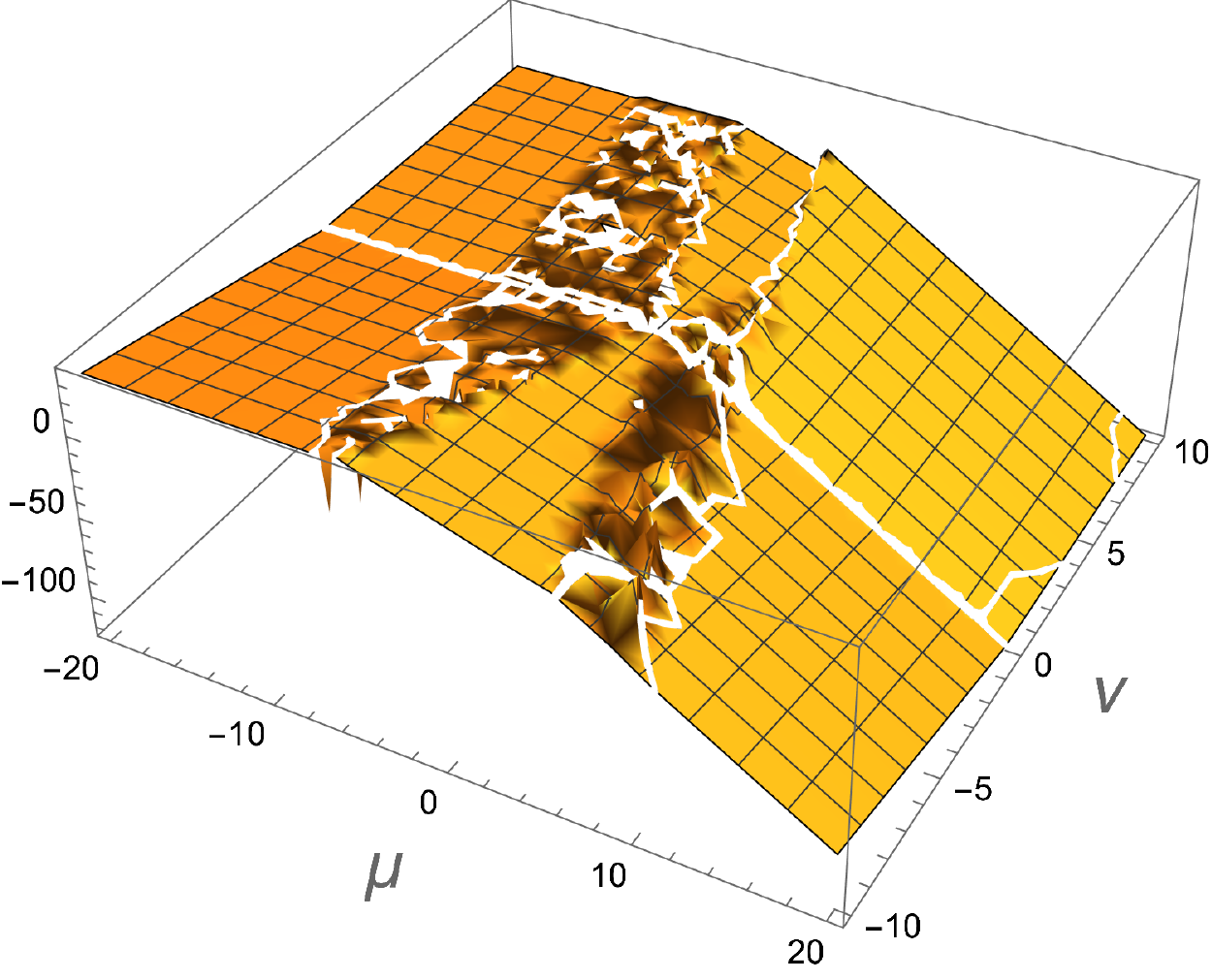}  \ \ \ \ \ \ \ \ \ \ \ \ \ 
    \includegraphics[width=5 cm] {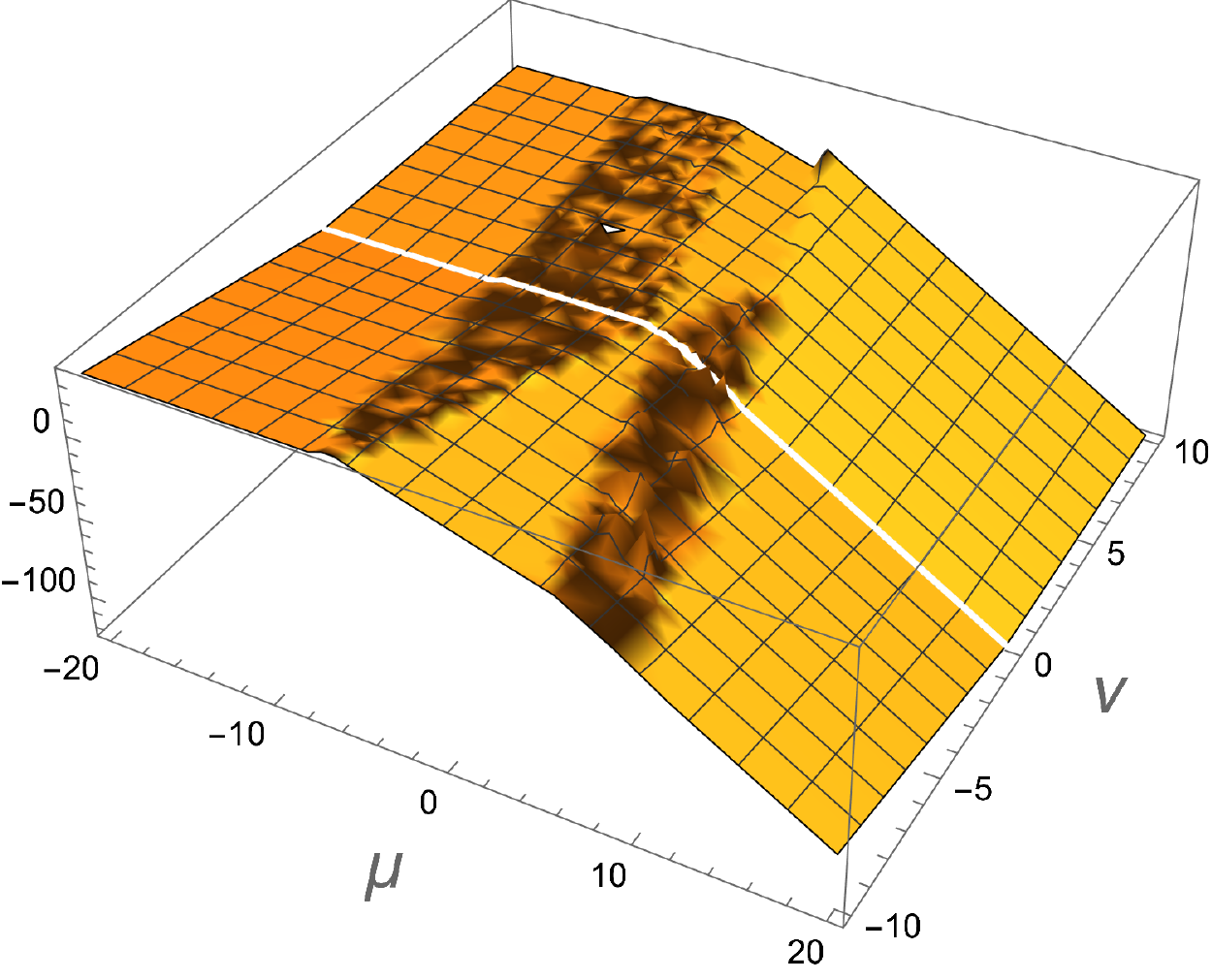} 
        \includegraphics[width=5 cm] {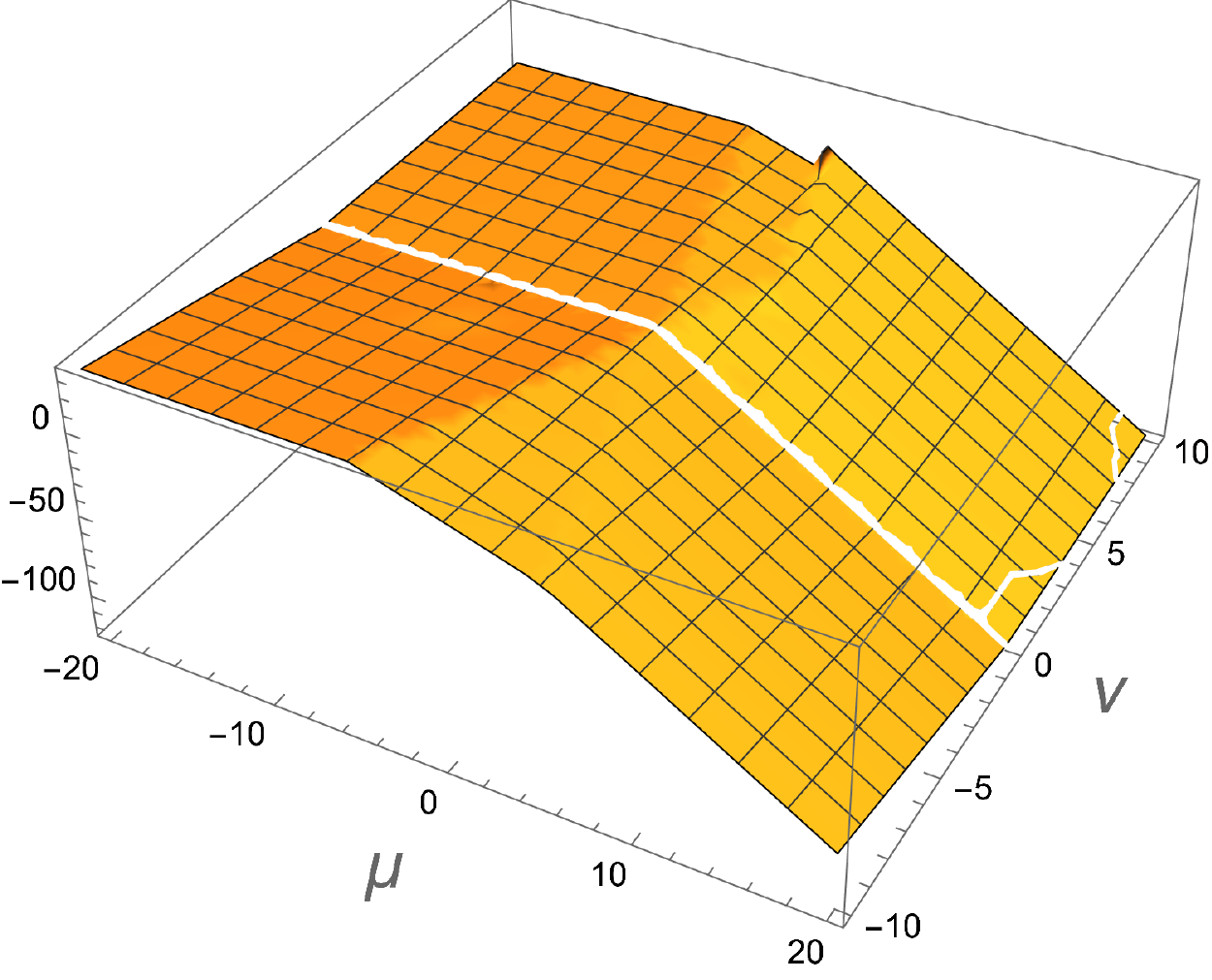}  \ \ \ \ \ \ \ \ \ \ \ \ \ 
            \includegraphics[width=5 cm] {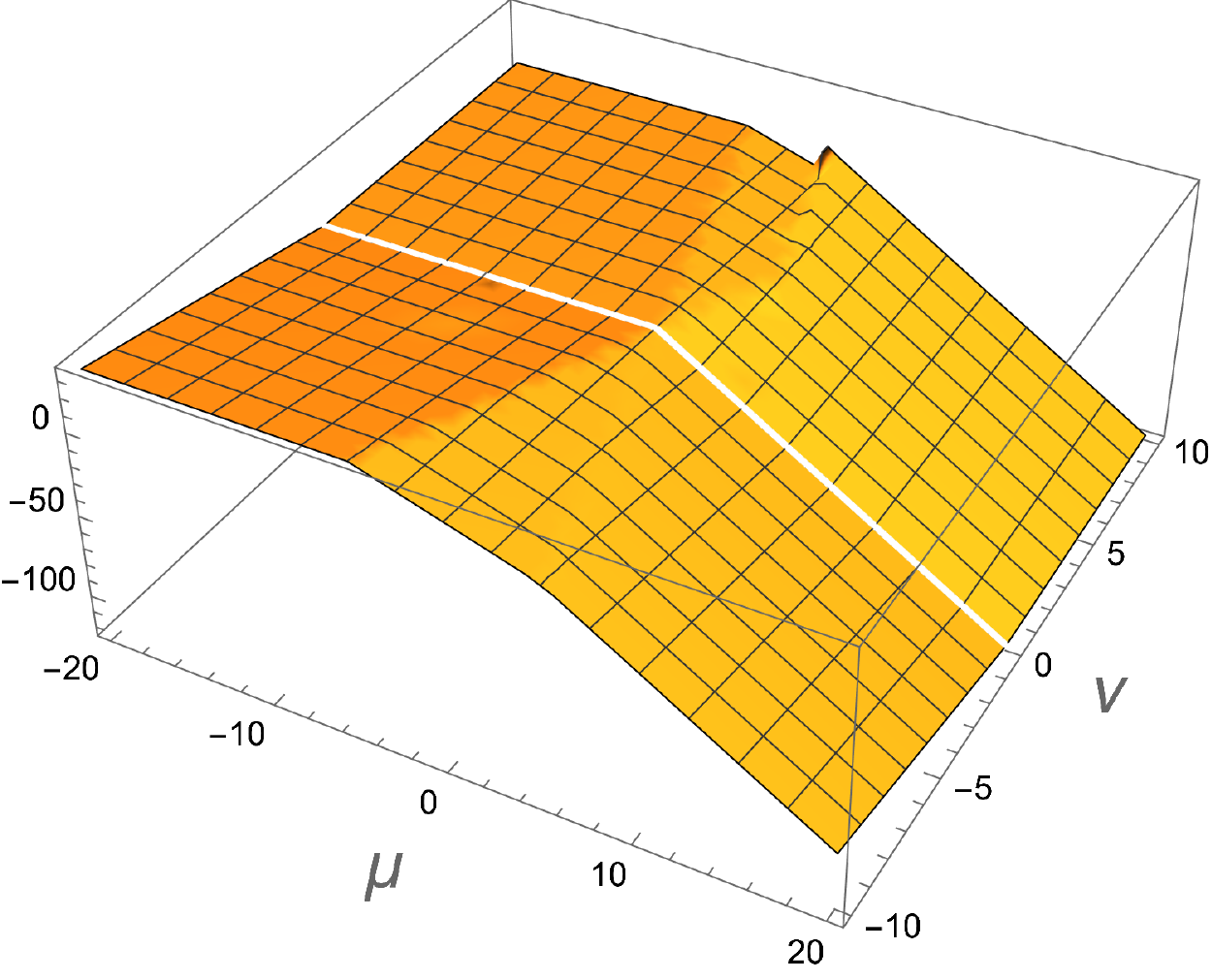} 
  \caption{The plots of solution of $b_c$ \textbf{before} the Page time. From left to right, they are the plot of $b_c(1)$, $b_c(2)$, $b_c(3)$ and $b_c(4)$. They show the behavior of the critical bath size $b_c$, in terms of the sizes of the two mixed system, i.e,  $\mu$ and $\nu$. Here $\beta$ and $l$ are set to one.}
 \label{fig:bcbeforePage}
\end{figure}

The real part of these four solutions are shown in figure \ref{fig:bcbeforePage}. It can be seen that their phase space become smoother from $b_c(1)$ to $b_c(4)$.

\begin{figure}[ht!]
 \centering
  \includegraphics[width=5.2cm] {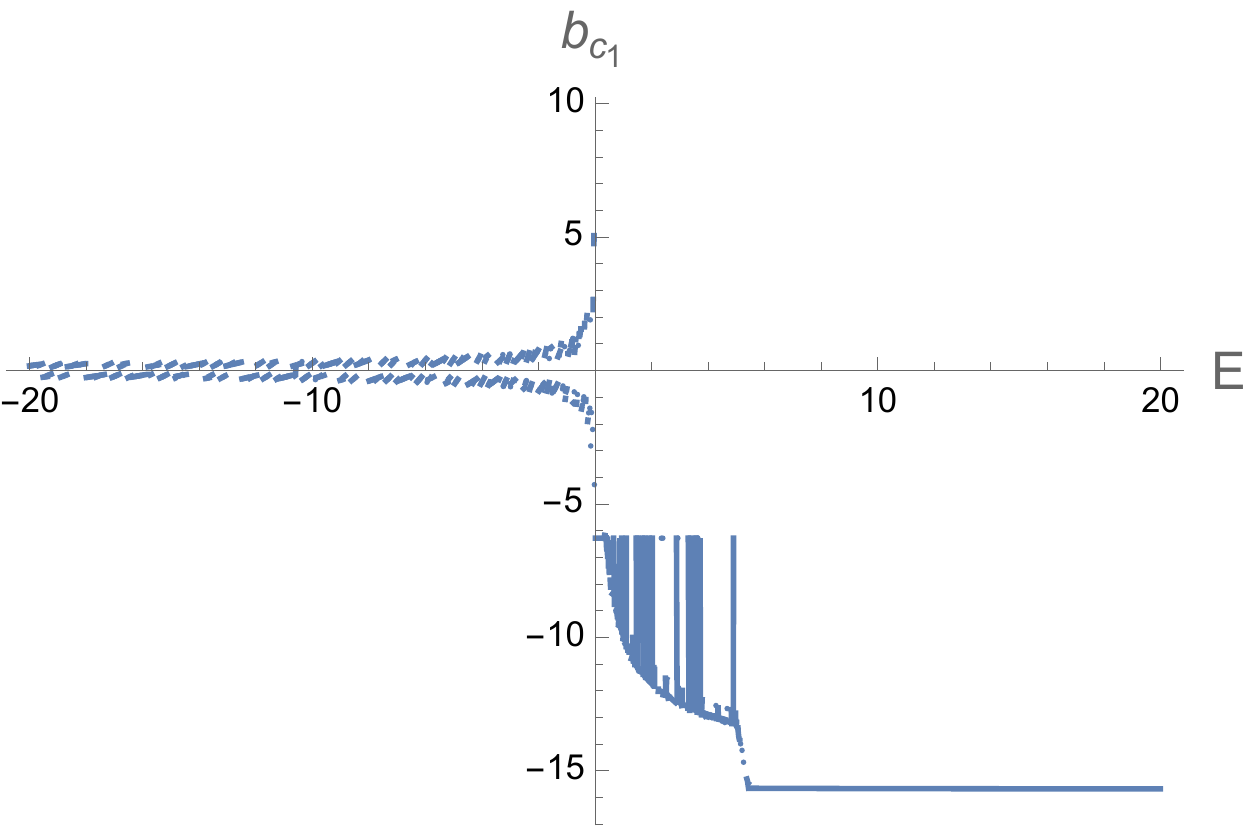} \ \ \ \ \ \ \ \ \ \ \ \ \ 
    \includegraphics[width=5.2 cm] {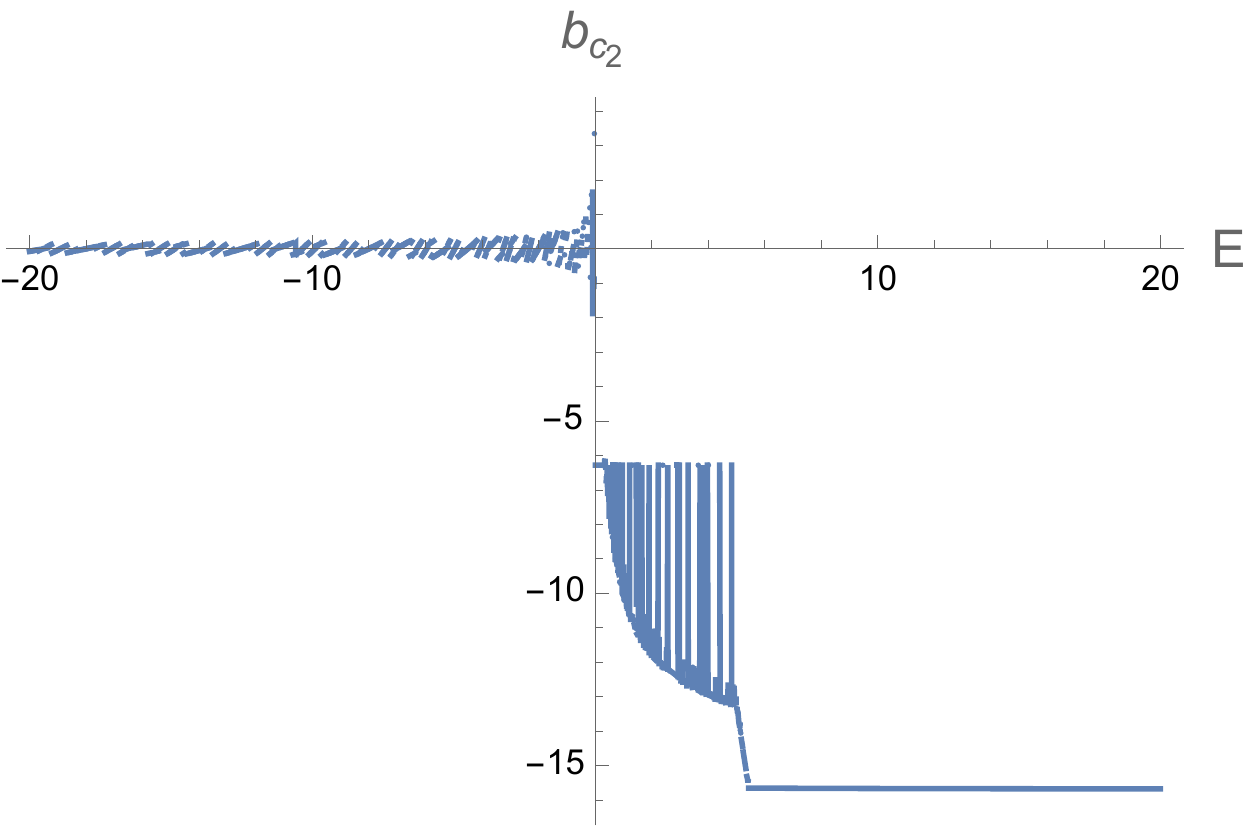} 
        \includegraphics[width=5.2 cm] {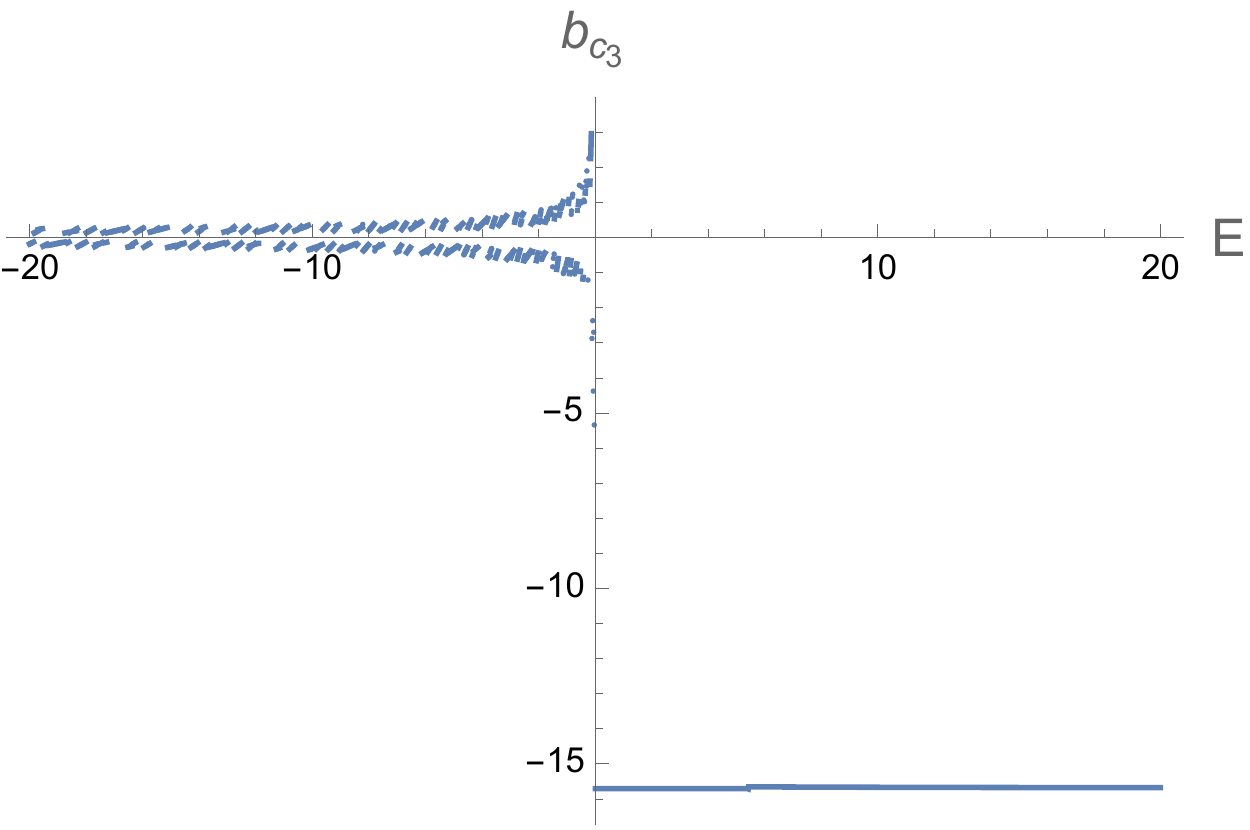} \ \ \ \ \ \ \ \ \ \ \ \ \ 
            \includegraphics[width=5.2 cm] {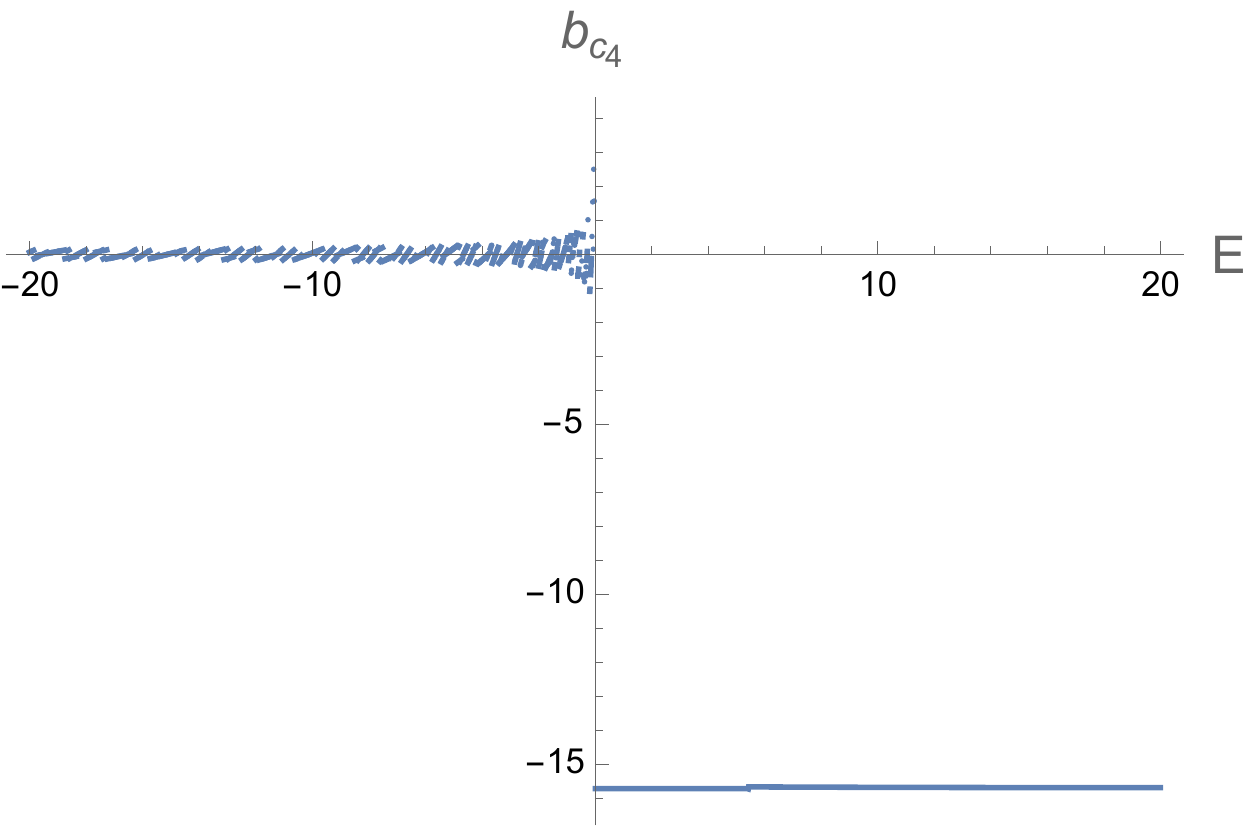} 
  \caption{The plots of solution of $b_c$ \textbf{before} the Page time versus energy, $E$. Here we set $\mu=\nu=3$, $l=1$ and $\kappa=2$. }
 \label{fig:bcEbeforePage}
\end{figure}

If we replace $\beta$ in the above relations \ref{eq:bcbefore} with $\beta= \pi / \sqrt{ \kappa E}$ which then leads to the temperature $T_H= \frac{1}{\pi} \sqrt{\frac{8 \pi G E}{ 2 \Phi_r} }$ \cite{Verheijden:2021yrb}, we get the plots shown in figure \ref{fig:bcEbeforePage} which show the behavior of $b_c$ versus $E$ in the four regimes. 

\begin{figure}[ht!]
 \centering
  \includegraphics[width=5.2cm] {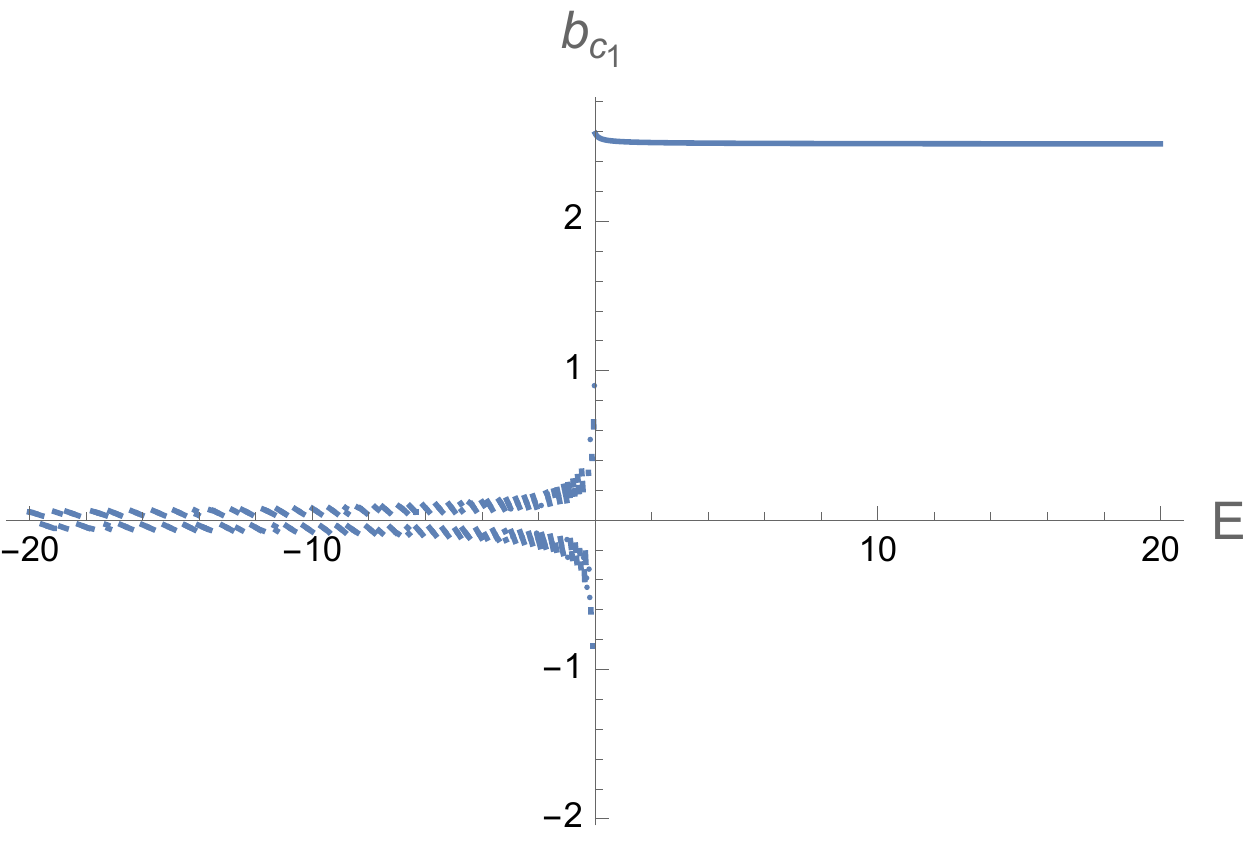}  \ \ \ \ \ \ \ \ \ \ \ \ 
    \includegraphics[width=5.2 cm] {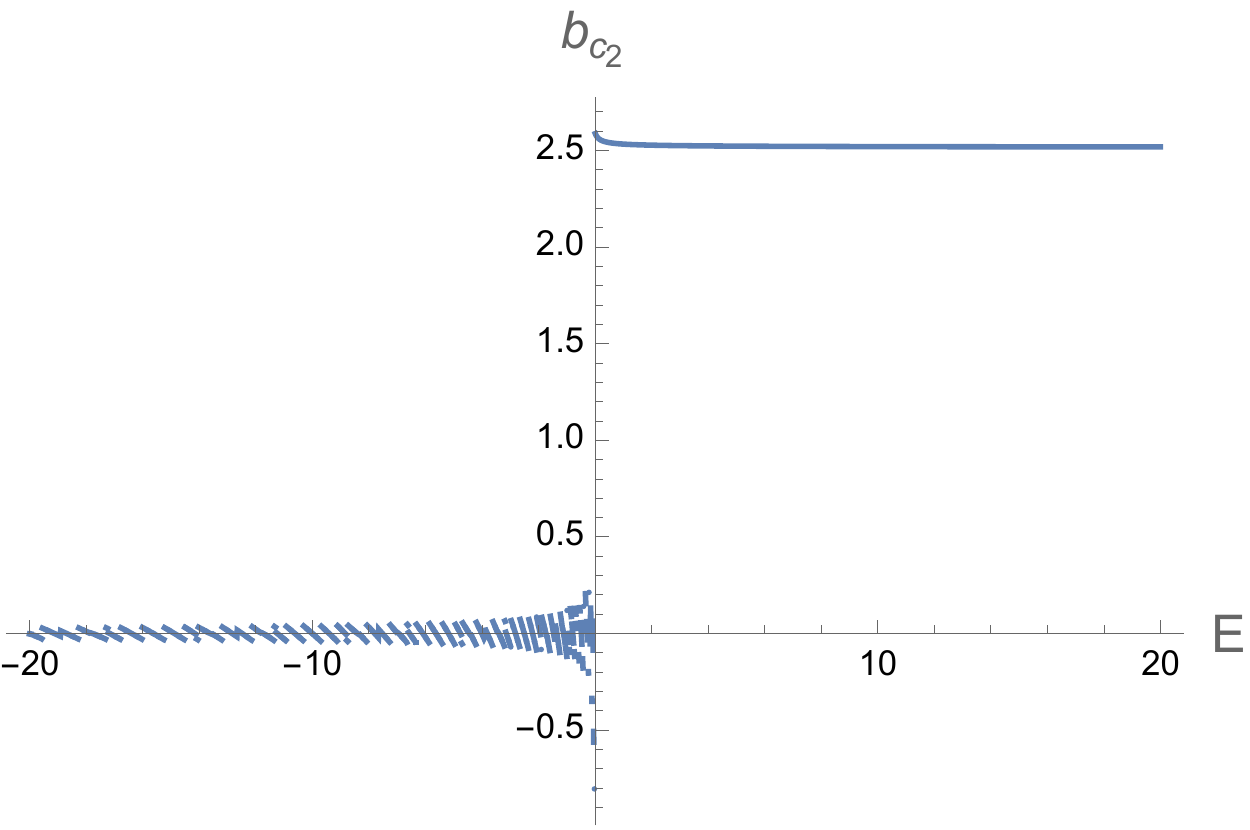} 
        \includegraphics[width=5.2 cm] {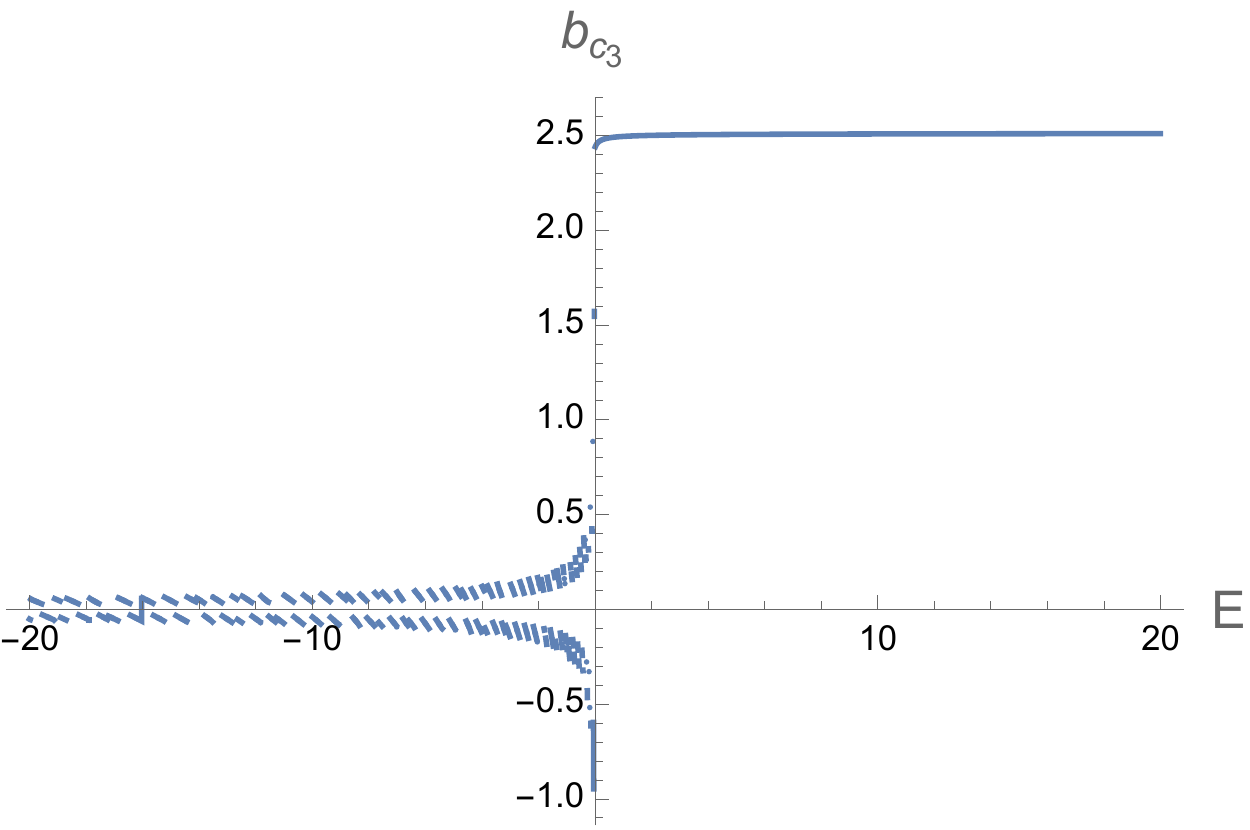}  \ \ \ \ \ \ \ \ \ \ \ \ 
            \includegraphics[width=5.2 cm] {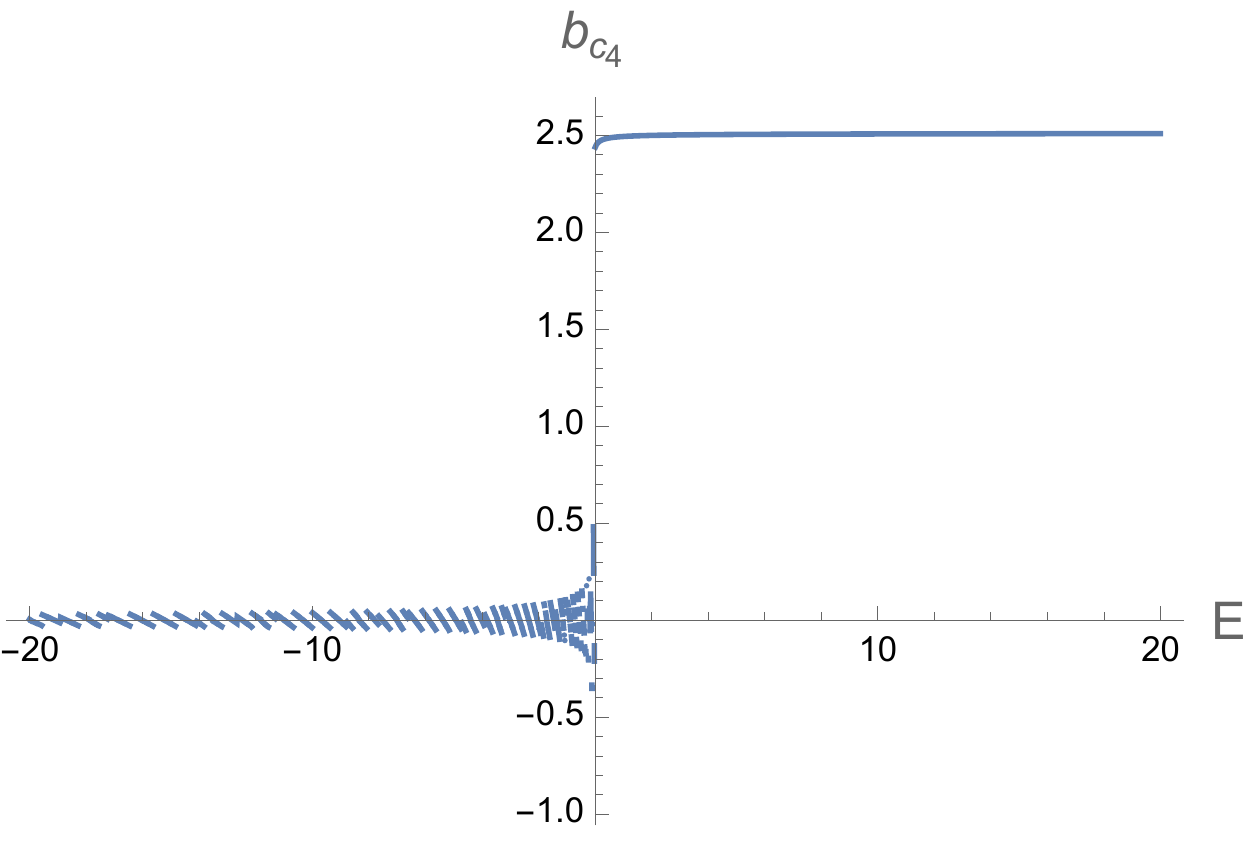} 
  \caption{The plots of solution of $b_c$ \textbf{before} the Page time versus energy, $E$. Here we set $\mu=0.1$, $\nu=0.2$,  $l=1$ and $\kappa=40$. }
 \label{fig:bcEbeforePage2}
\end{figure}

If we insist on getting positive $b_c$s, we need to tune the parameters, like increasing $\kappa$ and decreasing the intervals $\mu$ and $\nu$, which then the results are shown in figure \ref{fig:bcEbeforePage2}, where for positive $E$ we get an almost constant value for $b_c$ but the initial behavior for small values of $E$ looks different.

\begin{figure}[ht!]
 \centering
  \includegraphics[width=5.2cm] {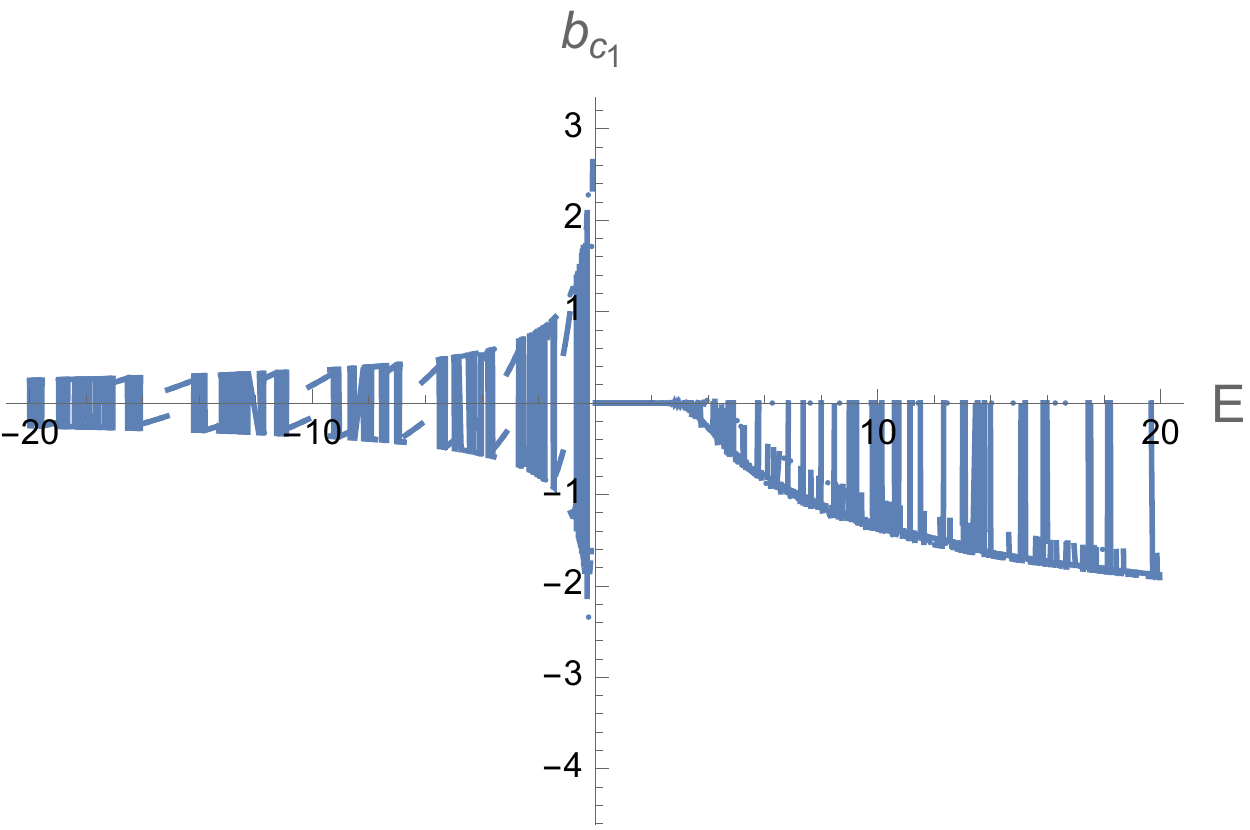}  \ \ \ \ \ \ \ \ \ \ \ \ 
    \includegraphics[width=5.2 cm] {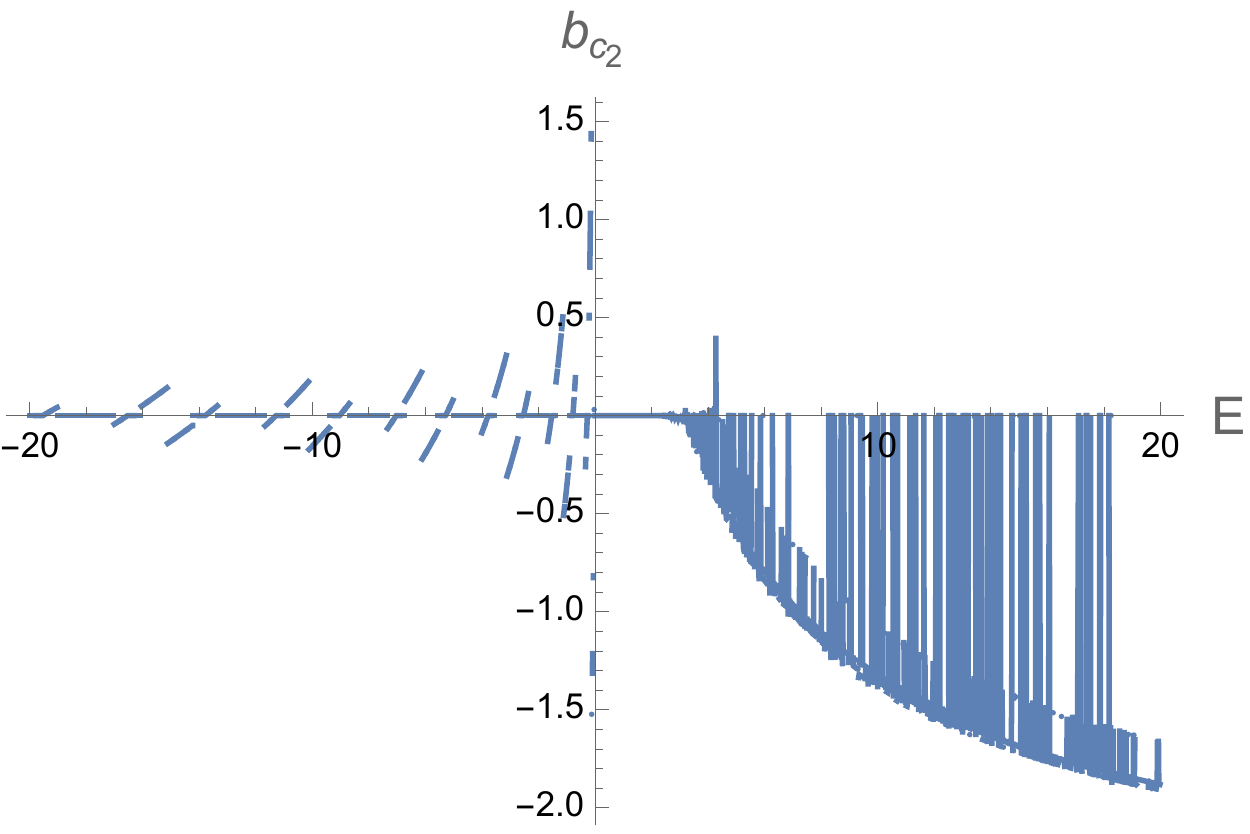} 
        \includegraphics[width=5.2 cm] {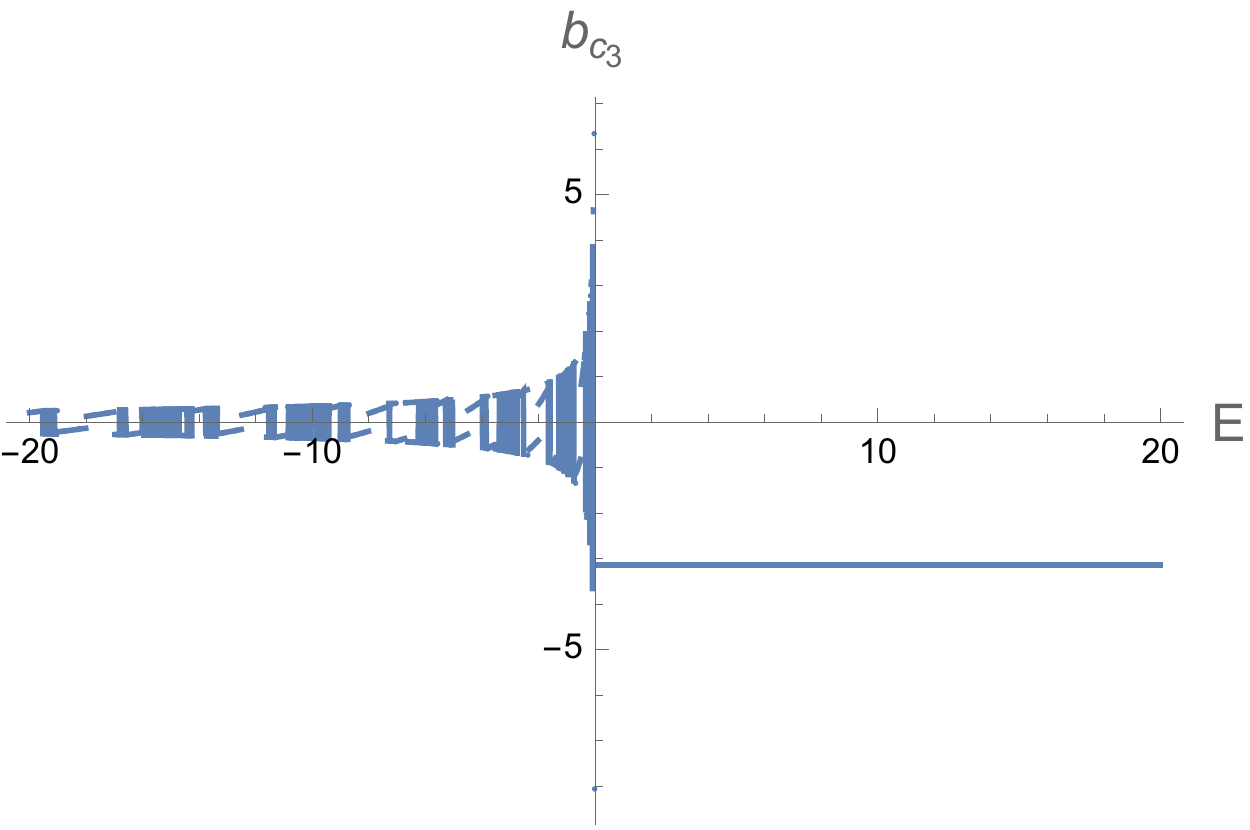}  \ \ \ \ \ \ \ \ \ \ \ \ 
            \includegraphics[width=5.2 cm] {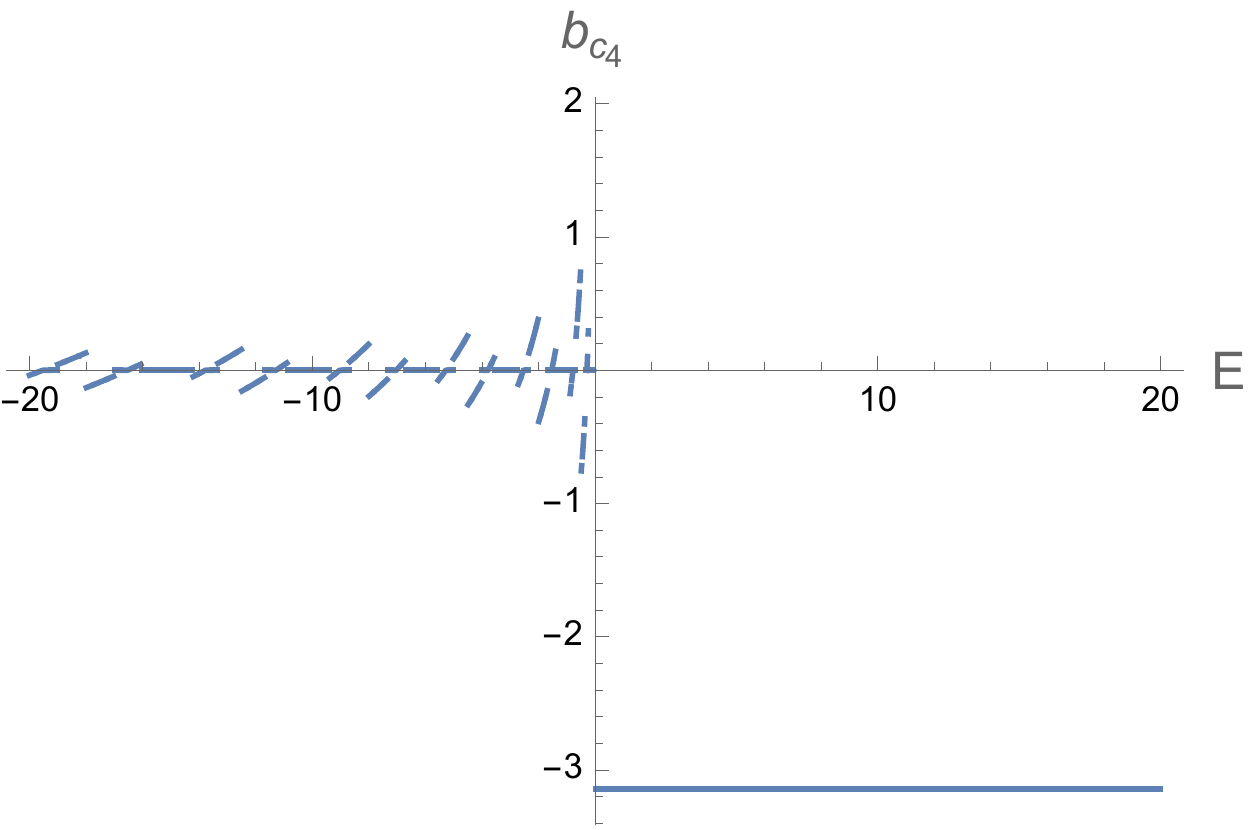} 
  \caption{The plots of solution of $b_c$ \textbf{before} the Page time versus energy, $E$. Here we set $\mu=\nu=1$,  $l=1$ and $\kappa=3$. }
 \label{fig:bcEbeforePage3}
\end{figure}

For the case of $\mu=\nu$ and by tuning $\kappa$ we can get another specific behavior which is shown in figure \ref{fig:bcEbeforePage3}.

\begin{figure}[ht!]
 \centering
  \includegraphics[width=6.5cm] {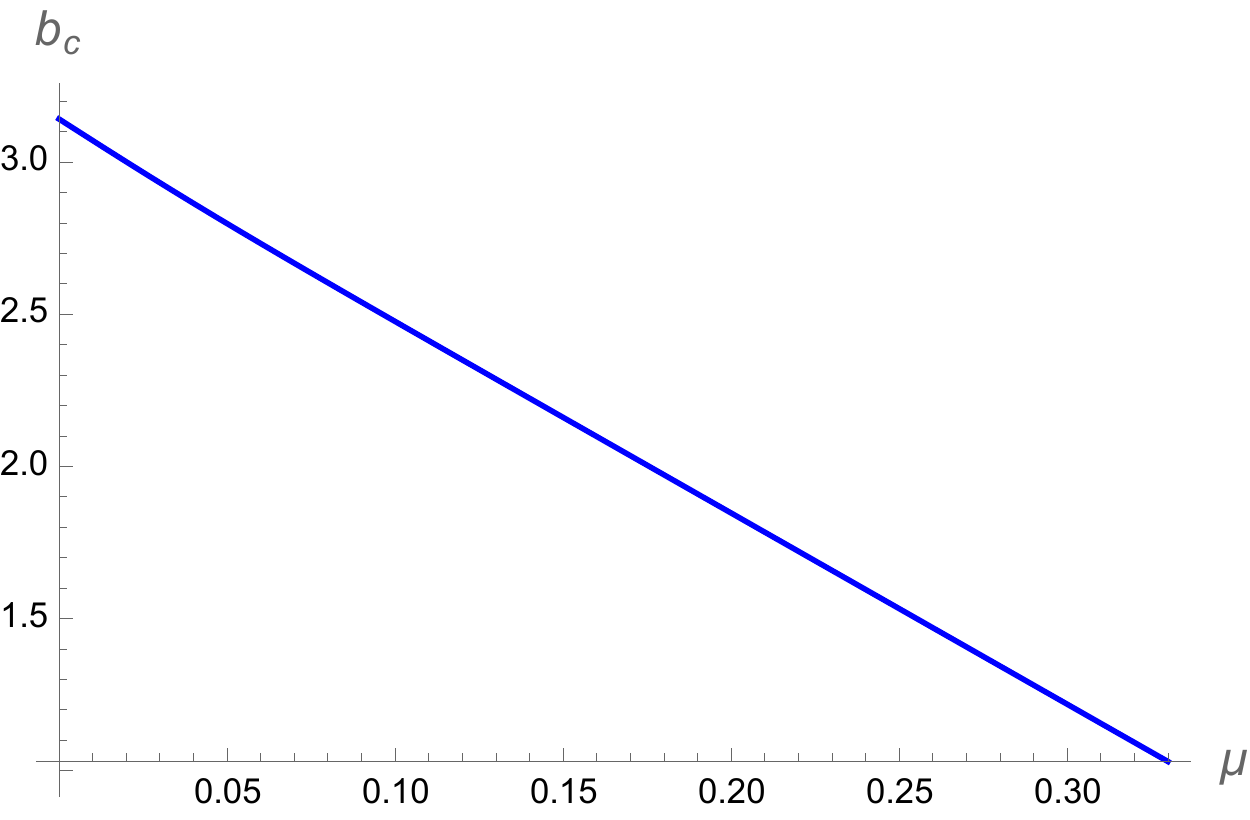} 
  \caption{The behavior of $b_c$ before the Page time versus the $\mu$, while $\nu=2\mu$.}
 \label{fig:bclinear}
\end{figure}

As expected from the Page curve and the results coming from negativity found in \cite{Shapourian:2020mkc}, for cases where the regions of $\mu$ and $\nu$ are very big, and also their ratio $\mu/\nu$ is big as in figure \ref{fig:bclinear}, the behavior would have a linearly decreasing function. However, when their sizes or the ratio between them is very small, the behavior is different. This difference is due to the fact that the mutual information relative to negativity, overestimates the contribution of classical correlations, which specifically for smaller size would have more significant effects. For bigger sizes, the effect of this difference would be negligible.

 \begin{figure}[ht!]
 \centering
  \includegraphics[width=6.5cm] {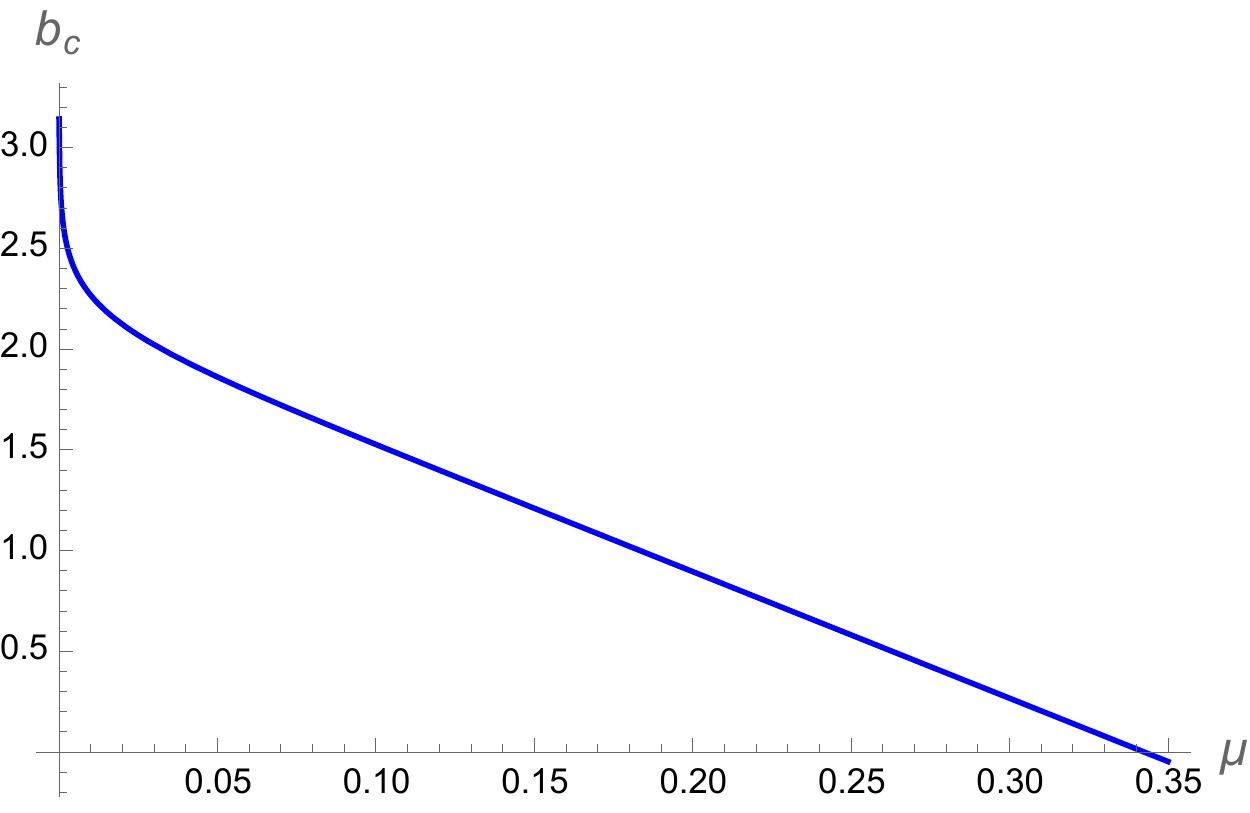}  \ \ \ \ \ 
    \includegraphics[width=6.5cm] {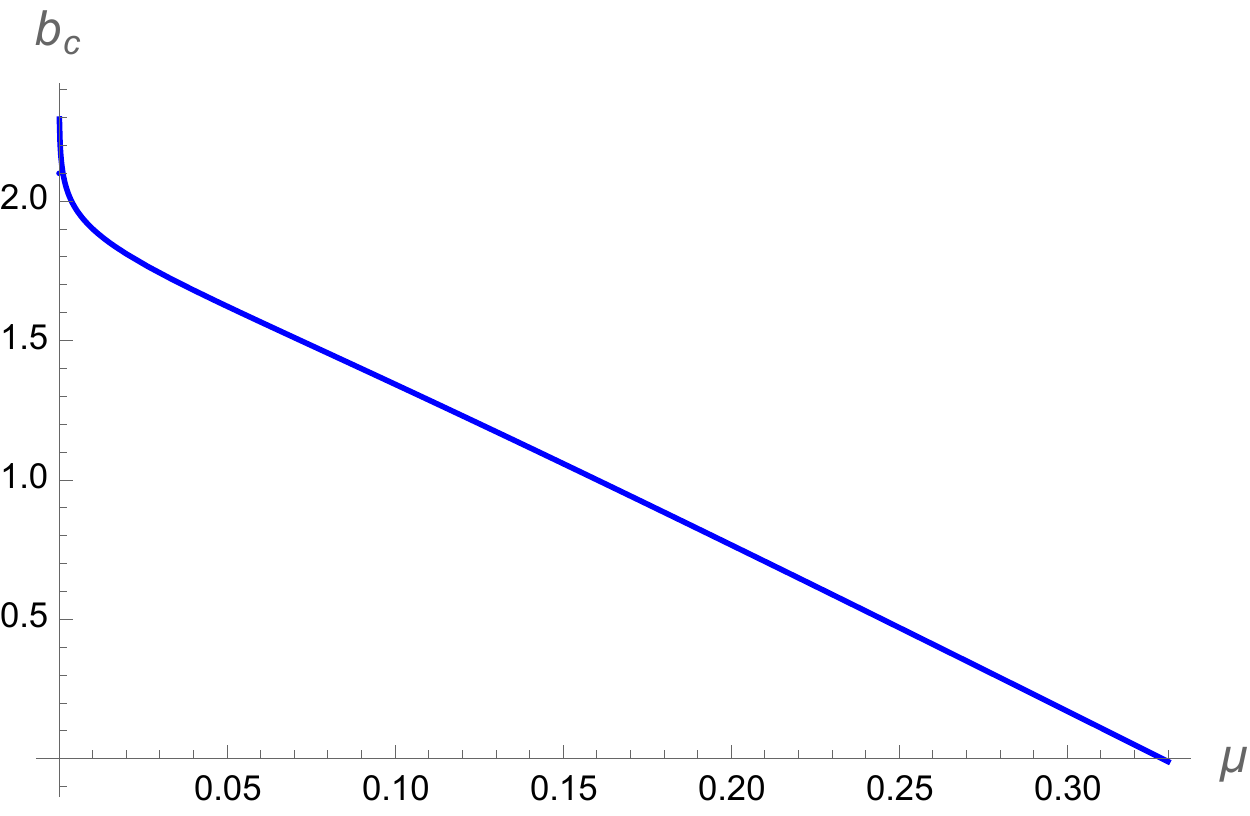} 
  \caption{The behavior of $b_c$ before the Page time, versus $\mu$, for very small size of $\nu=0.0000001$, in the left, and for the case where $\nu=0.0000001\mu$, in the right.}
 \label{fig:bcsmall}
\end{figure}

If we choose a small size for the subsystem angular size, $\nu$, then as shown in figure \ref{fig:bcsmall}, the critical size of the bath $b_c$ would decrease by increasing the size of the system $\mu$. By increasing the size of $\nu$, then the behavior of dropping of the critical bath size $b_c$ versus $\mu$, would change to a linear decreasing function.

In general, from these plots, one could find that \textit{before} the Page time, in order to get a vanishing mutual information and to get the disconnected phases, the critical size of the bath should be much bigger than the two system size $\mu$ and $\nu$, while as we see from figure \ref{fig:bcpage2}, this would not be true for the case \textit{after} the Page time.

So the case before the Page time corresponds to the case where  $L_B \gg L_A$ in \cite{Shapourian:2020mkc}, and there is no replica wormhole present in the gravitational interpretation of the random matrix theory \cite{Penington:2019kki}. The pattern of the dominant saddles could then be found by understanding the corresponding diagrams of \cite{Shapourian:2020mkc} in each phase.  For the case before the Page time, the dominant diagram for $ \langle \text{Tr} \rho^n \rangle = L_A^{1-n} \propto \mu^{1-n} $ is shown in equation 4.10 of \cite{Shapourian:2020mkc}. This case corresponds to the positive partial transpose (PPT) state with a vanishing logarithmic negativity (LN) as shown in the upper region of the phase diagram, region $I$ of figure. 2 in \cite{Shapourian:2020mkc}, which has a semicircular distribution.

Again it worths to mention here that as observed in \cite{Demulder:2022aij}, the mass of graviton which play a significant role in the model of Hawking radiation would depend on the angle between the island and the boundary. This angle is the parameter that the partial reduction is being performed over in \cite{Verheijden:2021yrb} which determines the phase and the Page curve.

In the model of \cite{Demulder:2022aij}, where its metric can been written as
\begin{gather}
ds_{10}^2 = L_4^2 ds^2_{\text{AdS}_4 }+f_1^2 ds^2_{S_1} +f_2^2 ds^2_{S_2} + 4 \rho^2 dz d\bar{z},
\end{gather}
where the warp factors are
\begin{gather}
L_4^8= 16 \frac{N_1 N_2}{W^2}, \ \ \ \ \rho^8= \frac{N_1 N_2 W^2} { h^4 \hat{h}^4}, \ \ \ f_1^8= 16 h^8 \frac{N_2 W^2}{N_1^3}, \ \ \ \ f_2^8 = 16 \hat{h}^8 \frac{N_1 W^2}{N_2^3},
\end{gather}
and 
\begin{gather}
W= \partial h \bar{\partial} \hat{h} + \bar{\partial} h \partial \hat{h} = \partial \bar{\partial} (h \hat{h}), \nonumber\\
N_1=2 h \hat{h} | \partial h |^2 - h^2 W,\nonumber\\
N_2 = 2 h \hat{h} | \partial \hat{h}|^2- \hat{h}^2 W,
\end{gather}
and $h$ are related to $\kappa$ as
\begin{gather}
h= - \frac{1}{2} e^z \kappa - N \log \text{th} \left ( \frac{i \pi}{4} - \frac{z}{2} \right ) + c.c\nonumber.\\
\hat{h} = \frac{1}{2} e^z \hat{\kappa} - N \log \text{th} \left( \frac{z}{2} \right ) + c.c,
\end{gather}
the four phases can be analyzed as well.

There, a new quantity has been defined named $\alpha$ as the ratio of $\alpha=N/K$, where $N$ is the number of D5 branes and also NS5 branes and there are $2N K $ semi-infinite D3 branes where also $K$ is a parameter in the function  $h$ and $\hat{h}$  in the metric. These parameters then can lead to the similar  phase structures we observed in the previous section.

The dilaton in fact alway affects the surfaces of the island greatly. Also, in the Karch-Randall brane setup, the mass of graviton $m_g^2$ would depend on the angle between the extremal surface, where the island is located inside the bulk and the boundary of $\text{AdS}_5$, i.e, $\partial(\text{AdS}_5)$. As we saw here, when the angle increases, the critical distance between the two intervals $b_c$ should decrease and this is compatible with the result of \cite{Ghodrati:2019hnn}, as there also when the mass of graviton increased the critical distance between the two subregion $D_c$ would decrease. This is because the mass of graviton breaks diffeomorphism and is a parameter for dissipation in the system, which as it increases,  the correlations diminish more and more, and therefore for keeping the mutual information between the two regions strong enough to be non-zero, the distance between them should decrease.

Also, in \cite{Demulder:2022aij}, it has been shown that there is a critical $\alpha_{\text{crit}}$ and a critical angle $\nu$ and therefore a critical mass for graviton $m_\text{g,crit}$, where below it, islands cannot exist and this is compatible with the results in our next section, where $b_c$ can get saturated in our diagrams there.

In \cite{Demulder:2022aij}, where the problem of island in higher dimension type IIB string theory has also been investigated,  the connections between the properties of the island and mass of graviton has also been explored. In their setup, there are two kinds of extremal surface which are either the Hartman-Maldacena ($S_{HM}$) or the quantum extremal island surface ($S_I$). They found that in the setup of type IIB, when the dilaton varies, the mass of graviton then cannot become arbitrary light. One subtlety worths to mention here is that only the funnel form of black holes can be embedded into $5d$ bulk space while ``droplet"-like solutions cannot be encoded in the $1d$ higher case, which then the contribution for each case in the several models we considered here could be further tracked.

\subsection{After the Page time}

For the case after the Page time, from the relation for the mutual information \ref{eq:MI} and entanglement entropies \ref{eq:Sbefore} and \ref{eq:Safter}, we can find
\begin{gather}
\sinh^2 \frac{\pi}{\beta} \Big ( 2 \pi \ell \mu + 2b_c \Big ) = \sinh \frac{\pi}{\beta} \Big ( 2\pi \ell ( 2 \mu+\nu) + 2b_c \Big) \sinh \frac{\pi}{\beta} \Big ( 2 \pi \ell \nu + 2b_c \Big), 
\end{gather}
which again at these specific size of the bath, $b_c$, the mutual information between the two intervals would vanish. The four solutions of the above equation are

\begin{flalign} \label{eq:bcafter1}
b_c(1) & \to \frac{\beta}{2\pi} \left ( 2 i \pi  c_1    +  \log \left(-\sqrt{\frac{e^{-\frac{4 \pi ^2 l (2 \mu +\nu )}{\beta }} \left(\left(e^{\frac{4 \pi ^2 l \mu }{\beta }}-1\right)^2 e^{\frac{4 \pi ^2 l \nu }{\beta }}-\eta \right)}{2 \left(e^{\frac{4 \pi ^2 l \nu }{\beta }}-1\right)}}\right) \right), \end{flalign}
\begin{flalign}\label{eq:bcafter2}
b_c(2) & \to \frac{\beta}{4\pi} \left (4 i \pi  c_1+ \log \left(\frac{e^{-\frac{4 \pi ^2 l (2 \mu +\nu )}{\beta }} \left(\left(e^{\frac{4 \pi ^2 l \mu }{\beta }}-1\right)^2 e^{\frac{4 \pi ^2 l \nu }{\beta }}-\eta \right)}{2 \left(e^{\frac{4 \pi ^2 l \nu }{\beta }}-1\right)}\right) \right ), 
\end{flalign}
\begin{flalign}\label{eq:bcafter3}
b_c(3) & \to \frac{\beta}{2\pi} \left ( 2 i \pi  c_1+ \log \left(-\sqrt{\frac{e^{-\frac{4 \pi ^2 l (2 \mu +\nu )}{\beta }} \left(\left(e^{\frac{4 \pi ^2 l \mu }{\beta }}-1\right)^2 e^{\frac{4 \pi ^2 l \nu }{\beta }}+ \eta  \right)}{2 \left(e^{\frac{4 \pi ^2 l \nu }{\beta }}-1\right)}}\right) \right), \end{flalign}
\begin{flalign}
b_c(4) & \to \frac{\beta}{4\pi} \left ( 4 i \pi  c_1+  \log \left(\frac{e^{-\frac{4 \pi ^2 l (2 \mu +\nu )}{\beta }} \left(\left(e^{\frac{4 \pi ^2 l \mu }{\beta }}-1\right)^2 e^{\frac{4 \pi ^2 l \nu }{\beta }} + \eta \right)}{2 \left(e^{\frac{4 \pi ^2 l \nu }{\beta }}-1\right)}\right)  \right ),
\end{flalign}
where in the above relation $\eta$ is
\begin{flalign}\label{eq:bcafter4}
\eta  & =\sqrt{-2 e^{\frac{8 \pi ^2 l (\mu +\nu )}{\beta }}+4 e^{\frac{4 \pi ^2 l (2 \mu +\nu )}{\beta }}+e^{\frac{8 \pi ^2 l (2 \mu +\nu )}{\beta }}-4 e^{\frac{4 \pi ^2 l (\mu +2 \nu )}{\beta }}-4 e^{\frac{4 \pi ^2 l (3 \mu +2 \nu )}{\beta }}+4 e^{\frac{4 \pi ^2 l (2 \mu +3 \nu )}{\beta }}+e^{\frac{8 \pi ^2 l \nu }{\beta }}}.
\end{flalign}

 \begin{figure}[ht!]
 \centering
  \includegraphics[width=5.2cm] {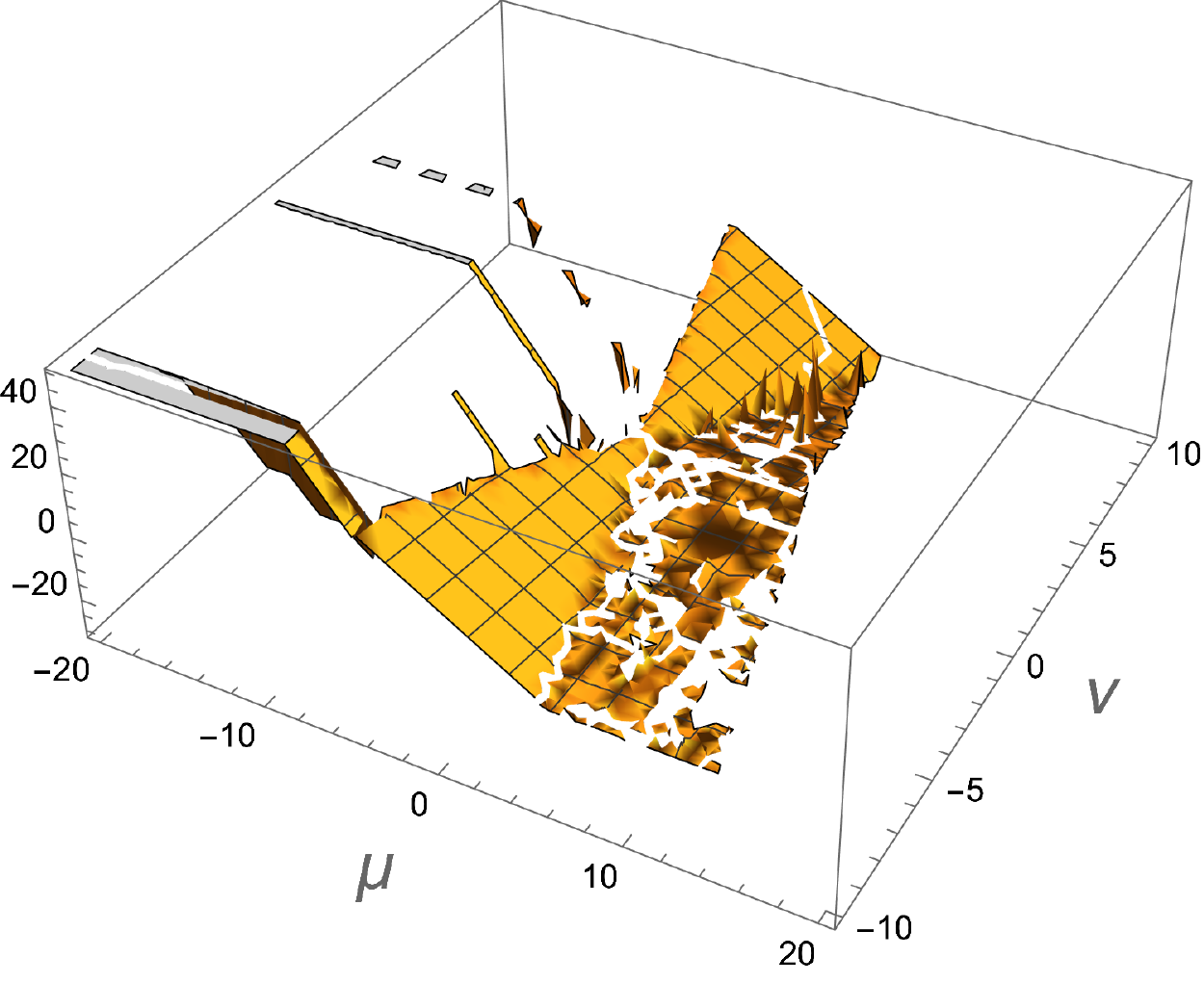} \ \ \ \ \ \ \ \ \ \ \ \ 
    \includegraphics[width=5.2 cm] {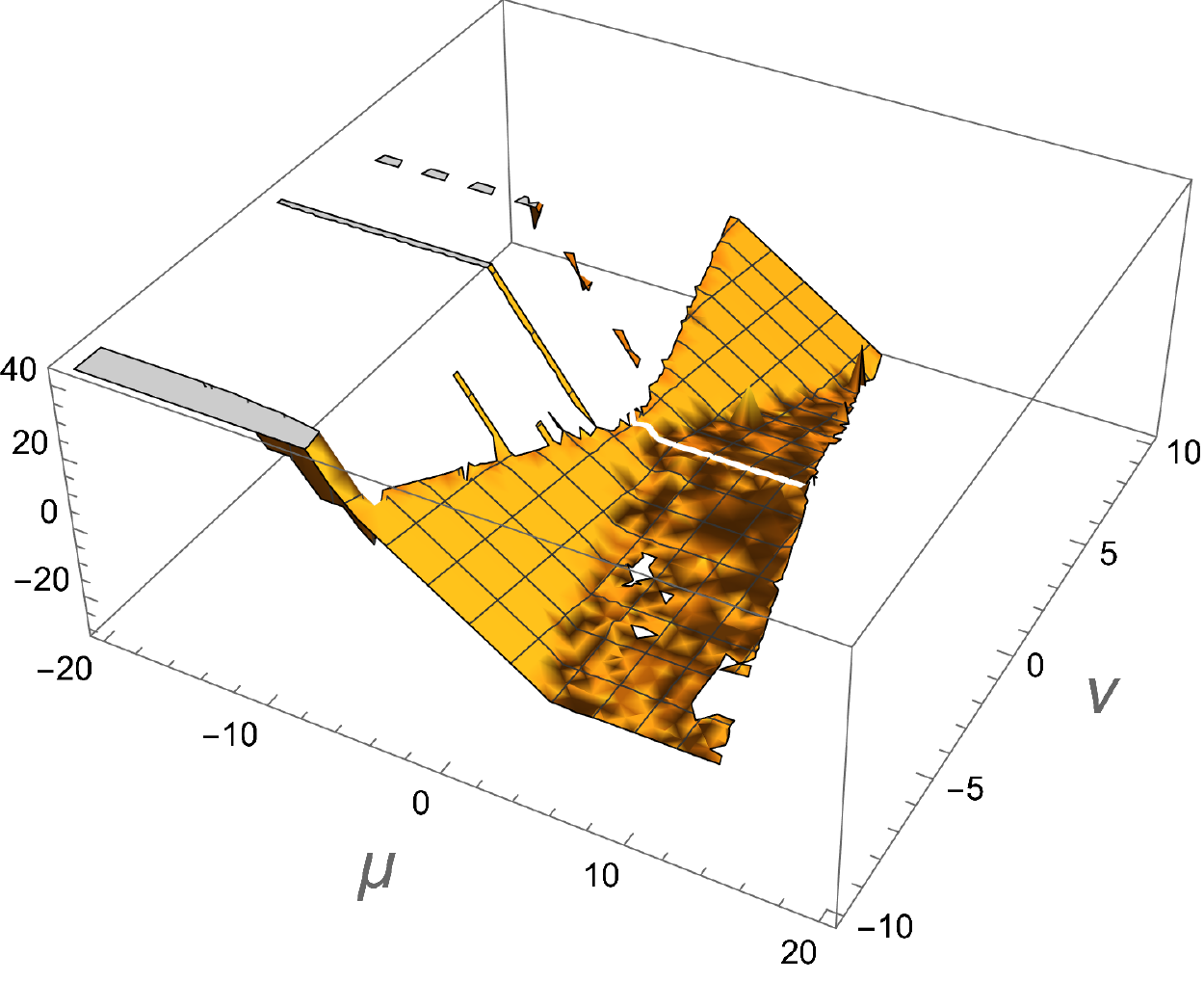} 
        \includegraphics[width=5.2 cm] {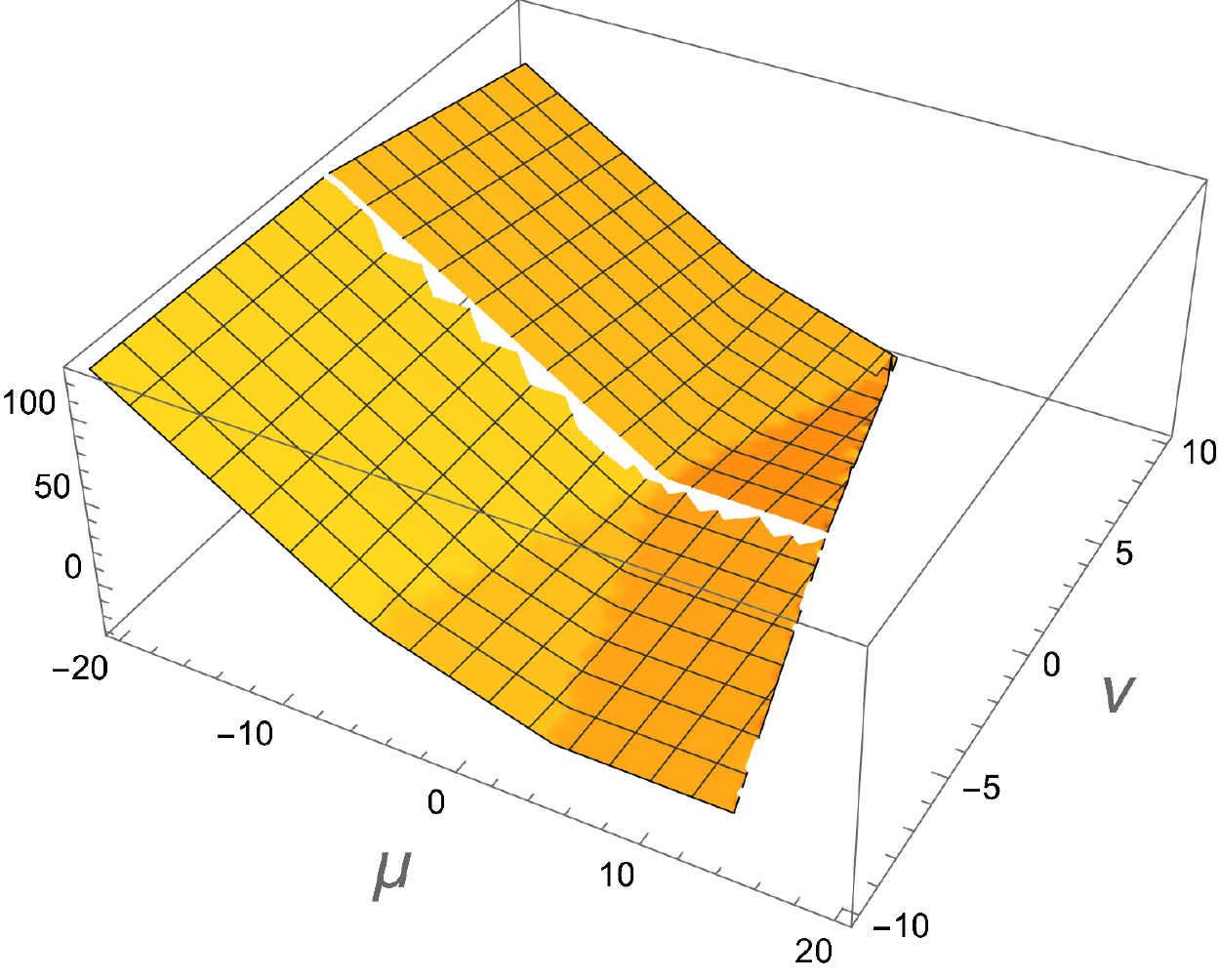} \ \ \ \ \ \ \ \ \ \ \ \ 
            \includegraphics[width=5.2 cm] {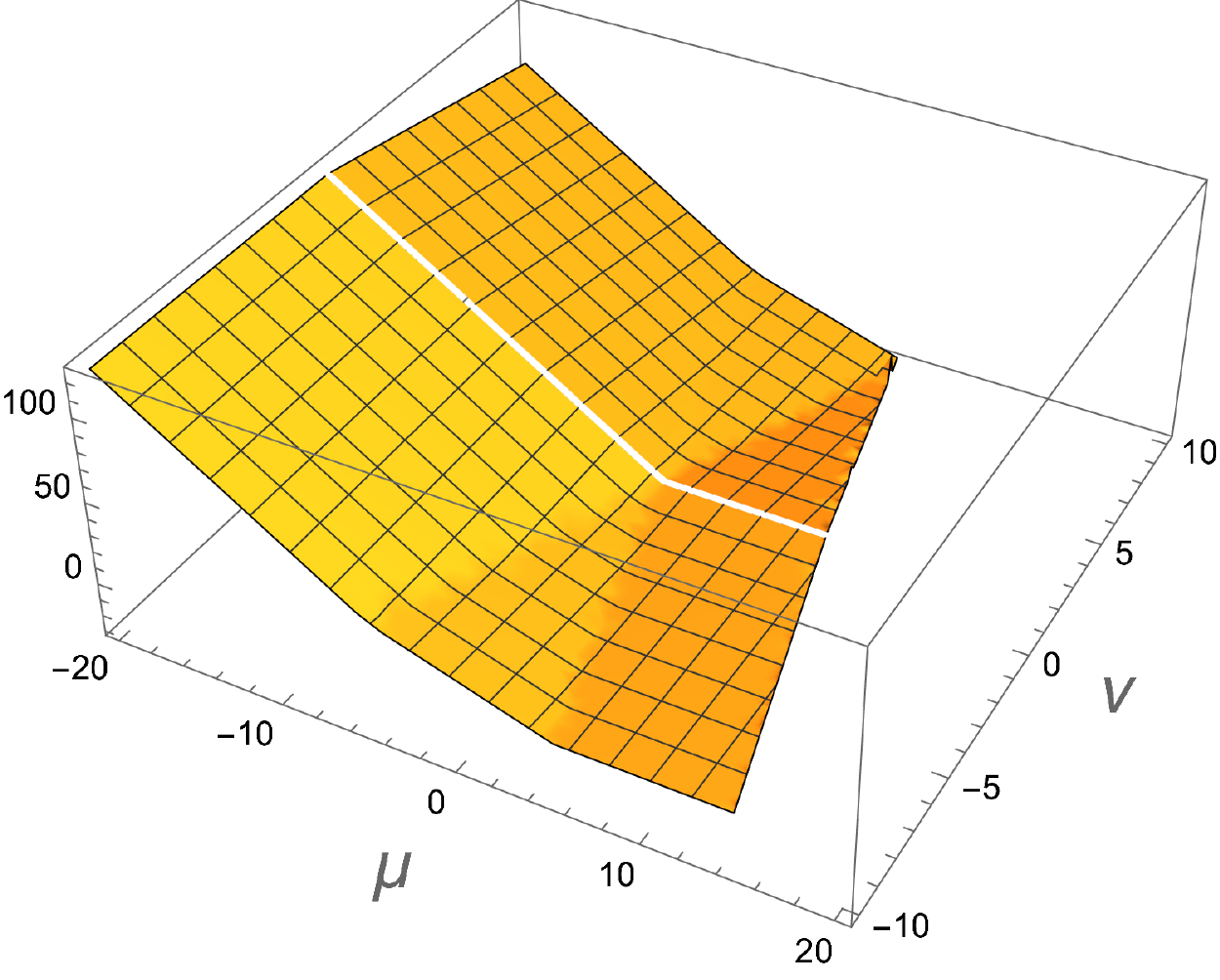} 
  \caption{The plots of solution of $b_c$ \textbf{after} the Page time are shown. From left to right are the plots of $b_c(1)$, $b_c(2)$,  $b_c(3)$ and $b_c(4)$. These plots show the behavior of the critical bath size $b_c$, in terms of the two mixed system size $\mu$ and $\nu$. Here $\beta$ and $l$ are set to one.}
 \label{fig:bcafterPage}
\end{figure}

The real part of these four solutions are shown in figure \ref{fig:bcafterPage}. Similar to the previous case, the phase space becomes smoother from the case of $b_c(1)$ to $b_c(4)$. If one fixes the ratio of the two system sizes as $\nu / \mu = 0.001$, the behavior of $b_c$ would be as sown in the right side of figure \ref{fig:bcafterPage}.

\begin{figure}[ht!]
 \centering
  \includegraphics[width=5.2cm] {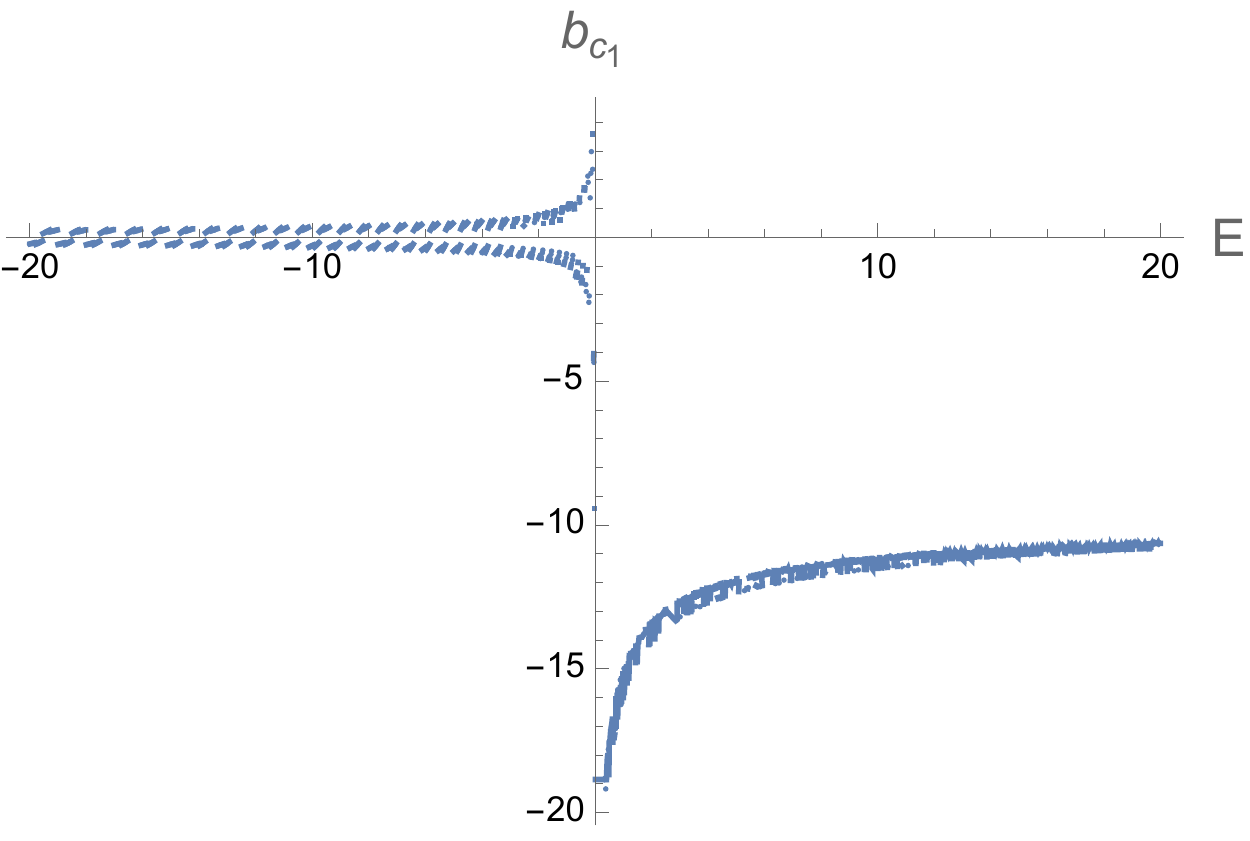} \ \ \ \ \ \ \ \ \ \ \ \ 
    \includegraphics[width=5.2 cm] {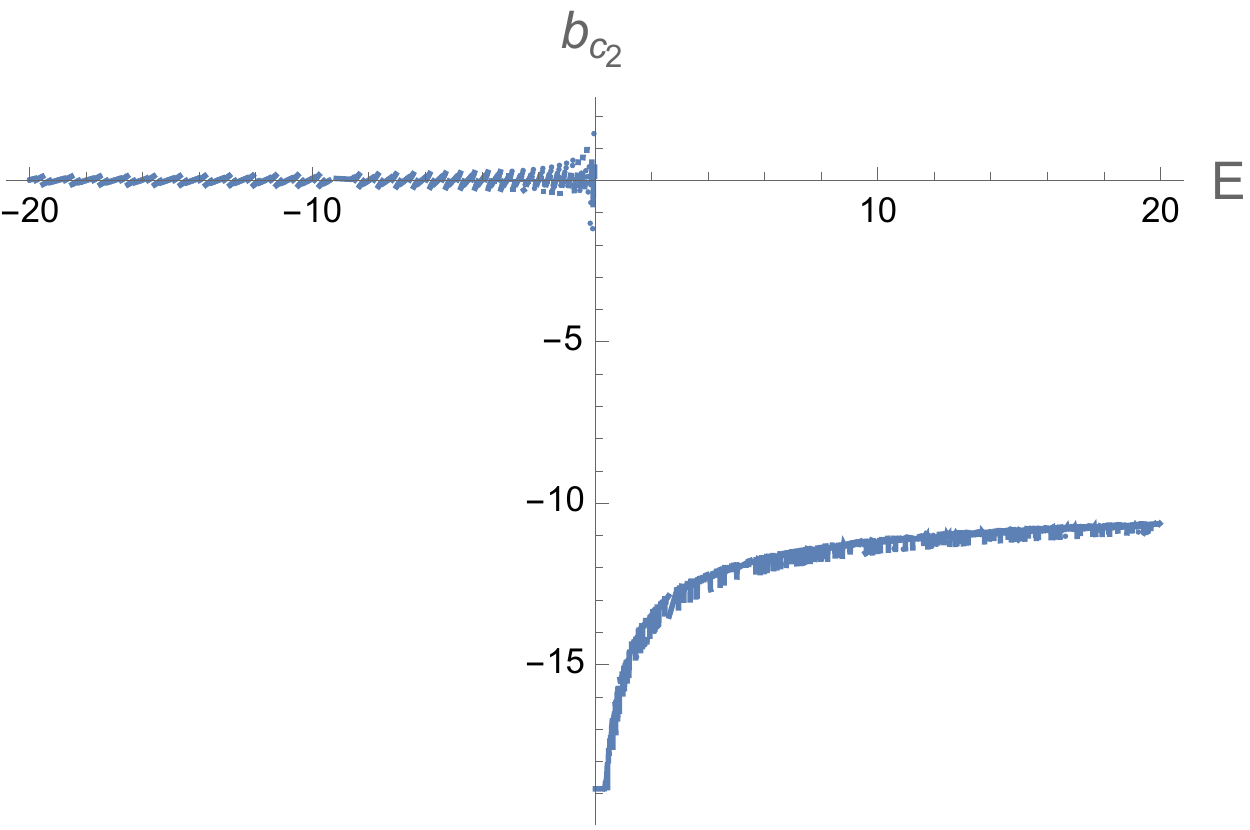} 
        \includegraphics[width=5.2 cm] {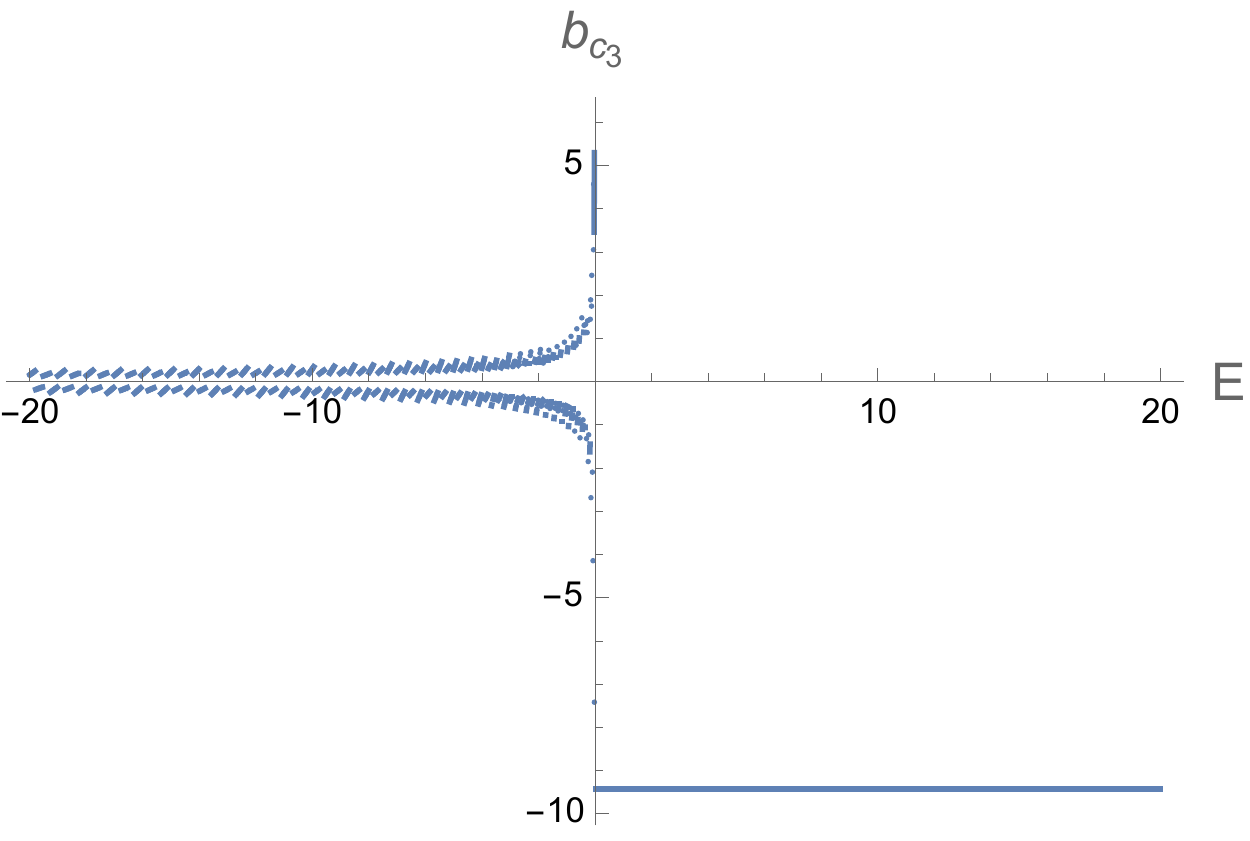} \ \ \ \ \ \ \ \ \ \ \ \ 
            \includegraphics[width=5.2 cm] {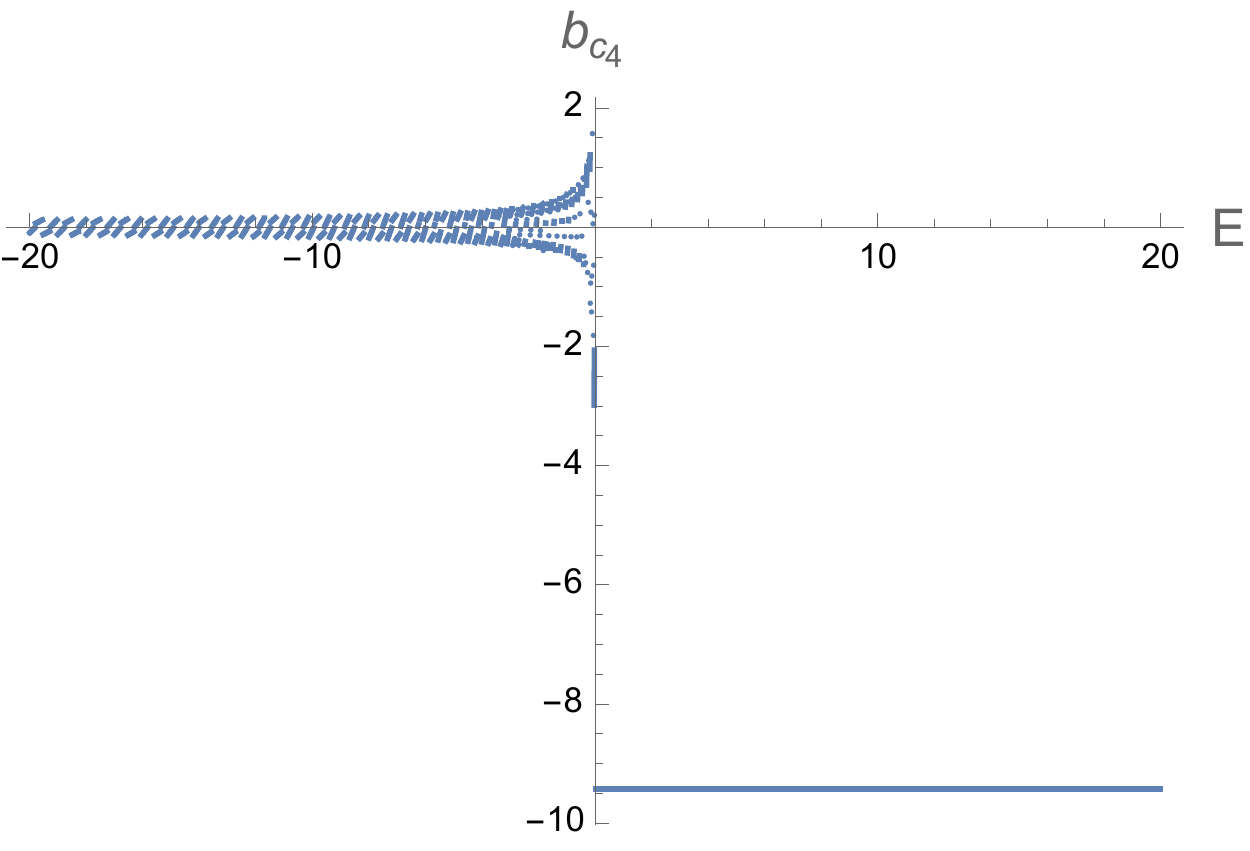} 
  \caption{The plots of $b_c$ versus energy $E$ for the cases \textbf{after} the Page time, for  $\mu=\nu=3$, $l=1$ and $\kappa=2$.}
 \label{fig:bcEafterPage}
\end{figure}

Inserting $\beta= \pi / \sqrt{ \kappa E}$ in the above relations \ref{eq:bcafter1}, \ref{eq:bcafter2}, \ref{eq:bcafter3} and \ref{eq:bcafter4} with $\beta= \pi / \sqrt{ \kappa E}$, we get the plots shown in figure \ref{fig:bcEafterPage} which show the behavior of $b_c$ versus $E$ in the four regimes.  One could see that for the first two solutions of $b_{c_1}$ and $b_{c_2}$, the behavior of $b_c$ versus $E$ are critically different.

\begin{figure}[ht!]
 \centering
  \includegraphics[width=5.2cm] {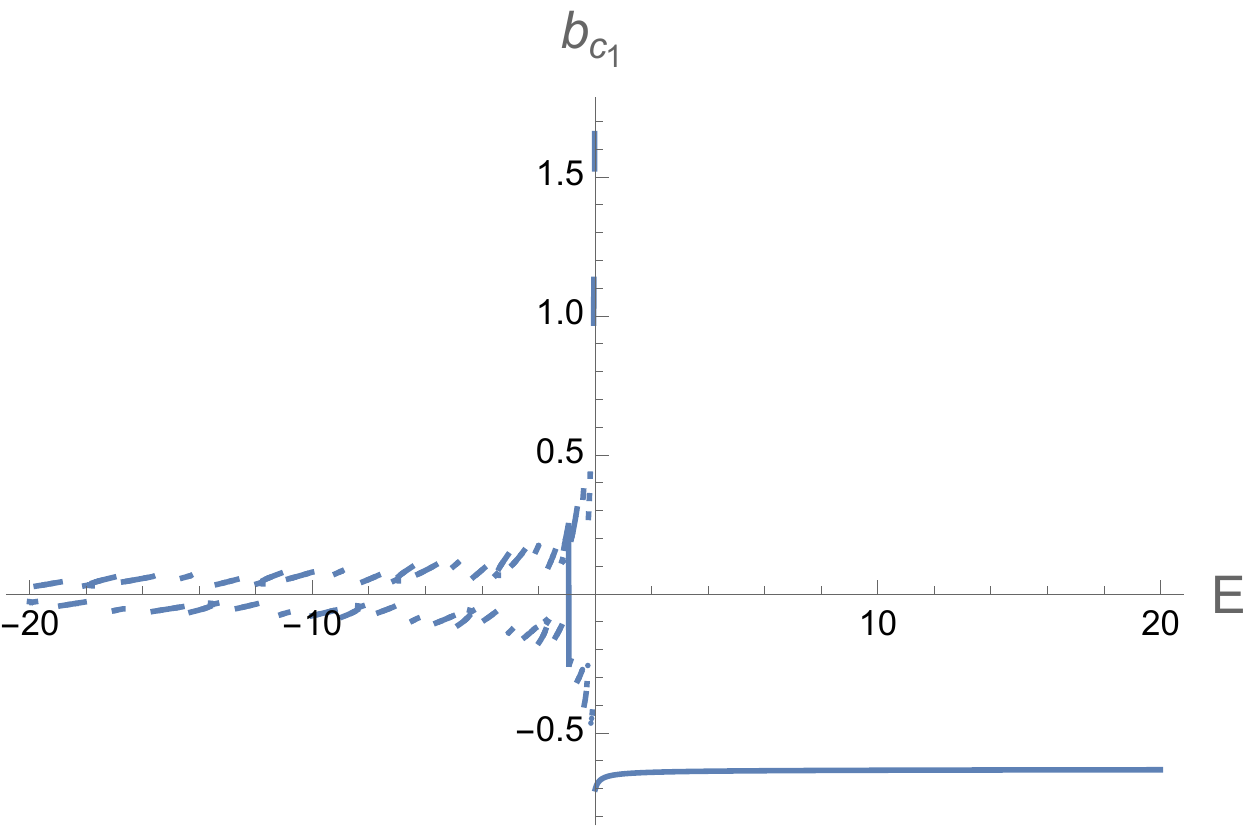} \ \ \ \ \ \ \ \ \ \ \ \ 
    \includegraphics[width=5.2 cm] {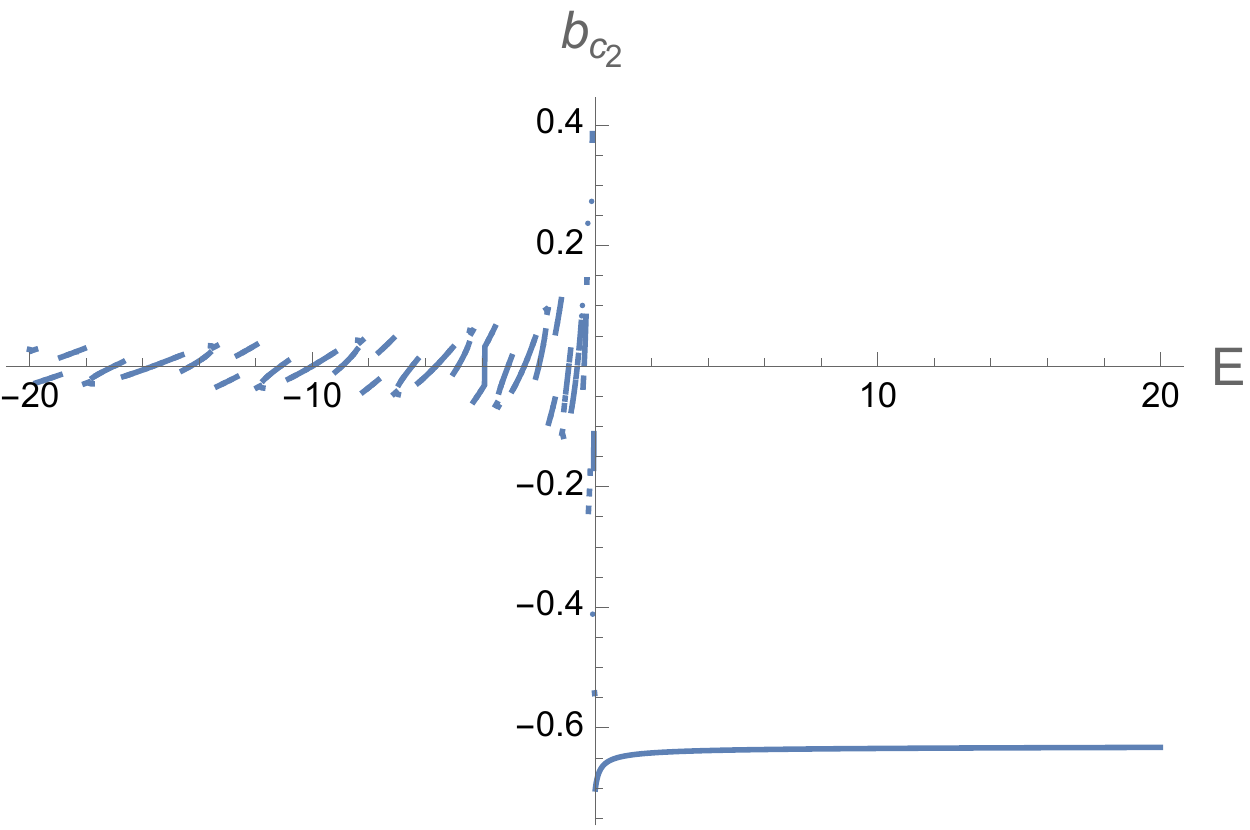} 
        \includegraphics[width=5.2 cm] {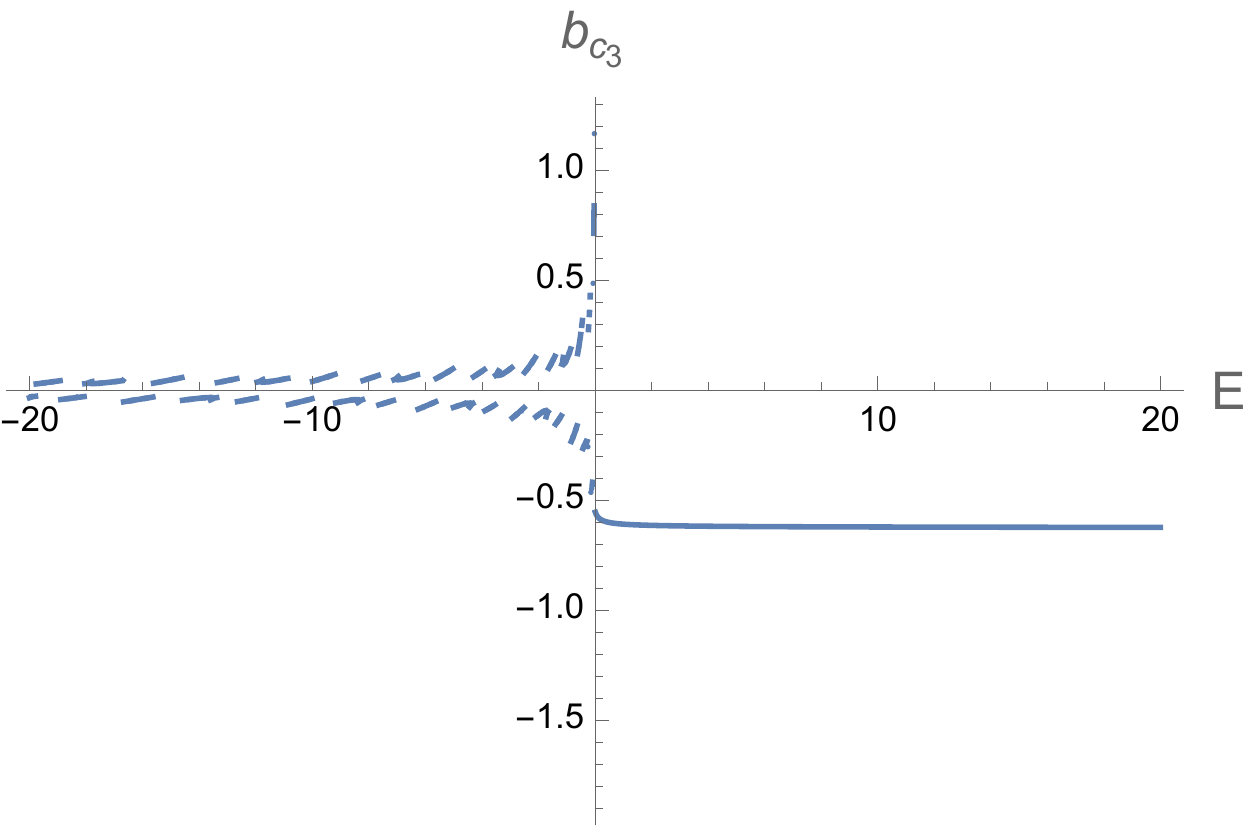} \ \ \ \ \ \ \ \ \ \ \ \ 
            \includegraphics[width=5.2 cm] {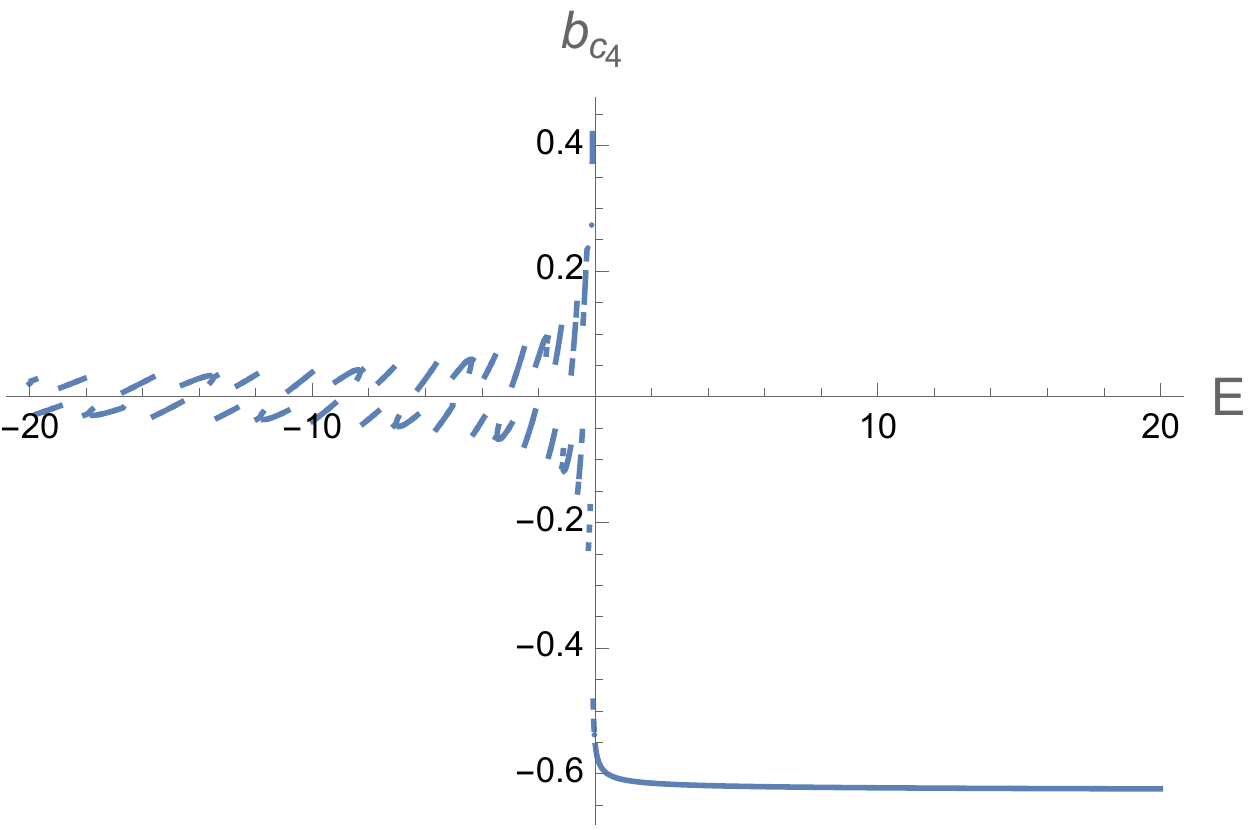} 
  \caption{The plots of $b_c$ versus energy $E$ for the cases \textbf{after} the Page time,  for  $\mu=0.1$, $\nu=0.2$, $l=1$ and $\kappa=40$.}
 \label{fig:bcEafterPage2}
\end{figure}

For the case after the Page time we could not tune energy to get positive values of $b_c$ but the results are shown in figure \ref{fig:bcEafterPage2}. One should note that for negative energies the results would not be physical.

\begin{figure}[ht!]
 \centering
  \includegraphics[width=5.2cm] {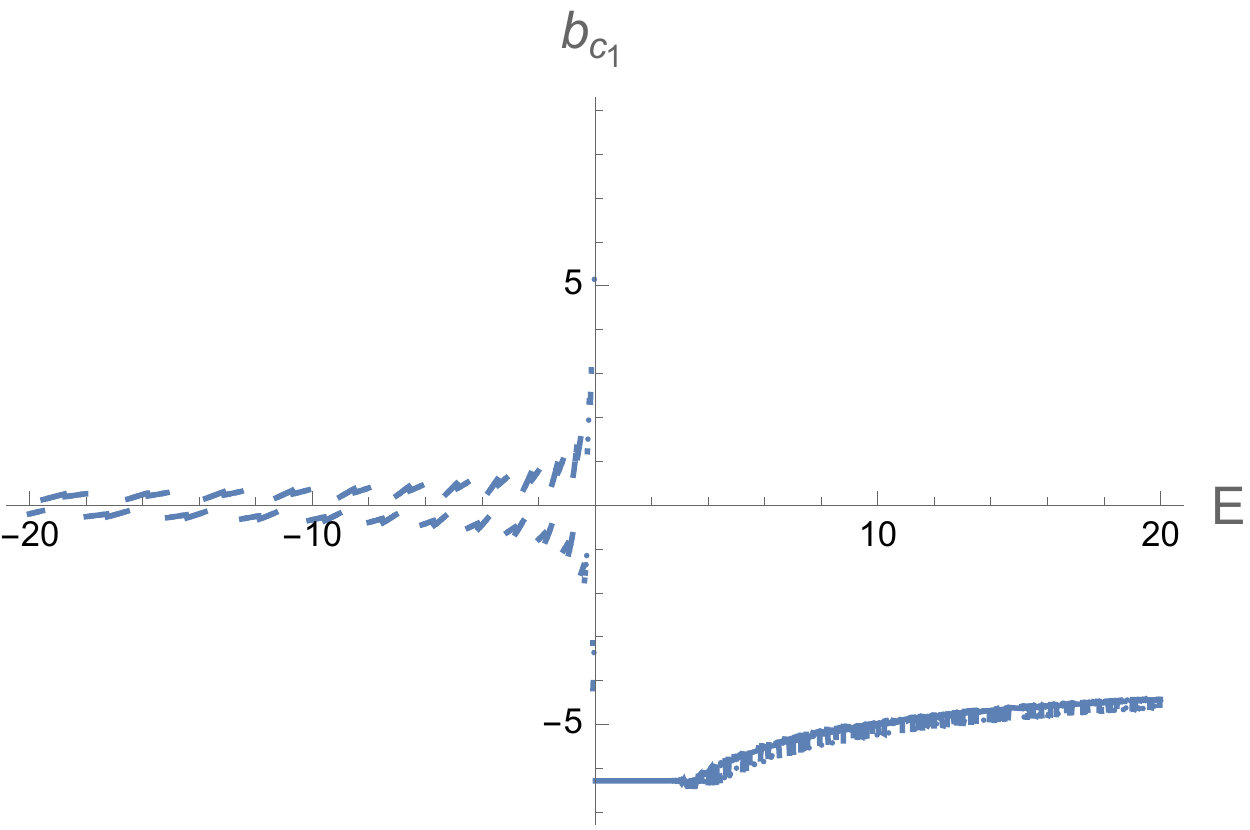}  \ \ \ \ \ \ \ \ \ \ \ 
    \includegraphics[width=5.2 cm] {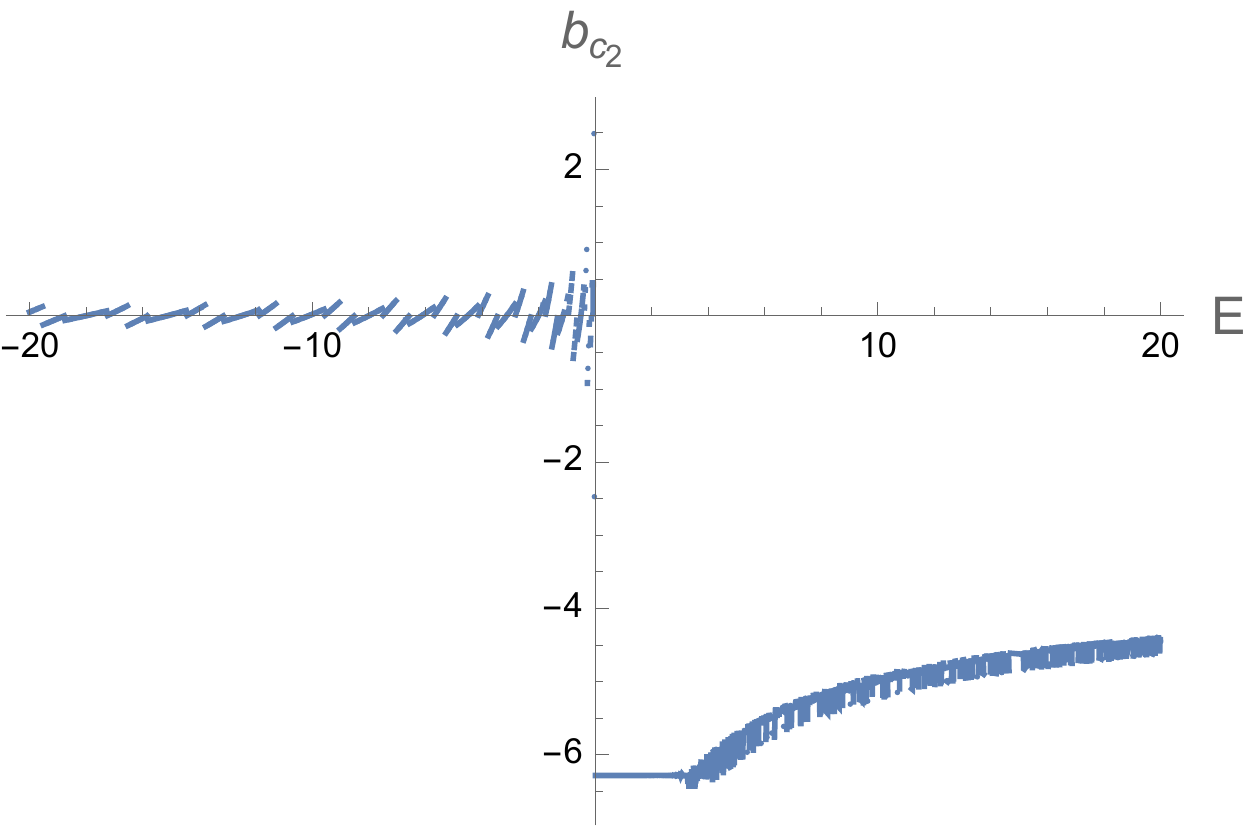} 
        \includegraphics[width=5.2 cm] {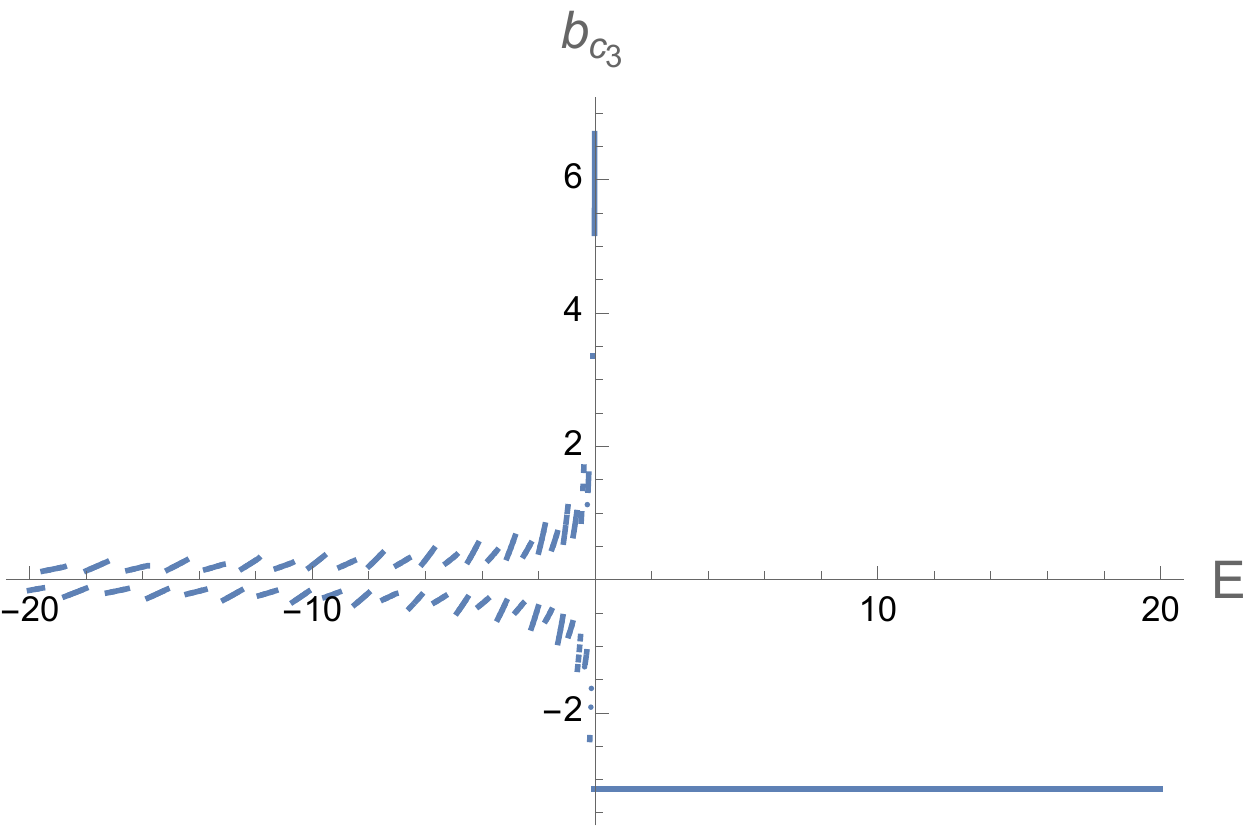}  \ \ \ \ \ \ \ \ \ \ \ 
            \includegraphics[width=5.2 cm] {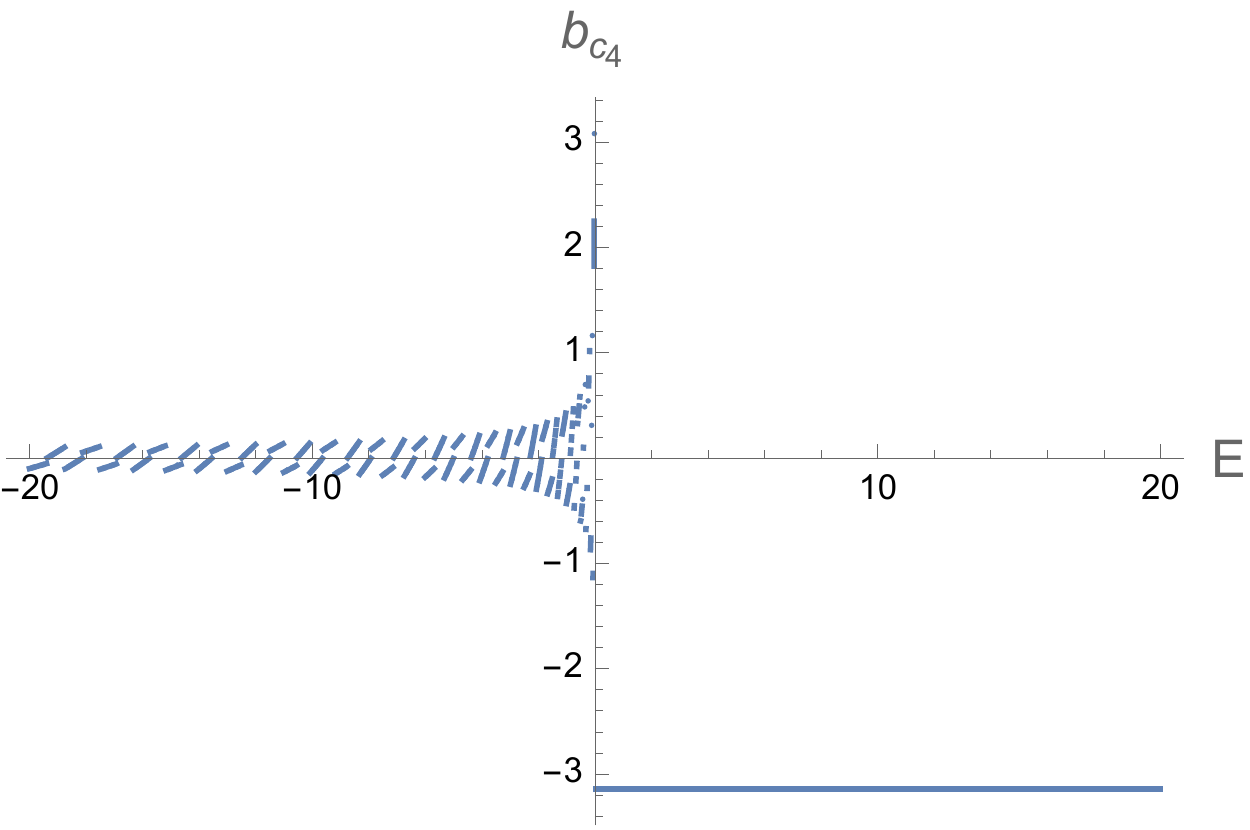} 
  \caption{The plots of $b_c$ versus energy, $E$,  for the cases \textbf{after} the Page time, for  $\mu=\nu=1$, $l=1$ and $\kappa=3$.}
 \label{fig:bcEafterPage3}
\end{figure}

For the case of $\mu=\nu=1$ and $\kappa=3$ we can get a third behavior shown in figure \ref{fig:bcEafterPage3}.

 \begin{figure}[ht!]
 \centering
  \includegraphics[width=6.5cm] {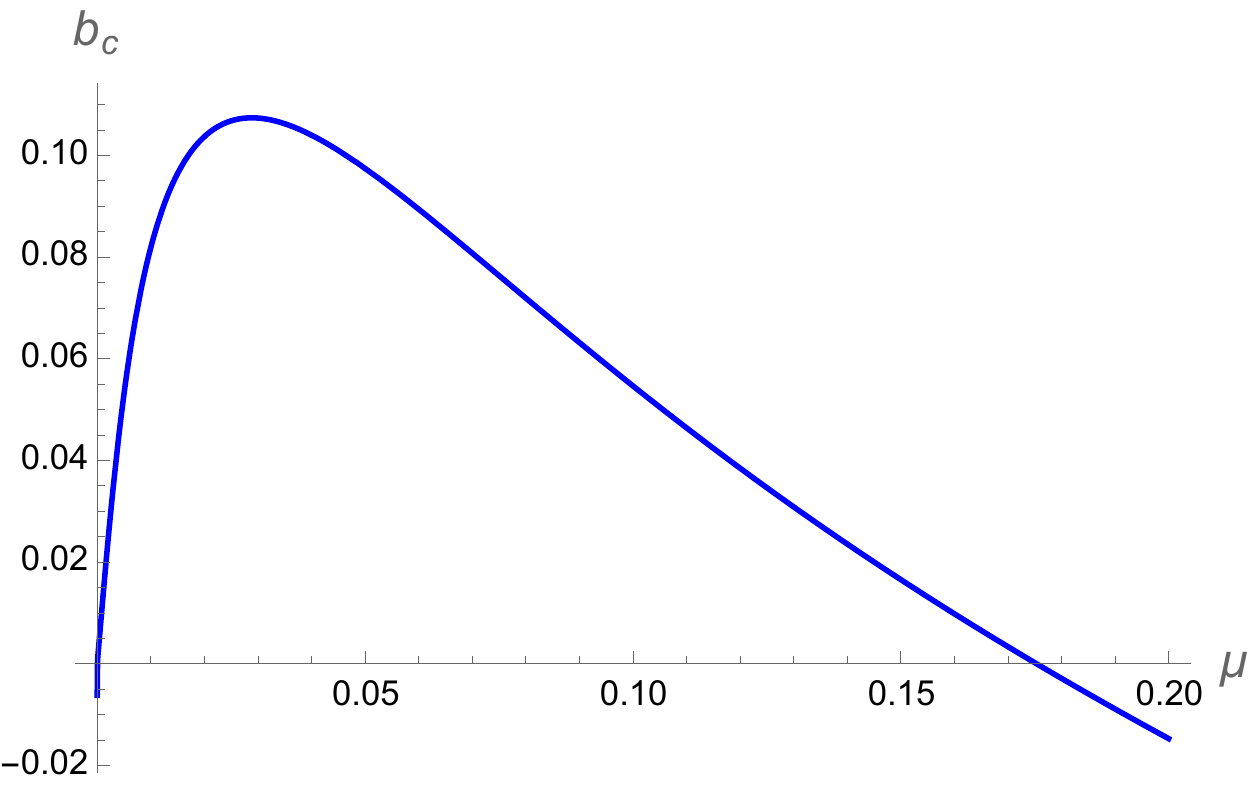} 
  \caption{The behavior of $b_c$ after the Page time, versus $\mu$, for $\nu=0.1\mu$. This behavior is very similar to phase diagram found using negativity in \cite{Shapourian:2020mkc}. }
 \label{fig:bcpage2}
\end{figure}

 \begin{figure}[ht!]
 \centering
  \includegraphics[width=6.5cm] {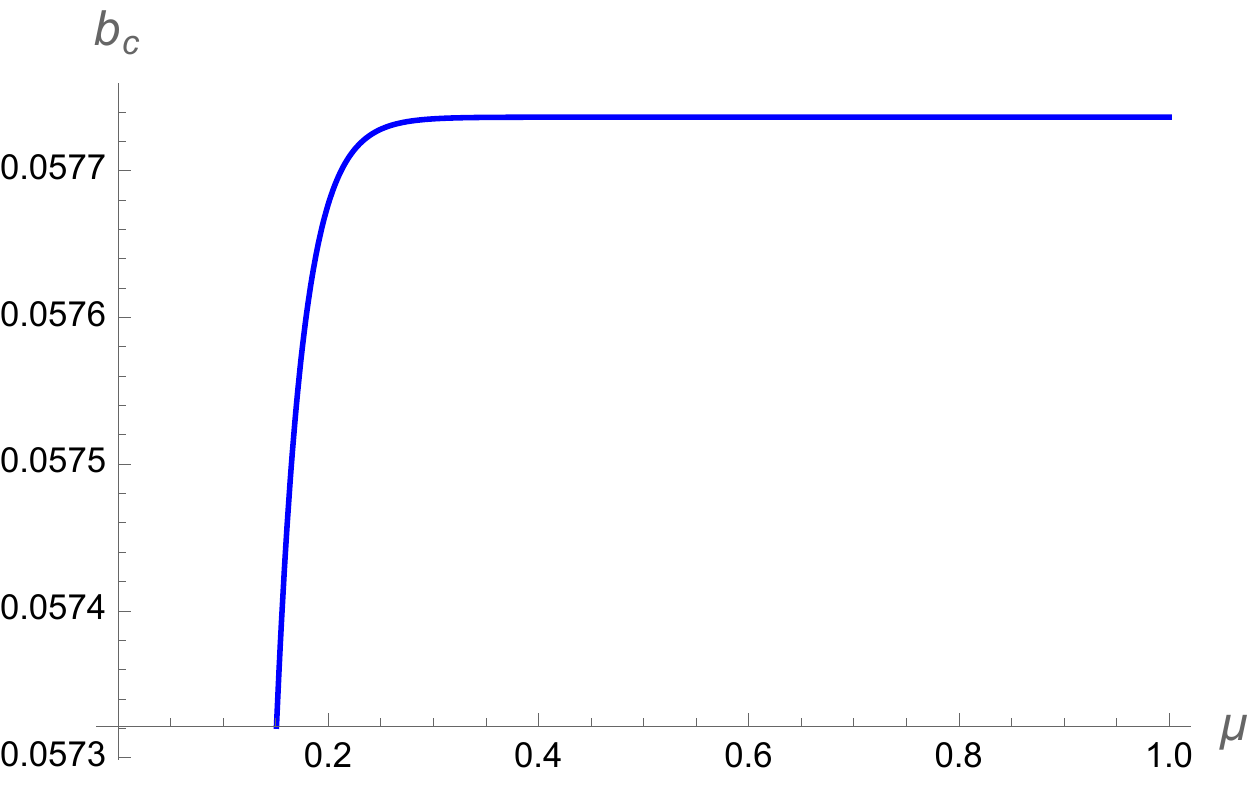}  \ \ \ \ \ \ \ \ \ \ \ 
    \includegraphics[width=6.5cm] {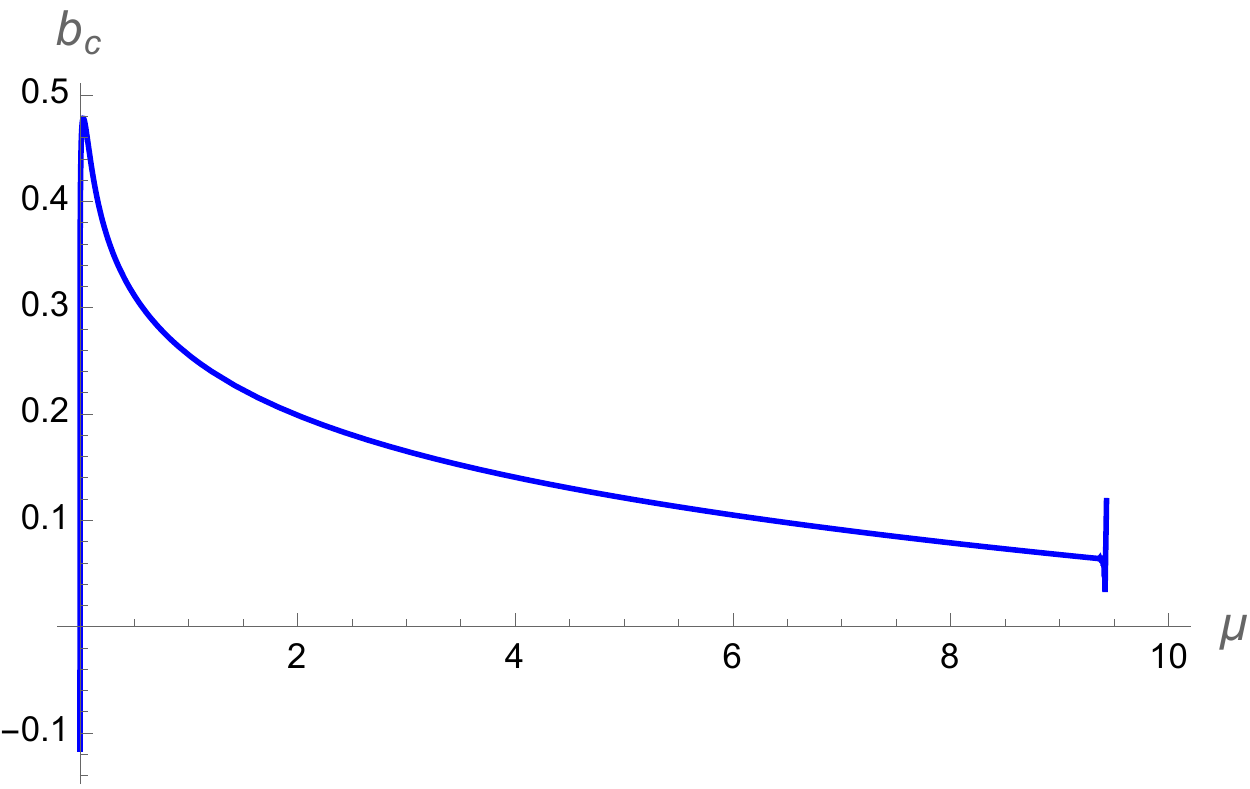} 
  \caption{The behavior of $b_c$ after the Page time, versus $\mu$, for very small size of $\nu=0.01$, in the left and for the ratio $\nu=0.001 \mu$, in the right.}
 \label{fig:bcpage2}
\end{figure}

The behavior of the critical size of the bath, $b_c$, versus the system size $\mu$, for the small size of the other system $\nu=0.01$ is shown in figure \ref{fig:bcpage2}. It can be seen that it would increase at the beginning and then, as shown in figure \ref{fig:bcsmall}, it becomes constant for the larger values of $\mu$, which is different from the case before the Page time.

Similar to \cite{Shapourian:2020mkc}, a dimensionless tuning parameter $q=b_c/ \mu$ could be defined which would probe the phase space in different setups.

It also worths to mention that for the cases that $b_c$ is positive, increasing $\beta$ would increase $b_c$. This is because when the temperature decreases, as also noticed in \cite{Ghodrati:2019hnn}, the correlation could sustain more, and therefore the critical distance between intervals could increase.  

Note that the first solution, $b_{c_1}$ is specially related to the kink mirror solution \cite{Akal:2022qei, Akal:2021foz}, where in addition to considerations of the symmetries, entanglement structure and the form of energy momentum tensor, the behavior of $b_{c_1}$ versus $\beta$ shows the same behavior.

In principle, in the classifications of the regimes, the factor $n$ in the mapping functions of \cite{Akal:2022qei}, would translate to the late time behavior of intervals $\mu$ and $\nu$ in our studies, as we discussed various scenarios above.

 \begin{figure}[ht!]
 \centering
  \includegraphics[width=5.5cm] {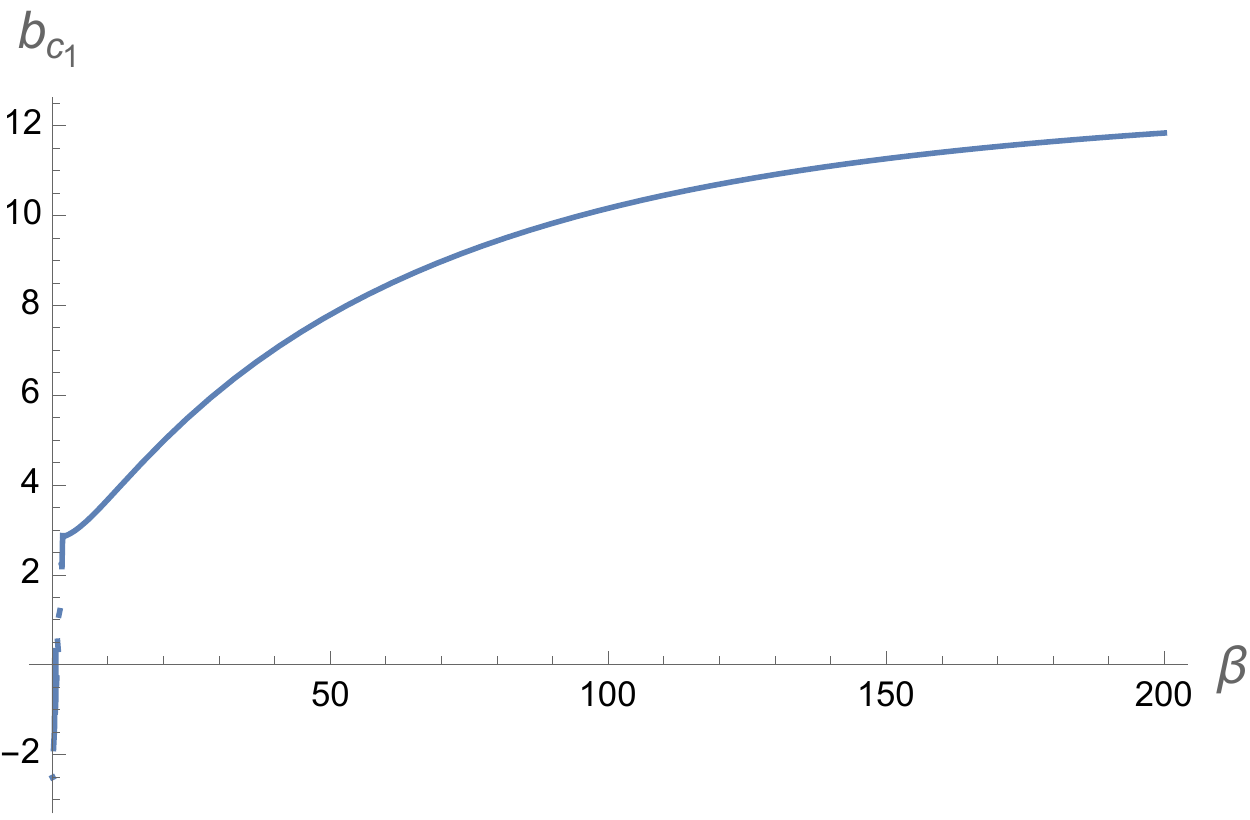}  \ \ \ \ \ \ \ \ \ \ \ 
    \includegraphics[width=5.5cm] {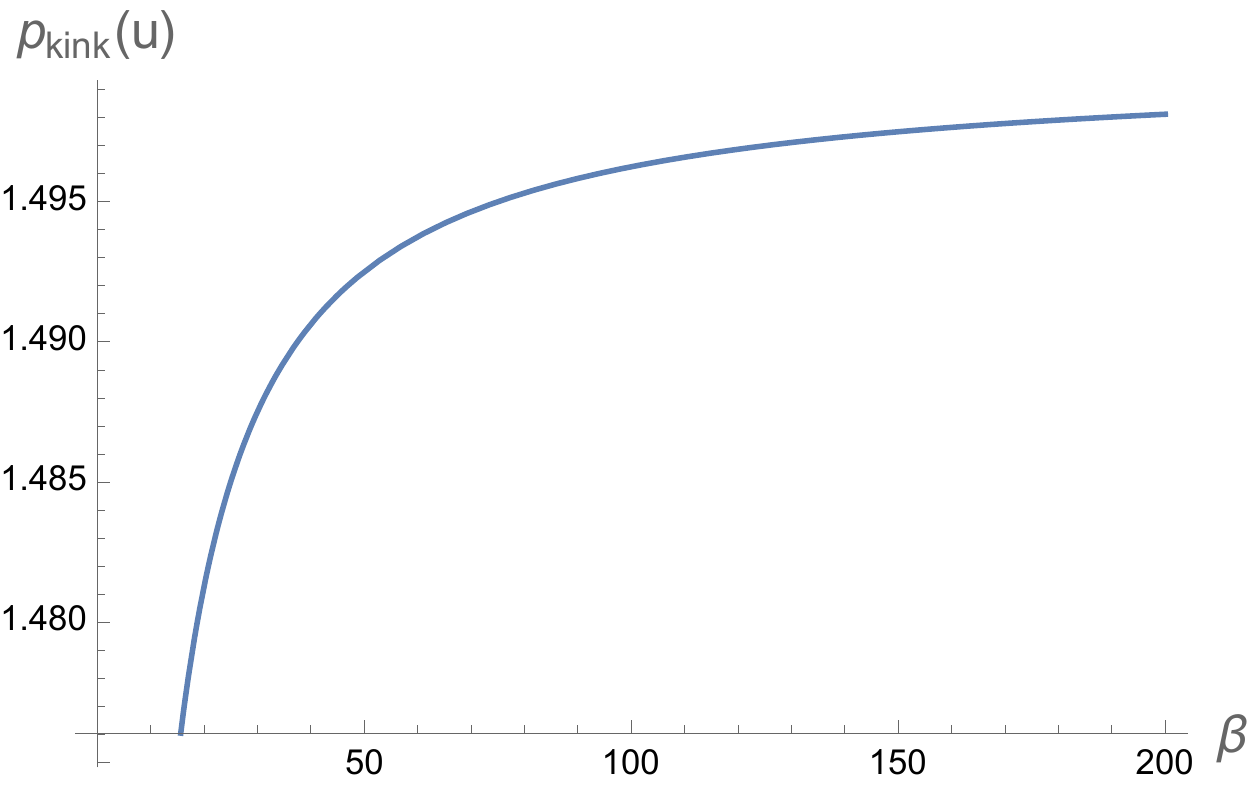} 
  \caption{The behavior of $b_{c_1}$ versus $\beta$ in the left, and $p(u)$ versus $\beta$, in the right, \cite{Akal:2022qei, Akal:2021foz}.}
 \label{fig:bcpage2}
\end{figure}

From figure \ref{fig:bcpage2}, one could also see that for the case after the Page time, the critical bath size, compared to $\mu$ and $\nu$, would be very small or more precisely, for getting a vanishing mutual information and the disconnected phases, the tuning parameter $q$ would be much smaller. This is due to the fact that after Page time most of the degrees of freedom of black hole are already evaporated and therefore a smaller size of bath would be enough for thermalizing the system and also there would be less quantum correlations between $\mu$ and $\nu$ in this case.

For this scenario of states after the Page time which corresponds to the case where  $L_A \gg L_B$ in \cite{Shapourian:2020mkc}, the replica wormhole would be present. Also, the dominant diagram for $ \langle \text{Tr} \rho^n \rangle = L_B^{1-n} \propto {b_c}^{1-n}$, would be the one shown in equation 4.9 of \cite{Shapourian:2020mkc}.
This case corresponds to the negative partial transpose (NPT) state with a non-vanishing logarithmic negativity (LN) as shown in the lower regions of the phase diagram, i.e, the regions $II$ and $III$ in figure. 2 of [4], which depending on the size of $\mu$ or $\nu$ would have different spectrum distributions, namely maximally entangled and entanglement saturation cases. In the first case when $\mu$ is small or $\nu$ is very large, we have two disjoint Marcenko-Pastur-like distributions, and when $\mu$ is large or $\nu$ is very small, there would be the semicircular distribution with partly negative domain, corresponding to the saturated entanglement. From our studies here though, these two cases cannot be distinguished.

In \cite{Grimaldi:2022suv}, the critical size of bath, where a phase transition occurs has also been discussed but in a different context,  where there is a brane and the corresponding defect for it. There, the critical size of the bath depends on the tension of the brane while here it depends on the relative size of the two subsystems $\mu $ and $\nu$ and also their sizes compared to the bath. There, also it has been confirmed that the Page time would depend on the ratio of the central charges of the defect brane versus the central charge of the system, which agrees with our results here.

In \cite{Alishahiha:2022kzc}, the complexity for a single-sided AdS black hole with an EOW brane has been calculated where the late time behavior of complexity is controlled by the parameter $e^{S_0}$. This parameter would completely depend on the tension of the EOW brane sitting at the end of the geometry with JT action.  In our relations \ref{eq:bcbefore} and also \ref{eq:bcafter1}, \ref{eq:bcafter2}, \ref{eq:bcafter3},  \ref{eq:bcafter4}, this term affects through $\beta$, as it follows the relation $\beta= \pi/ \sqrt{\kappa E}$, where $E=M$ is the mass of the black hole and $\kappa$ is a parameter related to the tension of the brane.

For our setup with such angles, for each interval we can write
\begin{gather}
\langle \mu(t) \rangle = - \underset{\Delta \to 0} {\text{lim}} \frac{\partial \langle \varsigma(t) \varsigma(0) \rangle_{\text{non-int} } }{\partial \Delta},
\end{gather}
where $\Delta$ is the scaling dimension for the operator $\varsigma$. So for two systems, with angular interval sizes of $\mu$ and $\nu$, the mixed correlation structure and critical size $b_c$, and also the time needed for the saturation in a dynamical setup, would all depend on $e^{S_0}$ and the tension of the brane.

Finally, it worths to mention here the four saddles have even been observed in new models of complexity constructed in \cite{Balasubramanian:2022tpr}, where the behavior of Krylov operator complexity or complexity-spreading and Krylov entanglement, or entanglement spreading (which they could be connected to each other through the exponential behavior), both show the same behaviors, having four regimes of ramp, peak, slope and plateau.

 \subsection{Comparing with results from partial transpose}
 
  The studies of \cite{Shapourian:2020mkc} has been done using partial transpose (PT) and logarithmic negativity (LN) and here we use mutual information and critical size of bath. By comparing our plots, one could deduce further results from various entanglement phase transitions which worths to mention here.
  
 Note that in general the mutual information relative to PT and LN would overestimate the entanglement of classically correlated states which are separable, relative to the case of LN and PT. 
 The separable state which is a completely classical state could be written as
 \begin{gather}
 \rho_{\text{sep}}= \sum_{i,j} p_{i,j} \rho_1^{(i)} \otimes \rho_2 ^{(j)} , \ \ \ \ \ \ \ p_{ij} > 0.
 \end{gather}

One then can use the random induced mixed states setup of \cite{Shapourian:2020mkc} where the random induced mixed states $\{ \rho_A \}$ in the Hilbert space of $\mathcal{H}_A = \mathcal{H}_{A_1} \otimes \mathcal{H}_{A_2}$ can be found by partial tracing of the whole pure states in $ \mathcal{H}_{A} \otimes \mathcal{H}_{B}$.
Then, in a tensor product basis, a random pure state could be written as
\begin{gather}
\ket{\Psi} = \sum_{i=1}^{L_A} \sum_{\alpha=1}^{L_B} X_{i\alpha} \ket {\Psi_A ^{(i)}} \otimes \ket{\Psi_B^{(\alpha)}},
\end{gather}
where $X$ are $L_A \times L_B$ rectangular random matrix with elements $X_{i\alpha}$ which are independent Gaussian random complex variables with the joint probability density as
\begin{gather}
P ( \{ X_{i \alpha} \} ) = \mathcal{Z}^{-1} \exp \{ -L_A L_B \text{Tr}(X X^\dagger) \},
\end{gather}
where here $L_A= L_\mu \times L_\nu$ in our case, and the density matrix $\rho_A$ could be written as $\rho_A= \frac{X X^\dagger}{\text{Tr} (X X^\dagger) }$.

Then, the spectral density of eigenvalues $\{ \lambda_i \}$ would follow the relation
\begin{gather}
P(\lambda)= \sum_{i=1}^{L_A} \delta (\lambda- \lambda_i) = \frac{q L_A^2}{2\pi} \frac{\sqrt {(\lambda_+-\lambda)(\lambda- \lambda_-) }}{\lambda}, \nonumber\\ \ \lambda_{\pm}=\frac{1}{L_A} ( 1 \pm 1 / \sqrt{q} )^2, \ \ \ \lambda \in \lbrack \lambda_-, \lambda_+ \rbrack, \ \ \ \ q=L_B/L_A \ge 1,
\end{gather}
which this behavior is compatible with the behavior of critical bath size $b_c$ in our setup. Similarly, for the case of $q <1$, the entanglement spectrum would have a delta-function at the origin which is exactly what we got here in the diagrams of $b_c$, for the case after the Page time where $q$ becomes much smaller than one.

\subsection{Quantum error corrections of black hole from $1d$ higher point of view} \label{subsec:QECPage}
In \cite{Balasubramanian:2022fiy}, for the JT gravity model,  the quantum error correction properties of the black hole interior which is entangled to a non-gravitational bath have been studied and it was shown that the interior is robust against the generic and low-rank operations.

They argued that after the Page time, the information of the interior degrees of freedom of the black hole and those in the bath are connected, and as the information of the black hole get encoded in the bath Hilbert space, so the noise in the  density matrix of the interior of black hole coming from quantum operations on the bath can be corrected. The bound on the noise would depend on the black hole entropy and the code subspace dimension.

So the holographic quantum channels in this case should satisfy the ``Knill-Laflamme" conditions
\begin{gather}
P_{\text{code}} E_m^\dagger E_n P_{\text{code}}= \alpha_{mn} P_{\text{code}}, 
\end{gather}
where $P_{\text{code}}= V V^\dagger$ is the projector onto the code subspace.

Since there is a phase transition in the behavior of mutual information and the saddles of Euclidean path integrals, the factorization and decoupling behavior of the total Hilbert space of the black hole and therefore the behavior of the holographic quantum error correction channels and codes would change for the case before and after the Page time.  Indeed, after the Page time the correctability of errors would significantly change.

For the JT gravity in \cite{Balasubramanian:2022fiy}, it was also found that the recovery channel $\mathcal{R}$ would satisfy the relation
\begin{gather}
\text{max}_\rho || \mathcal{R} ( \mathcal{E} (\rho)) - \rho ||_1 \le \epsilon , \ \ \ \ \ \epsilon \sim e^{-S_0/4},
\end{gather}
so the bigger the entropy of the higher-dimensional extremal black hole (where the JT theory would be the result of its two-dimensional reduction), the lower the errors of the holographic quantum recovery channels would be and so the more robust the black hole would be against the noises in the bath. The parameter $S_0$ here could be considered as the ground state entropy of the JT system.  This relation then led to $2(t-1) \le ( \mathcal{S} - \log_2 d)$ \cite{Balasubramanian:2022fiy}, where $t-1$ is the maximum number of qubits in the physical Hilbert space which can tolerate the error, $d$ is the dimension of the code space and $2^{\mathcal{S}}$ is the dimension of the physical Hilbert space.

The parameter $t$ also determines how much of the information of the $3d$ BTZ black hole would actually penetrate to the lower dimensional $2d$ JT gravity case, which we will investigate this point further in the next section.  This penetration of information between various dimensions can also be understood by studying the partition functions, the logarithmic negativity, and also similar to \cite{Balasubramanian:2022fiy}, the R\'enyi entropy, i.e,  $S^{(n)}= \frac{1}{1-n} \log \text{Tr} \rho^n$ from the R\'enyi mutual information.

If similar to \cite{Balasubramanian:2022fiy}, we write the reduced density matrix on the interior and exterior code subspace, and also on the environmental part, with the normalization constant $\mathcal{N}$, one would get the relation $\mathcal{N}= d_i d_e e^{S_0} Z_1$,  where $d_i$ and $d_e$ are the dimensions of the interior and exterior factors in the code subspace and $Z_1$ is the exponential of the  on-shell, JT gravitational action on a disc which is capped off by an EOW brane, and some additional non-code-subspace bulk field theory modes in the Hartle-Hawking vacuum state. So, one could consider the information of the $2d$ case gathered in the $Z_1$ part, times the factor of $d_i d_e e^{S_0}$ leading to the higher $3d$ case where the non-robust information are gathered in the non-code subspace section.

Then,  according to the result of \cite{Balasubramanian:2022fiy} and based on the discussions on erasures of errors far from the phase transition points, the $\mathbb{Z}_n$-breaking bulk geometries cannot dominate. The two dominant geometries out of the four are the fully disconnected and the fully connected ones. Both of them preserve the $\mathbb{Z}_n$ symmetry of the asymptotic boundary conditions. The preservation of this symmetry is the reason that from the $3d$ BTZ point of view in \cite{Verheijden:2021yrb}, the phase structure of $2d$ JT gravity can be obtained.

So, approximately, the reduced density matrix of the $3d$ case can be written as 
\begin{gather}
\text{Tr} \rho'^n_{\mathfrak{R}}=\frac{1}{k^n \mathcal{N}^n} \sum_{i_1, ..., i_n} \sum_{i'_1, ..., i'_n} \sum_{\alpha_1, ..., \alpha_n} \langle \psi^{\alpha_1}_{i_2, i'_1} |   \psi^{\alpha_1}_{i_1, i'_1} \rangle_B ... \langle \psi^{\alpha_n}_{i_1, i'_n} |   \psi^{\alpha_n}_{i_n, i'_n} \rangle_B,
\end{gather}
which could be considered as an ensemble average of the microstates of the $2d$ JT gravity plus the EOW branes. When the parameters $k$ and $S_0$ become larger, more and more information of the $3d$ geometry is encoded in the $2d$ JT section.

Based on the quantum error correction idea of \cite{Balasubramanian:2022fiy}, the connected contribution of the partition function would dominate the disconnected one when
\begin{gather}
\frac{ (\frac{\ell}{k} )^{n-1} + k^{n-1}  }{1+ \ell^{n-1} } \approx \left (\frac{k}{\ell}   \right ) ^{n-1} \ll e^{(1-n) (S_{BH} + \log d_i ) },
\end{gather}
where $d_i$ and $d_e$ are the dimensions of the interior and exterior factors in the code subspace, and $n$ is the order of the Renyi entropy.

Here the Renyi entropy is defined over the random variables of the set $\mathcal{A}= \{x_1, x_2, . . . , x_n \} $ where $S_n$ is a non-increasing function of $n$. Also, $\ell$ is the dimension of the environmental factor and $k$ controls the dimension of the radiation subspace that is entangled with the black hole microstates.

This relation then could be extended for the four phase scenarios of the mixed systems, and therefore the quantum error correction relations could be applied there, as we try to explain next.

So using the results of \cite{Dong:2021oad}, for the case of transitioning from $g=I$ to $g=\tau$, for the even values of $m$, we should have the inequality
\begin{gather}
\frac{Z_2^m}{Z_1^{2m} } \sim e^{(1-m) S_{BH} } > \frac{d_e^{m-1} }{k^m C_m },  
\end{gather}
and for the odd case the condition would be
\begin{gather}
\frac{Z_2^{m-1} }{Z_1^{2m-2} } > \frac{d_e^{m-1} }{ k^{m-1} C_{m-1} (2m-1)  }.
\end{gather}

Note that here $k=k_1 k_2$, and also $C_m= \frac{1}{m+1} \binom{2m}{m}$ is the Catalan number which is the number of non-crossing pairings here. For the disconnected and pairwise connected case, $d_e^{1-n}$ controls the behavior of the partition functions.

For going from $g=X$ to $g=\tau$, for the odd case then we should have the relation
\begin{gather}
\frac{Z_{2m-1} }{Z_1 Z_2 ^{m-1} } < \frac{(2m-1) C_{m-1} }{d_i ^m d_e^m k^{m-1} },
\end{gather}
while for the even case, the inequality
\begin{gather}
\frac{Z_{2m} }{Z_2^m} < \frac {k_2 ^{2m-2}  C_m}{d_i^m d_e^m k^{m-1} },
\end{gather}
should be satisfied, which unlike the odd case is independent of the $Z_1$.

For going from $g=X^{-1}$ to $g=\tau$, one only needs to swap $k_2$ to $k_1$ in the above relations.
 Note that $g=I$ cannot reach to $g=X$ or $g=X^{-1}$, and also $g=X$ is further away from $g=X^{-1}$, which can also be seen from the relations between the dimensions of the code subspaces.

 Therefore,  for a mixed setup, the lower bounds for each saddle, would depend in a rather complicated way to the number of microstates in each case, i.e, $k_i$, and the dimension of the code subspace of the interior or exterior, depending whether we are in the connected or the disconnected phase.

\section{Comparing partition functions and density states of $3d$ versus $2d$}\label{sec:dimensionsPage}

In \cite{Penington:2019kki}, for the computation of the purity, using the gravity amplitudes in $2d$ JT gravity, the density matrix and also the R\'enyi 2-entropy have been calculated where it was argued that for the $2d$ case, there are two ways of filling the boundary condition. In this section we would like to compare this procedure for the case of $3d$ gravity and specifically for the $\text{AdS}_3$ and BTZ solutions and then compare the result with the case of dimensional reduction from BTZ to $2d$ JT gravity of \cite{Verheijden:2021yrb}, in order to see how much of the information, phase structures, and patterns of entanglement would project from the higher dimensions to the lower dimensions, and to connect the different parameters of these theories.

Note that the theories in $3d$ and $2d$ have been connected together through either holography or dimensional reduction, where these connections have been extensively studied in \cite{Mertens:2018fds}.
The $3d$ gravity would lead to $2d$ \textit{JT gravity} by dimensional reduction and is connected to $2d$ \textit{Liouville theory} by holography. The $2d$ Liouville CFT would then lead to $1d$ \textit{Schwarzian theory} by dimensional reduction.

 \begin{figure}[ht!]
 \centering
  \includegraphics[width=6cm] {dimHol2}  
  \includegraphics[width=6cm] {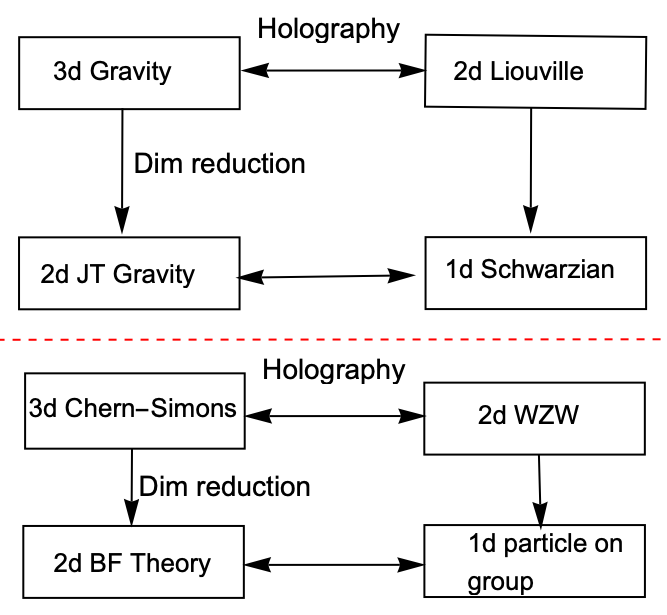}  
  \caption{The connections between low dimensional quantum gravity models through holography and dimension reduction.}
 \label{fig:dimHoltheories}
\end{figure}

One the other hand the $3d$ Chern-Simons theory would be related to $2d$ BF theory by dimensional reduction and to $2d$ WZW model by holography. The $2d$ WZW model after the dimensional reduction would lead to the model for $1d$ particle on group.

An example of $1d$ case is also the BFSS matrix model which arises in nearly $\text{AdS}_2/ \text{CFT}_1$ models. Note that one can even go to lower dimensions,  down to $D(-1)$-branes in type IIB string theory knows as IKKT matrix model. In \cite{Suzuki:2021zbe}, also the connections between $2d$ JT gravity and the $c <1$ limit of $2d$ string theory where its world-sheet theory consists of a spacelike Liouville CFT, coupled to a non-rational time-like Liouville CFT, has been studied.
Also, the $1d$ Schwarzian model can be connected to $1d$ random matrix model such as BFSS or $1d$ SYK model.

The dynamics governed completely by the Schwarzian derivative action
\begin{gather}
S=- C \int d\tau \{ F, \tau \},
\end{gather}
where $C=\frac{a}{16\pi G}$, and
 $\{F,\tau \}=\frac{F'''}{F'}-\frac{3}{2} \left (\frac{F''}{F'} \right)^2 $, where $F(\tau)$ is the time reparametrization which then becomes the physical and dynamical degree of freedom, and $a$ is a constant \cite{Blommaert:2020yeo}. This relation is completely identical to the Chern-Simons/WZW topological duality where the large gauge transformation of Chern-Simons theory becomes the dynamical degrees of freedom of WZW model. Now the goal is to understand the interconnections between these dynamics, the island formation and black hole evolutions in each case. One way to see these connections is to compute the boundary correlators of the thermal JT theory \cite{Blommaert:2020yeo}
\begin{gather}
\langle \mathcal{O}_{h_1}  \mathcal{O}_{h_2} ... \rangle_\beta = \frac{1}{Z} \int_{\mathcal{M}} \lbrack \mathcal{D} F \rbrack \mathcal{O}_{h_1} \mathcal{O}_{h_2} ... e^{C \int_0 ^\beta d\tau \{F, \tau \}}
\end{gather}
and compare with the corresponding ones in other theories.

When one writes the JT gravity in terms of the first order formulation of $SL(2, \mathbb{R})$ BF formulation of JT gravity, the operators $\mathcal{O}_{h_i}$ can be considered as the boundary anchored Wilson line. The Wilson line of this operator can be written as 
\begin{gather}
\mathcal{O}_h ( \tau_1, \tau_2) \equiv \left ( \frac{F'(\tau_1) F'(\tau_2)}{(F(\tau_1)- F(\tau_2) )^2} \right)^h,
\end{gather}
where $F(\tau)$ is time reparametrization for each end point of the Wilson line. The value of $h$ for the case where both of these end points are on the manifold $M$ (or $Q$), or one of them ends on $M$ and the other on $Q$, would be different. Therefore, this boundary two-point function and its holographic dual can probe the phase structures of black hole radiation and island formation.
The genus-zero partition function of JT gravity for each scenario is
\begin{gather}
\frac{1}{Z} \int dE_2 e^{- \beta E_2} \rho_0 (E_2) \int dE_1 \rho_0(E_1) e^{- \tau (E_1- E_2) }\frac{\Gamma (h \pm i \sqrt{E_1} \pm i \sqrt{E_2} ) }{\Gamma (2h)}, \ \ \ \ \ \tau= \tau_2 - \tau_1,
\end{gather}
where $Z$ is the Schwarzian disk partition function, i.e, genus $0$, and $\rho_0 (E) = \frac{1}{2\pi^2} \sinh 2 \pi \sqrt{E}$. 
However, for the mixed systems, or for the case where the island appears (or it becomes outside of the black hole horizon,) this relation is not complete. Specifically the order of the weight of the operators $h_i$ for the case ending on manifold $M$ which is the physical subsystem would be completely different from the case when the Wilson line ends on the manifold $Q$ where the island forms. In addition, this shows that only the ``massive" fields (graviton, scalar or fermion, etc) can detect the island formation as also shown in \cite{Demulder:2022aij}, from other methods. Therefore, the above relation for the mixed case, and also for the evolutions after the Page time become more complicated. We propose that, for the first order of approximation, this relation would become

\begin{multline}
\frac{1}{Z_M} \frac{1}{Z_Q} \frac{1}{Z_{QM}}\times \nonumber\\
\int (dE_2)_M e^{- \beta_M (E_2)_M} \rho_0 (E_2)_M \int d(E_1)_M \rho_0((E_1)_M) e^{- \tau ((E_1)_M- (E_2)_M) }\frac{\Gamma (h_M \pm i \sqrt{(E_1)_M} \pm i \sqrt{(E_2)_M} ) }{\Gamma (2h_M)}\times \nonumber\\
 \int (dE_2)_Q e^{- \beta_Q (E_2)_Q} \rho_0 (E_2)_Q \int d(E_1)_Q \rho_0((E_1)_Q) e^{- \tau ((E_1)_Q- (E_2)_Q) }\frac{\Gamma (h_Q \pm i \sqrt{(E_1)_Q} \pm i \sqrt{(E_2)_Q} ) }{\Gamma (2h_Q)}\times \nonumber\\
  \int (dE_2)_{QM} e^{- \beta_{QM} (E_2)_{QM}} \rho_0 (E_2)_{QM}\times \nonumber\\
   \int d(E_1)_{QM} \rho_0((E_1)_{QM}) e^{- \tau ((E_1)_{QM}- (E_2)_{QM}) }\frac{\Gamma (h_{QM} \pm i \sqrt{(E_1)_{QM}} \pm i \sqrt{(E_2)_{QM}} ) }{\Gamma (2h_{QM})},
\end{multline}
where the label $M$ specifies the parameters of the operators which are inserted on the manifold $M$, and $Q$ labels those which end on the end of the world brane where the island is being created, as shown in figure \ref{fig:islandBCFT1}.

Similar to \cite{Suzuki:2021zbe}, the connections between $2d$ WZW model and $2d$ BF theory can then be constructed. In addition, The model of $1d$ particle on group could be related to some matrix-like model.

So similar to the story depicted in \cite{Verheijden:2021yrb} which embedded the JT gravity in $3d$ BTZ black hole geometry, the $1d$ Schwarzian model can be embedded in $2d$ Liouville theory and the corresponding flow of information could get tracked. On the other hand the story of  \cite{Verheijden:2021yrb}  can be connected to compact group construction using the pictures in  \cite{Mertens:2018fds}.

Now, using the figure \ref{fig:dimHoltheories} which depicts the connections between various theories, we can study the problem of dynamical evaporation of black holes similar to \cite{Verheijden:2021yrb} for the case of $3d$ Chern-Simons, $2d$ WZW and $1d$ particle on group, as one of the motivations of this work is to investigate the island formation, black hole evolution, and saddles of mixed states of black hole radiation using gauge theory, WZW model and its connections to the JT gravity.

\subsection{The $3d$ Chern-Simons gravity to $1d$ particle on group}
In \cite{Verheijden:2021yrb} the $1d$ Schwarzian action has been derived from three dimensional gravity. We wish here to do the same for the Chern-Simons $3d$ gravity leading to $1d$ particle on a group.

So, first we study the dimensional reduction from $3d$ to $2d$ WZW. The  Chern-Simons action is
\begin{gather}
CS \lbrack A \rbrack = \frac{k}{4\pi} \int_{\mathcal{M}} \text{tr} \left ( A \wedge dA + \frac{2}{3} A \wedge A \wedge A \right ).
\end{gather}

The total gauge-invariant action for this theory is
\begin{gather}
\widetilde{CS} \lbrack A, g \rbrack = CS \lbrack A \rbrack + WZNW \lbrack g \rbrack + \frac{k}{4\pi} \int_{\partial \mathcal{M}} \text{tr}  \left ( 2 A_z \partial_{\bar{z}} g g^{-1}+ A_{\bar{z}} A_z \right ) d^2 z,
\end{gather}
where the Wess-Zumino-Novikov-Witten (WZNW) action is 
\begin{gather}
WZNW \lbrack g \rbrack = \frac{k}{4\pi} \int_{\partial \mathcal{M}} \text{tr} \left ( \partial_z g g^{-1} \partial_{\bar{z}}g g^{-1} \right) d^2 z - \frac{k}{12 \pi} \int_{\mathcal{M} } \text{tr} \left ( dg d^{-1} \right )^3.
\end{gather}

 The simplest $2d$ gravity model would be de-Sitter, which is characterized by $D= \Phi$, $V \propto \Phi$ and $Z \equiv 0$. In terms of the non-abelian gauge theory as in BF-type, it could be written as 
 \begin{gather}
 L_{deS} \lbrack A, B \rbrack = -2 \int_{\mathcal{M}} \text{tr} (BF),
 \end{gather}
where $F= dA + A \wedge A$ is the curvature two-form of the standard gauge field $A$, and $B$ is a function on the $2d$ manifold $\mathcal{M}$.

The boundary term which similar to the Gibbons-Hawking term would be necessary to add to the above action to make it gauge invariant can be written in the following form
\begin{gather}
S_{\text{boundary}}= 2 \int_{\partial \mathcal{M}}  \text{tr}  \left ( B(A+ dg g^{-1} ) \right).
\end{gather}

So the whole action would be
\begin{gather}
\widetilde{L_{\text{deS}}} \lbrack A, B, g \rbrack = -2 \int_{\mathcal{M}} \text{tr} (BF)+2 \int_{\partial \mathcal{M}}  \text{tr}  \left ( B(A+ dg g^{-1} ) \right).
\end{gather}

This corresponds to the Gibbons-Hawking term, i.e, equation 2.24 of \cite{Verheijden:2021yrb}. This action is invariant under $g \to g^h= h^{-1} g$.

Also, one can write the $3d$ CS theory as
\begin{gather}
S_{CS} \sim \int_{M_3} d^3x \epsilon^{ijk} A_i \partial_j A_k,
\end{gather}
which by setting $A_\phi \chi$ and $\partial_\phi$ to zero, one can get the $2d$ BF theory as
\begin{gather}
S= \int_M d^2 x \chi F + \frac{1}{2} \oint_ {\partial M} dt \chi A_0.
\end{gather}

For the matter action in \cite{Mertens:2018fds}, the boundary term has been found as
\begin{gather}
S_{\text{matter} } = - \oint_{\partial M} dt \sigma J^r = - \oint_{\partial M} dt \ \sigma (J_+ - J_-),  
\end{gather}
which is the net inward flux of charge which corresponds to the relation 2.24 of \cite{Verheijden:2021yrb}, 
\begin{gather}
S_{GH}= \frac{2\pi \alpha \ell}{8 \pi G^{(3)} } \int dt \sqrt{-h_{tt} } \phi_b \left ( K^{(3) } + \frac{2}{\ell} \right ),  
\end{gather}
representing the flux of dilaton through the curvature of the geometry.

Additionally, similar to the $3d$ to $1d$ case of \cite{Verheijden:2021yrb} for the Einstein theory,  the non-abelian BF theory  could be written as
\begin{gather}
S=\int_M d^2 x \text{Tr} \chi F + \frac{1}{2} \oint_{\partial M} dt \text{Tr} \chi A_0,
\end{gather}
which after applying the boundary condition reduces to a particle on a group manifold action as
\begin{gather}
S= \frac{1}{2} \oint_{\partial M} dt \text{Tr} ( g^{-1} \partial_t  g)^2. 
\end{gather}

The only difference that this boundary term can create would be in the Hamiltonian part $H= H_{\text{grav} } + H_{CS}$  leading to just a shift in the energy
as
\begin{gather}
T_{tt} = \frac{\dot{\sigma}^2 }{2}, 
\end{gather}
which its effects on the mixed correlations, similar to the case of \cite{Verheijden:2021yrb} can be investigated. This injection or shift in the energy momentum tensor could be considered as the quench in BCFTs and similar to \cite{Kawamoto:2022etl}, the entropy and energy could be related by the first law of entanglement as 
\begin{gather}
T_{tt} (x_A, t)=  \underset{| x_A - x_B|    \to 0}{\lim} \ \frac{3}{ \pi | x_A - x_B|^2} \ . \ \Delta S_{A, B} ( x_A, x_B, t).
\end{gather}

So additional entropy $\Delta S= \frac{\pi}{6}  \frac{\Delta \dot{\varphi}^2 }{ \underset{| d\varphi|  \to 0}{\lim} \frac{1}{d\varphi^2}}$ would be created due to this injection of energy, which for two strips and for the phase structure of the mixed system would have a significant effect. This energy momentum tensor also affects the flux to the bath and therefore the resulting linear and exponential evaporation of the black hole, such as those studied in \cite{Verheijden:2021yrb}

The story of the black hole evolution, Page curve and island formations could also be studied through CS/WZW duality  instead of $2d$ JT gravity model.  The WZW is more string theory based, where its symmetry algebra is an affine Lie algebra, i.e, $G(z) \times G(\bar{z})$ symmetry.  In string theory, the $\text{SL}(2, \mathbb{R})/ U(1)$ gauged WZW model have been interpreted in \cite{PhysRevD.44.314} as the Witten's two-dimensional Euclidean black hole. Geometrically, the WZ term describes the torsion of the manifold, which then affects the phase transitions and mixed correlations.

The black hole and the island both can be defined from the $\text{SL}(2, \mathbb{R})$ group manifold \cite{Ashok:2021ffx} which could be written as the hyperboloid as 
\begin{gather}
x_{-1}^2+x_0^2 -x_1^2 - x_2^2=1,
\end{gather}
where the black hole corresponds to the topological circle in the $(x_{-1} , x_0)$  plane that never shrinks to zero. The island also should be defined on the same plane. Both the BTZ black hole and the island would be obtained by dividing the universal covering group by ``two" $\mathbb{Z}$ orbifold actions.  The group elements of $\text{SL}(2, \mathbb{R})$ would act as
\begin{gather}
g \to e^{\pi (r_+ - r_-) \sigma_3} g e^{\pi (r_+ + r_-) \sigma_3} ,
\end{gather}
where $\sigma_3$ is the diagonal Pauli matrices.  

The symmetries in the WZW model could define the generators by exponentiating $\sigma_3$ which then can lead to the energy and angular momentum generators. This could also be written using the group element parametrization
\begin{gather}
g= e^{\frac{r_+ - r_-}{2} (t+\phi) \sigma_3 } e^{\rho \sigma_1} e^{-\frac{r_+ +  r_-}{2} (t-\phi) \sigma_3 },
\end{gather}
where the radial coordinate $\rho$ is related to the BTZ radial coordinate $r$ through
\begin{gather}
\cosh^2 \rho = \frac{r^2 - r_-^2}{r_+^2 - r_-^2}. 
\end{gather}

The geodesic length and therefore the entanglement entropy and mixed correlation measures corresponding to the WZW case then can be defined using these $g$ elements.
On the other hand, similar to the usual case of the JT gravity action coupled to a CFT bath, the corresponding ``gauged" model contains two BF systems coupled to each other. Then, the arising of the Page curve in the group analysis and the orbifolded manifold could be found by analyzing the geodesics.

Note that the JT gravity itself can be written as an equivalent Schwarzian quantum mechanical theory on the holographic boundary. The JT gravity can be considered as the s-wave dimensional reduction of the $3d$ pure gravity with $\Lambda <0$.

\subsection{The connections between JT model and Liouville gravity}

First, we review the change of partition function during black hole evolution and Page transition in $2d$ JT model as studied in \cite{Penington:2019kki}, and then compare it with the case of $2d$ BF theory and $2d$ WZW model.

 The action in \cite{Penington:2019kki} was for a black hole in JT gravity with an EOW brane behind the horizon which is a $\mathbb{Z}_2$ quotient of the two-sided black hole. A particle which has the mass  $\mu$, ($\mu \ge 0 $) would be considered behind the horizon. For this system, the Euclidean action can be written as
\begin{gather}
I=I_{JT}+ \mu \int_{\text{brane}} ds,
\end{gather}
where  the JT action is
\begin{gather}
I_{JT}= - \frac{S_0}{2\pi}  \left \lbrack \frac{1}{2} \int_{\mathcal{M}} \sqrt{g} R + \int_{\partial \mathcal{M}} \sqrt{h} K \right \rbrack -\left \lbrack \frac{1}{2} \int_{\mathcal{M}} \sqrt{g} \phi ( R +2 ) + \int_{\partial \mathcal{M}} \sqrt{h} \phi K \right \rbrack.  
\end{gather}

The partition function is
\begin{gather}
Z_n=e^{S_0} \int_0^\infty d\ell_1 ... d\ell_{2n} e^{\frac{\ell_1+ ...+ \ell_{2n}}{2} } I_{2n} (\ell_1,..., \ell_{2n}) \varphi_\beta (\ell_1) e^{-\mu \ell_2} . . . \varphi_\beta(\ell_{2n-1}) e^{-\mu \ell_{2n}},
\end{gather}
where $I_{2n}$ are
\begin{gather}
I_{2n}( \ell_1, . . . , \ell_{2n}) = 2^{2n} \int_0^\infty ds \rho(s) K_{2is}(4 e^{- \frac{\ell_1}{2})} ... K_{2is} ( 4 e^{- \frac{\ell_{2n}}{2} }), \ \ \ \ \ \ \ \rho(s)=\frac{s}{2\pi^2} \sinh (2\pi s),
\end{gather}
and $\varphi_\beta ( \ell)$ is
\begin{gather}
\varphi_\beta ( \ell) = 4 e^{-\frac{\ell}{2} } \int_0^\infty ds \rho (s) e^{-\frac{\beta s^2}{2}} K_{2is} ( 4 e^{-\frac{\ell}{2} } ).
\end{gather}
Here, $K$ is the modified Bessel function and $\varphi_\beta$s are the Hartle-Hawking state in the geodesic basis, and $I_{2n}$ consists of $n$ geodesics on the EOW branes and $n$ geodesics that should be glued to form the Hartle-Hawking states.

If one assumes that the EOW brane has a very large number of internal and orthogonal states $k$, the Page transition would be between $\log(k)$ and $S_{BH} \sim S_0+2\pi \phi$.  So the island formulation for the JT predicts
\begin{gather}
S(R)=\text{min} \{ \log (k), S_{BH} \}.
\end{gather}
So the transition is between the two and a function of $k$, where then also after dimension reduction it would change, such as those examples in \cite{GonzalezLezcano:2022mcd}.

So the relations we can observe between these theories are dual to the equivalence between the three definitions of the boundary entropy explained in \cite{Akal:2020twv,Miyaji:2021ktr}, i.e, between the disk amplitude, $S_{\text{bdy} (\alpha)  } = \log g_\alpha$, where $g_\alpha \equiv \langle 0 | B_\alpha \rangle$, and cylinder amplitude $Z_{(\alpha, \beta)}^\text{cylinder}= \langle B_\alpha | e^{ - H L} | B_\beta \rangle \approx g_\alpha g_\beta e^{-E_0 L} $, where $g_\alpha g_\beta$ is related to the boundary part and $e^{-E_0 L }$ is related to the bulk part, and finally to the usual entanglement entropy relation of $2d$ CFT case,  $S_A = \frac{c}{6} \log \frac{l}{\epsilon} + \log g_\alpha$.

The level $k$ in the WZW model is actually related to this $g_\alpha g_\beta$ of the boundary part, which then after the dimension reduction further, would be related to the term of the $1d$ particle on a group manifold, i.e, $\text{Tr} ( g^{-1} \partial_t g)^2$.

In the $\text{AdS}_3/ \text{BCFT}_2$ setup, the boundary entropy which is a measure of degrees of freedom at the boundary can be written as $g=e^{S_{\text{bdy}} }$ where $g$ is a function that is monotonically decreasing under the boundary RG flow and is a function of the boundary conditions labeled by $\alpha$. The movement of the island from inside of the black hole toward the horizon will decrease the degrees of freedom on the boundary brane, as it enforces this flow leading to this monotonic behavior.  Another interesting connection between the wormhole and island is that in the replica trick calculation of the von Neumann entropy, by taking $n \to 1$ the mouth of the wormhole would become the island itself, which again leads to the result found before that by changing the topology and by increasing the genus, the island can move inside of the horizon.

\begin{figure}[ht!]   
\begin{center}
\includegraphics[width=0.45\textwidth]{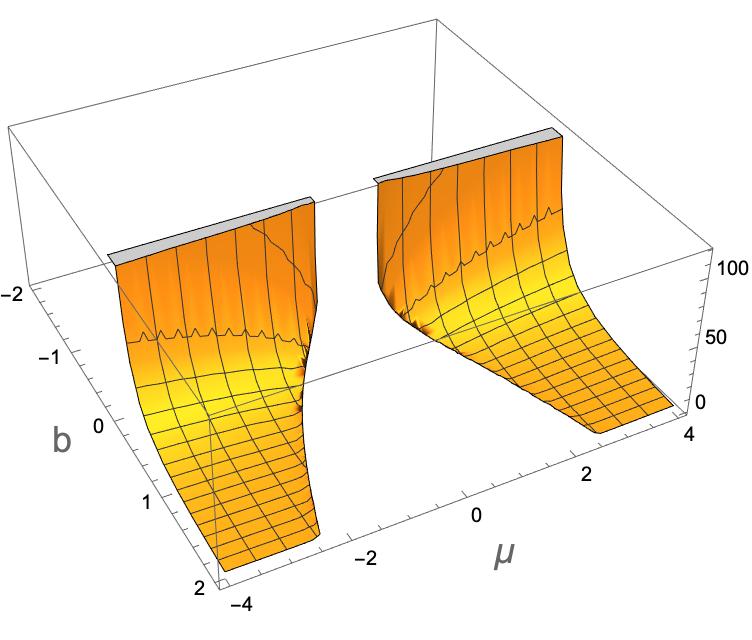}  
\caption{The relations between the entanglement entropy of the island and the parameters of the Liouville potential, where we fix $y_a^- =y_+^a=2, c=2$.}
\label{fig:island1.jpeg}
\end{center}
\end{figure}

Then, the $2d$ Liouville gravity which can be interpreted as the quantum (q) deformation of JT gravity, and also a specific model of the sinh dilaton gravity \cite{Mertens:2020hbs} can be written as
\begin{gather}\label{eq: liouville}
S_L= \frac{1}{4\pi} \int_\Sigma \left \lbrack  (\tilde{\nabla} \phi)^2 + Q \tilde{R} \phi + 4 \pi \mu e^{2 b \phi} \right \rbrack, \ \ \ Q= b+b^{-1}, \ \ c_L=1+Q^2,
\end{gather}
where $b$ is the parameter which defines the Liouville theory, and $\mu$ here is the boundary cosmological constant. For building the Fateev-Zamolodchikov-Zamolodchikov-Teschner (FZZT)-brane boundary model, a boundary term as \begin{gather}
\frac{1}{2\pi} \oint_{\partial \Sigma} \left \lbrack Q \tilde{K} \phi+ 2 \pi \mu e^{b \phi} \right \rbrack,
\end{gather}
is being added to the Liouville action \ref{eq: liouville}, where $\mu_B$ is the boundary cosmologicall constant. The metric and the field $b$ are related as $ds^2 e^{2 b \phi{ dz d\bar{z}}} $, and the boundary length is $\oint e^{b \phi} $, and these parameters could be looked at from one-dimension higher and the compactification point of view.  For the Liouville theory, the entanglement entropy with the presence of an island has been calculated in \cite{Tian:2022pso}, where as shown in figure \ref{fig:island1.jpeg}, it can be seen that by increasing the parameters of the potential, $\mu$ and $b$, the entanglement entropy of the island would decrease, and since $b$ is in the exponent, it has a higher effect on breaking the correlations. This Liouville potential $C \mu e^{b \phi}$ would reflect the energy eigenvectors before reaching $\phi = + \infty$, and also the background charge is $Q=b+ \frac{1}{b}$. Note that this exponential potential also breaks the momentum conservation which its effect during black hole evolution and island entanglement can be seen. In Liouville theory, the correlation functions just depend on $b$ and the momenta, therefore the interpolations between the saddles of entanglement and evolution of island, also just depend on these two parameters.

For several generalized dilaton theories, in \cite{Tian:2022pso}, the island formulation for higher dimensional spherical black holes with asymptotically flat spaces have been studied. By correcting the black hole solution that is being used, the authors found that the island formulation can always solve the information paradox, including for the case of ``Liouville black hole solution'', and the island always appear barely outside of the horizon.  There, the island structures for several charged black holes have also been studied. Specially, for the case of charged dilaton black hole with the $4d$ metric
\begin{gather}
ds^2= -r^2 \left ( 1- \frac{2M}{r^2} + \frac{Q_c^2}{4r^4} \right ) dt^2 + \left( 1- \frac{2M}{r^2} + \frac{Q_c^2}{4r^4} \right) ^{-1} dr^2 + r^2 ( dx^2 + dy^2),
\end{gather}
the ``effective" $2d$ model would be
\begin{gather}
ds^2= - H(r) dt^2 + r^2 H(r)^{-1} dr^2, \ \ \ \ X=r^2, \ \ \ \ \ H(r)= r^2 \left ( 1- \frac{2M}{r^2} + \frac{Q_c^2}{4r^4} \right ),
\end{gather}
and then their results for the position of island match with the result of \cite{Ahn:2021chg}. 

In the late times, the island appear barely outside of the horizon, while in the early times of radiation the island is ``inside" the black hole horizon as it moves towards the horizon gradually by the Hawking radiation. So the islands are not stationary but in fact they effectively move by the effect of the momentum of radiation. The speed of the movement of the island through the process of radiation and at each step can be studied using the momenta of Hawking quanta at each step.  

One way to understand the movement of island inside the black hole is using the renormalization flow. So, for the metric of the form 
\begin{gather}
ds^2= e^{2 A(\rho)} \lbrack - f(\rho)^2 dt^2 + d \vec{x}^2 \rbrack + d \rho^2,
\end{gather}
the a-function would have the form of 
\begin{gather}
a_T (\rho) = \frac{\pi^{d/2}  }{ \Gamma  \left (\frac{d}{2} \right)  \ell_P^{d-1} } \left \lbrack \frac{f(\rho) }{A'(\rho) }  \right \rbrack^{d-1},
\end{gather}
where the speed of the movement of the island toward the horizon from the inside, for the general black hole, would be proportional to the derivative of $a_T$ as
\begin{gather}
v_{\text{island} } \propto  \frac{da_T}{d \rho} = \frac{ (d-1) \pi^{d/2}  }{\Gamma \left (\frac{d}{2} \right ) \ell_P^{d-1}   }  \frac{ f(\rho)^{d-2}  }{ A' (\rho)^d} \times \lbrack f'(\rho) A'(\rho) - f(\rho) A''(\rho) \rbrack.
\end{gather}

\begin{figure}[ht!]   
\begin{center}
\includegraphics[width=0.45\textwidth]{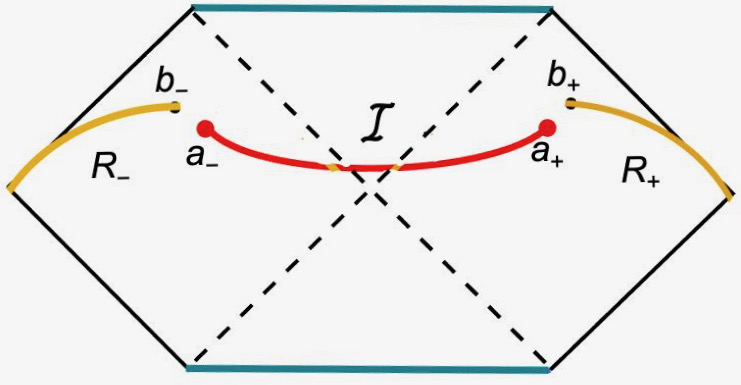}  
\caption{Penrose diagram with island configuration for two sided black hole.}
\label{fig:penroseisland}
\end{center}
\end{figure}

 \begin{figure}[ht!]   
\begin{center}
\includegraphics[width=0.45\textwidth]{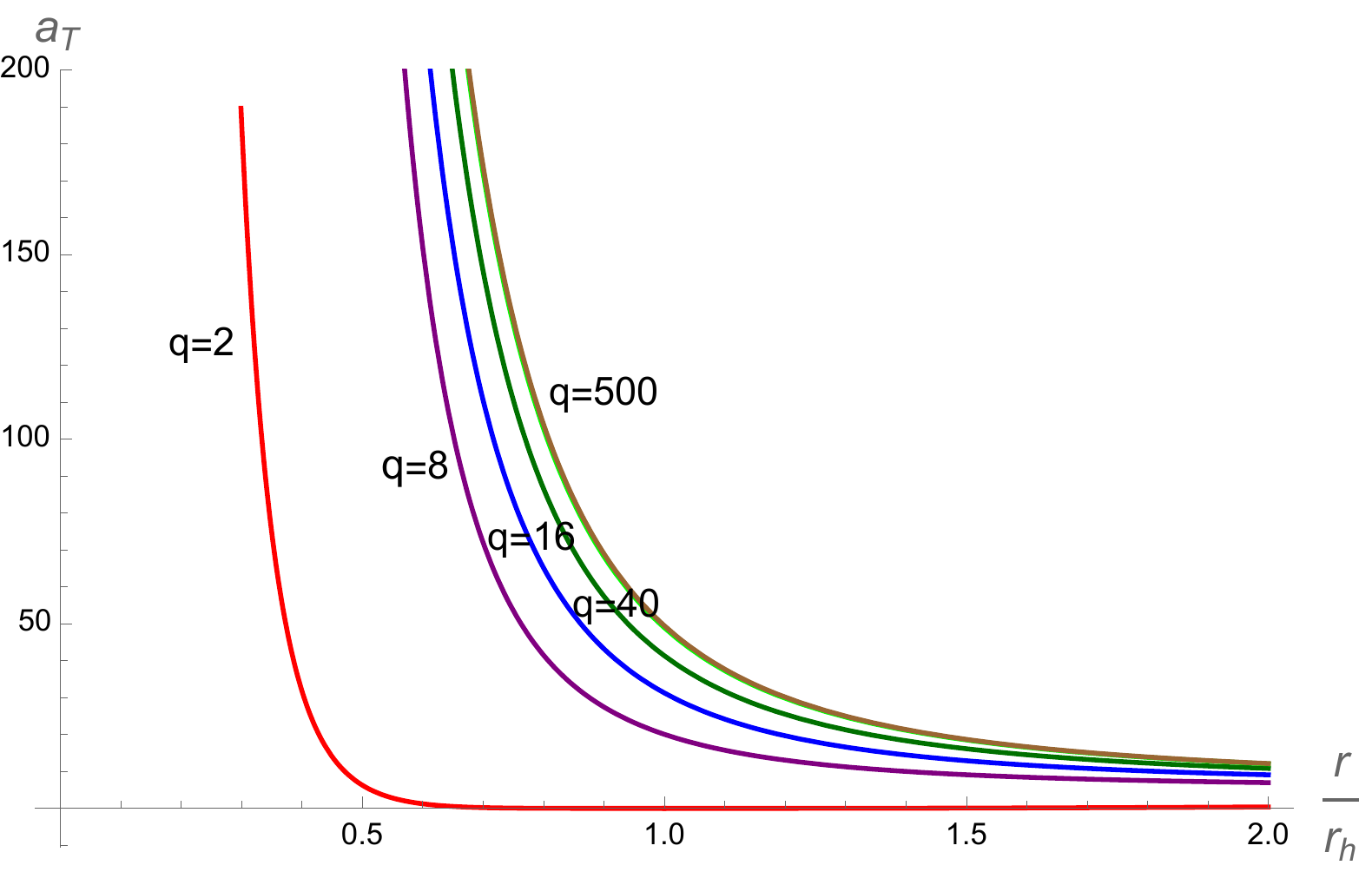}  \ \ \ \ \ \ \ \ \ \ \ \ 
\includegraphics[width=0.45\textwidth]{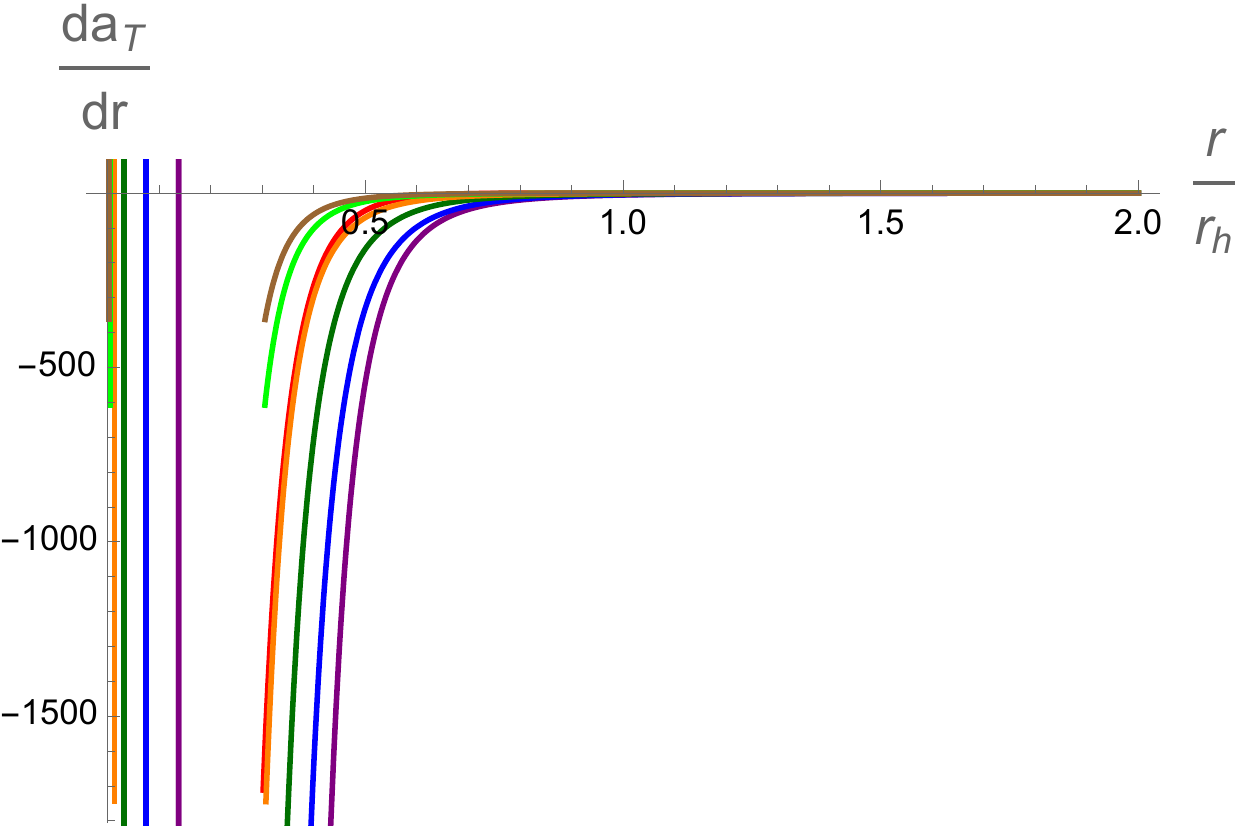}  
\caption{The behavior of a-function versus $r$ for Schwarzschild black hole.}
\label{fig:penroseisland}
\end{center}
\end{figure}

 \begin{figure}[ht!]   
\begin{center}
\includegraphics[width=0.45\textwidth]{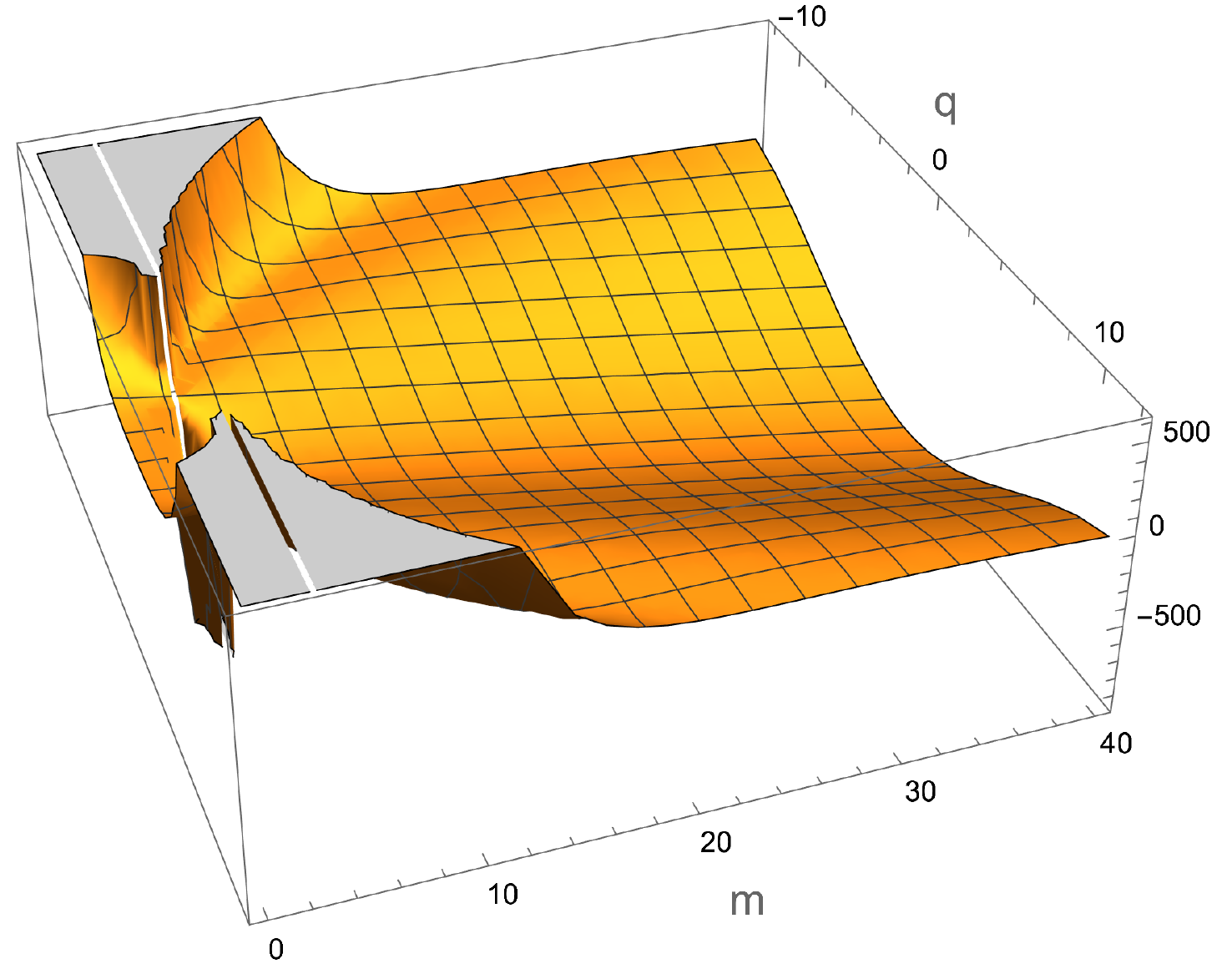}  
\caption{The behavior of the a-function versus $q$ and $m$, at $r=2$, for Schwarzschild black hole.}
\label{fig:aTmq}
\end{center}
\end{figure}

The behavior of the $a_T$ which determines the behavior of the renormalization flow versus $m$ and $q$ are shown in figure \ref{fig:aTmq}, where one could notice that the mass decreases the flow and the charge would increase it, which also agrees with the expectation that the mass can slow down the movement of the island and charge can increase its velocity, as another evidence that the velocity of the island moving inside of the black hole toward the horizon is proportional to $a_T$.

Note also that there is another velocity dubbed \textit{``entanglement velocity"}, $v_E$, which has an instantaneous bound $|v_E(t)| \le 1$.

This effect could also be analyzed from the geometric perspective and the $3d$ point of view similar to \cite{Verheijden:2021yrb}. The shape of the figure \ref{fig:island1.jpeg} demonstrates that for Liouville theory the results are compatible as well.

The Wess-Zumino-Witten (WZW) model and the Liouville theory are also related to each other by a quantum Drinfeld-Sokolov reduction of the former. Also, since the correlation functions of the Euclidean $SL_2 (\mathbb{R})$ WZW model (dubbed $H_3^+$ model), the $2d$ black hole $SL_2/U_1$ and the theories which continuously interpolate between Liouville and $H_3^+$ model, could be written in terms of the correlation functions of Liouville theory, one would expect that the behavior of the island in these theories would be the same, with no way of distinction.  Also, Liouville theory with $c\ge 25$ can be mapped exactly to some log-correlated random energy models with a random potential that is logarithmically correlated.  The correlation functions then would get mapped to the correlation functions of the Gibbs measure of the particle, where in $2d$, it becomes the Gaussian free field model. Again, we expect that the entanglement entropy cannot distinguish such theories and island would behave the same way. 

\begin{figure}[ht!]   
\begin{center}
\includegraphics[width=0.45\textwidth]{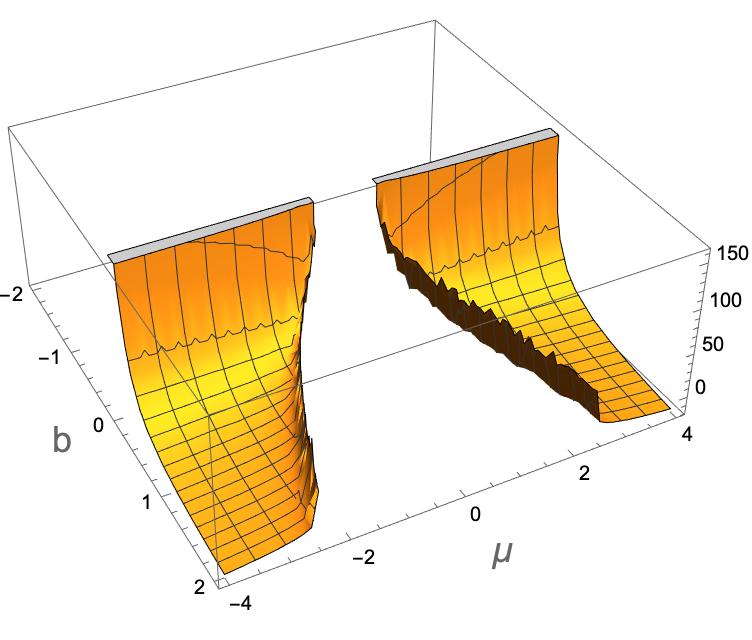}  
\caption{The relations between the entanglement entropy of the island and the parameters of the log-correlated random energy models, for large central charge and for $y_a^- =y_+^a=2, c=25$. The oscillatory wall starts to appear around $c=5$.}
\label{fig:largecIsland.jpeg}
\end{center}
\end{figure}

For large $c$, i.e, $c=25$, we did not detect a particular specific change, and only for bigger $c$s, close to zero, a wall is taking shape which is noticeable after $c=5$. For $c=25$, its plot is shown in figure \ref{fig:largecIsland.jpeg}. 

\begin{figure}[ht!]   
\begin{center}
\includegraphics[width=0.45\textwidth]{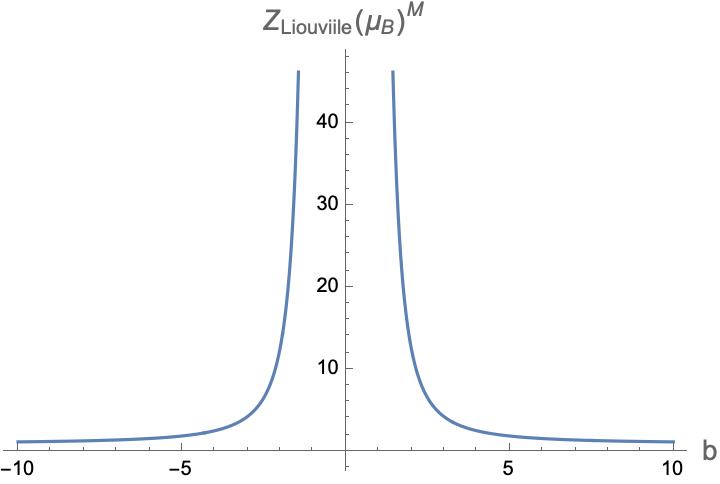}  
\caption{The relations between the partition function and $b$ in Liouville theory, where we fix $s=2$.}
\label{fig:Liouvillepartition.jpeg}
\end{center}
\end{figure}

The effects of $b$ on the black hole evolution in Liouville theory, can also be seen from the disk partition function of Liouville \cite{Seiberg:2003nm} with an FZZT brane boundary as well, which has the relation 
\begin{gather}
Z(\mu_B)^M \sim \mu^{\frac{1}{2b^2} } \cosh \frac{2\pi s}{b},
\end{gather}
where $\mu_B (s)= \kappa \cosh 2\pi b s $ is the FZZT brane parameter, and $\kappa \equiv \frac{\sqrt{\mu} }{\sqrt{\sin \pi b^2} }$. The plot is shown in figure \ref{fig:Liouvillepartition.jpeg}, where again one can see that increasing $b$ would reduce the partition function.

\begin{figure}[ht!]   
\begin{center}
\includegraphics[width=0.46\textwidth]{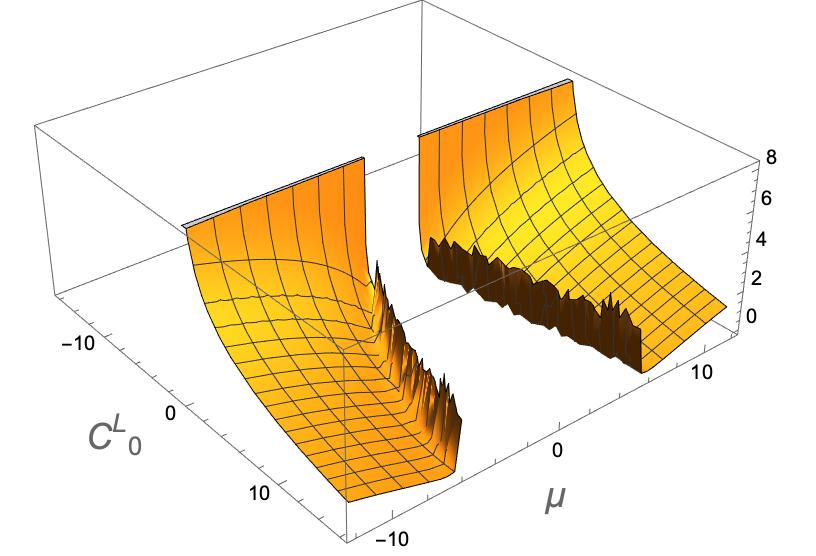} \ \ \ \ \ \ \ \ \ 
\includegraphics[width=0.46\textwidth]{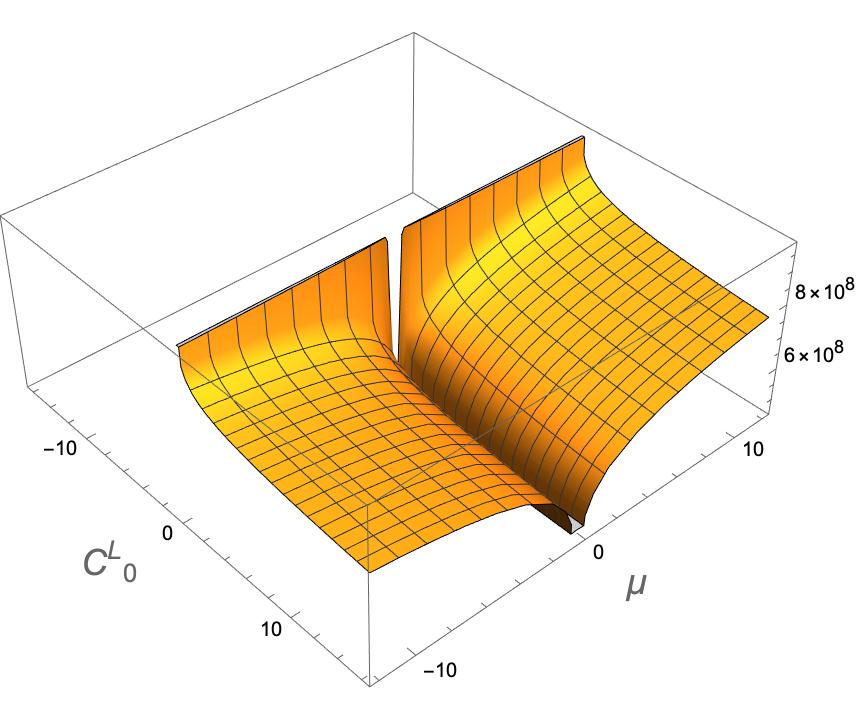}   
\caption{In the left, the entanglement entropy with island in Liouville gravity, with parameters $c=2$ and $\beta=2$ is shown, and in the right we set $\beta=10^{-6}$ and $c=2$, where the transformation to a smoother JT theory can be observed.}
\label{fig:smallBeta}
\end{center}
\end{figure}

As shown in \cite{Mertens:2020hbs}, the Liouville gravity amplitude and JT gravity partition functions are also related to each other by a double scaling limit, i.e, $b \to 0$ where the boundary length go to infinity, i.e, $\ell \sim \frac{\ell_{JT} }{\kappa b} \to + \infty$. Setting $\beta$ to a constant  value such as 2 or $10^{-6}$, one could see the difference between the entanglement entropy of the island in these two limits in figure \ref{fig:smallBeta}. Note that here $C_0^L$ is the constant coming from the integral which we can vary, and $\mu$ is the factor in the Liouville potential. The limit of $b \to 0$ is also related to the genus zero of Weil-Petersson volume.

These results can also be checked from the group interpretations, i.e, BF theory for the JT gravity, and WZW model for the Liouville theory.

In \cite{Verheijden:2021yrb}, using partial dimension reduction from $3d$ to JT, the authors concluded that the black hole energy decays exponentially.  For the non-evaporating black holes, they found 
\begin{gather}
E= - \frac{\Phi_r}{8\pi G} \left \{ \tau, t \right \} = \frac{2\pi^2}{\beta^2} \frac{\Phi_r^0}{8\pi G}\equiv E_0,
\end{gather}
while for the evaporating case which the dilaton behaves as $\tilde{\Phi}_r = 2 \pi \ell \alpha(\tilde{t}) \equiv \Phi_r^0 \alpha( \tilde{t})$ and $\alpha (\tilde{t})$ decreases from 1 to 0, the temperature of the $2d$ black hole remains the same, which is directly inherited from the $3d$ BTZ black hole.

In the evaporating case where the dividing line between the JT gravity and CFT part of BTZ moves, the equation of motion changes and the dilaton becomes time dependent as $\tilde{\Phi}_r = \Phi_r^0 \alpha( \tilde{t})= \Phi_r^0 \left ( 1-\frac{A}{2} \tilde{t} \right )$, where $\frac{A}{2} = \epsilon \frac{c}{6} \frac{G} {\Phi_r^0}$ is the evaporation rate, and the energy decreases as $\frac{d \tilde{E}}{d \tilde{t}} = - \epsilon \frac{c}{24\pi} \frac{2 \pi^2}{\beta^2} = - E_0 \frac{A}{2}$. Using this picture, the partition function can also be time dependent and the time evolution would induce the interpolation between the JT and Liouville in the IR and modify the boundary in the UV case.

So considering the UV corrections of \cite{Mertens:2020hbs}, and by changing the dimensional reduction parameter $\alpha$, the evaporation of black hole changes from the exponential behavior to a power law, specifically this could be observed in the model of the geometric evaporation.

As for the case of Liouville and JT models, by interpreting $\ell = \beta$ as inverse temperature, the partition function would be
\begin{gather}
Z(\beta) \sim \int_\kappa^\infty dE e^{-\beta E} \rho_0 (E), \ \ \ \ \ \rho_0 (E) = \sinh \left ( \frac{1}{b^2} \text{arccosh} \frac{E}{\kappa} \right ),
\end{gather}
where in the IR limit, i.e, $E= \kappa+ \epsilon$, it gives the JT regime,  as it can also be seen from figure \ref{fig:dimHoltheories}.  In the UV, the Cardy scaling limit of JT gravity is being modified from $\rho_0 (E) \sim e^{2\pi \sqrt{E}}$ into power-law $\rho_0 (E) \sim E^{1/b^2} $. The thermodynamic saddle would also be $\sqrt{E^2 - \kappa^2}= \frac{1}{b^2 \beta}$, where in the IR gives the first law of JT black hole, i.e, $\sqrt{E_{JT}} \sim \beta^{-1} $ and in the UV shows that the boundary of the bulk geometry is being modified, and it would not be asymptotically AdS anymore, which has not been considered in \cite{Verheijden:2021yrb}. Therefore, for the full solutions, such as the case of fast rotations and a UV complete theory, to derive the Page curve, this modifications from exponential to power law should be considered, i.e, additional term should be added to the Schwarzian action $S=\frac{1}{8\pi G} \int dt \Phi_r \{ \tau, t \}$ which becomes dominant in the UV limit.

\subsection{The $3d$ Wess-Zumino-Witten model}

Now we would like to find the corresponding parameters for the WZW model in the $3d$ case,  and the corresponding behavior of the islands using this model.

In general, qualitatively one could notice that $\sqrt{h} \phi K$ in $2d$ JT model corresponds to $\chi A_0$ in $2d$ BF model and the term $\sqrt{g} \phi R$ in JT would correspond to the term $\chi F$ in the BF model. Also, the Liouville momentum which is a continuous parameter labeled by $k$ is related to continuous irreducible representation of $SL(2,\mathbb{R})$ labeled by $R$. 

Another way to see the connections could be done through the study of \cite{Mertens:2019tcm}, using the deformation of each theory. The defect in the $2d$ JT model (such as conical defects or wormholes) is related to the deformation of the $1d$ Schwarzian theory where the reparametrization mode is integrated over different co-adjoint orbits of the Virasoro group. Geometrically, there are two parameters, $\Phi_h$ which is the horizon area operator, and $L(\gamma)$ which is the geodesic length operator, which then as shown in \cite{Mertens:2019tcm} is structurally related to the deformation of the particle-on-a-group quantum mechanics where a ``chemical potential" or a ``magnetic monopole"  is being added. These chemical potentials then would change the partition functions the way derived in \cite{Mertens:2019tcm}.

Note that the effects of these defects in BF theory are similar to the case of confining backgrounds with the wall at the end of the geometry which can increase the number of saddles for the mixed system, i.e, for the two symmetric subsystem, it changes the number of phases from two to four as shown in \cite{Jain:2020rbb, Ghodrati:2021ozc, Ghodrati:2022kuk}.

For the partition function of the $3d$ case instead of the $2d$ JT gravity, we could consider the $SL(2, \mathbb{R})$ WZW model as
\begin{gather}
S_{\text{WZW}} \lbrack g \rbrack = \frac{k_L}{8\pi} \int d\tau d \sigma \sqrt{-h} \text{Tr} ( \partial_a g^{-1} \partial^a g ) + k_L \Gamma_{\text{WZ}} \lbrack g \rbrack,\nonumber\\
\Gamma_{\text{WZ}} \lbrack g \rbrack = - \frac{1}{12\pi} \int \epsilon^{abc} \  \text{Tr} \ ( \partial_a g \ g^{-1} \partial_b g \ g^{-1} \partial_c g  \ g^{-1} ),
\end{gather}
 where $k_L$ is generally a complex number and is the level on the Riemann surface $\Sigma$. However, for the compact manifolds we have $k_L \in \mathbb{Z}$, and by comparing with the JT gravity, it could be seen that $k_L$ corresponds to the entropy of black hole at zero temperature.

The fields of this model, $g(z, \bar{z})$, are matrices which are the (faithful) representation of the Lie-group G, where $g(z, \bar{z})$ is a map from the manifold to the group, as $g: S^2 \longrightarrow G$.
 
 The left and right currents of this theory would be defined as
 \begin{gather}
 J_-= k_L g^{-1} \partial_- g, \ \ \ \ \ \ \ \ J_+ = - k_L \partial_+ g g^{-1},
 \end{gather}
 and the worldsheet stress tensor is
 \begin{gather}
 T_{\pm \pm}=  \frac{1}{2k_L} \text{Tr} ( J_{\pm} J_{\pm}) \sim - \frac{k_L}{4} m^2,
 \end{gather}
 where $m$ is  the mass of a point-like probe particle. Then, according to the first law of entanglement entropy, this energy momentum tensor would cause the shift in the $S_{EE}$.

Note also that in string theory, the $SL(2,\mathbb{R}) / U(1)$ gauged WZW model would be interpreted as the Witten's $2d$ Euclidean black hole. The spectrum and partition function of this $3d$ BTZ black hole have been derived in \cite{Israel:2003ry, Nippanikar:2021skr}. This partition function for $\text{AdS}_3$ has been found as \cite{Israel:2003ry}
\begin{gather}
Z= 4 \sqrt{\tau_2} ( k_L-2) ^{\frac{3}{2}} \int^1_0 d^2 s \int^1_0 d^2t \frac{e^{\frac{2\pi}{\tau_2} (\text{Im} (s_1 \tau-s_2) )^2 }  }{|\vartheta_1 (s_1 \tau-s_2 | \tau) |^2 } \nonumber\\
\times \sum_{m, w, m' , w' \in \mathbb{Z}} \zeta \begin{bmatrix}
w+s_1-t_1\\
m+s_2-t_2
\end{bmatrix} (k_L) \  \zeta \begin{bmatrix}
w'+t_1\\
m'+t_2
\end{bmatrix} (-k_L), 
\end{gather}
where the free boson conformal blocks that appear in this expression is defined as
\begin{gather}
 \zeta \begin{bmatrix}
w\\
m
\end{bmatrix} (k_L)  = \sqrt{\frac{k_L}{\tau_2}} \text{exp} \left ( - \frac{\pi k_L}{\tau_2} | w \tau -m |^2  \right ),
\end{gather}
and the $\vartheta_1$ function is 
\begin{gather}
\vartheta_1 ( v | \tau) = \sum_{p \in \mathbb{Z}} e^{\pi i \tau (p+ \frac{1}{2})^2+ 2 \pi i (v+\frac{1}{2})(p+\frac{1}{2}) }.
\end{gather}

By taking $q=e^{2 \pi i \tau}$,  it can be written as an infinite product as
\begin{gather}
\vartheta_1 ( v | \tau) =- 2 q^{1/8} \sin \pi v \prod_{p=1}^\infty (1- e^{2 i \pi v}  q^p) (1- q^p) ( 1- e^{-2 i \pi v} q^p). 
\end{gather}

In \cite{Nippanikar:2021skr}, this partition function has been expanded as
\begin{flalign*}
Z&  =  8 i (k_L-2) \sum_{w, w', n , n' \in \mathbb{Z}\ \& N, \bar{N} } \delta_{n, n'}  \int_0^1 dt_1 \Bigg \{ \int_0^\infty ds \rho(s) \text{exp} \left \lbrack -2 \pi \tau_2 \left ( \frac{2s^2 + 1/2}{k_L-2} \right) \right \rbrack & \nonumber\\ &
+ \sum_{q, \bar{q}} \delta_{n, \bar{q} -q} \text{exp}\left \lbrack -2 \pi \tau_2 \left ( \frac{-2 j (j+1) }{k_L-2} \right ) \right \rbrack_{\frac{1}{2} < -j = \frac{q+\bar{q}}{2} + \frac{k_L}{2} (w-t_1) < \frac{k_L-1}{2} } \Bigg \}  & \nonumber\\ & 
\times \text{exp} \lbrack 2 \pi i \tau_1 ( n ( w-t_1) + n' ( w'+t_1) + N - \bar{N} ) \rbrack &  \nonumber\\ &
\times \text{exp}  \left \lbrack  -2 \pi \tau_2  \left ( \frac{n^2}{2k_L} - \frac{n'^2 }{2 k_L} + \frac{k_L (w- t_1)^2}{2} - \frac{k_L (w'+t_1)^2 }{2} - \frac{3k_L}{12(k_L-2) }+ N + \bar{N} \right ) \right \rbrack,
\end{flalign*}
where the density of states is
\begin{gather}
\rho(s) = \frac{1}{\pi i } \text{Re} \left \lbrack \sum_{q, \bar{q} }^{\ \ \ \ +} \frac{\delta_{n, \bar{q} - q} }{ 2 i s + q + \bar{q} +1 + k_L(w-t_1)}- \sum_{q, \bar{q}}^{\ \ \ \ \ -} \frac{\delta_{n, \bar{q} - q } }{2i s + q + \bar{q} -1 + k_L(w-t_1) }  \right \rbrack. 
\end{gather}

Similar to \cite{Verheijden:2021yrb}, we can connect the field $g$ in $3d$ WZW model, (which is a function of the $2d$ Riemann surface $\Sigma$, onto a Lie group $G$, i.e, $g : \Sigma \to G$) to the $2d$ JT gravity model as 
\begin{gather}
\Phi= 2\pi \alpha g,
\end{gather}
where $\Phi$ is the dilaton in the JT gravity model, and $\alpha \in  (0, 1 \rbrack $ is a parameter which controls the partial reduction.  Note that here $\Phi$ in the $2d$ case is a scalar while $g(z, \bar{z})$s are matrices.

When the level parameter $k_L$ is a positive integer, the affine Lie algebra is the unitary highest weight representations which would be the dominant integral. This dominant part then can be connected to the $2d$ JT part easier.  This representation decomposes into finite-dimensional sub-representations  with respect to the sub-algebra spanned by each root. If the level is non-integer, the Lie group $SL(2, \mathbb{R})$ is non-compact and its homotopy group, $\pi_3 (SL(2, \mathbb{R}))$, is trivial.

In \cite{Verheijden:2021yrb}, the time dependence of black hole evolution has been controlled by the partial reduction of the $3d$ case which divides the geometry into the black hole part and the bath section. The energy of the black hole decreases linearly when the diving line moves slowly. From the lower $2d$ dimension point of view, i.e, JT case, some of the degrees of freedom would be hided in the one dimension higher, i.e, the bath which could be tracked through the partition function.

For the case of reduction from $\text{AdS}_3$ which is a solution of ``Einstein gravity" to $2d$ JT gravity, in \cite{Verheijden:2021yrb}, the following relation has been used
\begin{gather}
K^{(3)} =h^{\mu \nu}K_{\mu \nu}= K^{(2)} + h^{\varphi \varphi} K_{\varphi \varphi},
\end{gather}
which would lead to the relation
\begin{gather}
h^{\varphi \varphi}K_{\varphi \varphi}= - \frac{1}{\ell}.
\end{gather} 

In this case, for the connection between WZW and JT we can write
\begin{gather}
-\frac{k_L}{8\pi} K^{(3)}(\gamma^{-1} \partial^\mu \gamma) - \frac{k_L}{24\pi} \epsilon^{ijk} K^{(3)}(\gamma^{-1} \partial_i \gamma)=
-\frac{S_0}{2\pi} \left( K^{(2)}+h^{\varphi \varphi} K_{\varphi \varphi}\right).
\end{gather}

For the BTZ case, similarly one would have $h^{\varphi \varphi}K_{\varphi \varphi}=- \frac{1}{\ell}$.

So in the WZW model, instead of $\log k$ which specifies the number of orthonormal states, we should take $\log k_L$, which is related to $S_0$ of JT gravity.

Then, the question is that whether in this case, contributions from other modes than the s-waves, or the effects such as grey-body factor or Schwinger effects, should be considered or not. If one assumes that the distance between points of one's interest is much larger than the length scale of their size, then one can only consider the s-wave contribution and use the $2d$ CFT results.

 In \cite{Eberhardt:2022wlc}, the off-shell partition function for the chiral-gravity which is related to PSL(2, R) Chern-Simons theory has been calculated, and they found that there is a scaling limit where their partition function reduces to JT gravity using the equivariant localization. The level $k$ in chiral gravity is proportional to the dimensionless ratio $\ell_{\text{AdS}}/G_N$ and the dual CFT has a left-moving central charge $c=24k$. The higher genus partition functions become complicated due to the oscillatory relation to the Newton's constant, which lead to the result that gravity can indeed gives non-smooth contribution to the partition function.

 The higher genus corrections can actually construct the structure of $1/N$ corrections in the dual symmetric orbifold CFT. These higher genus and multi-boundary amplitudes are actually very important in understanding very late-time correlators behavior and the Page curve.

  In \cite{Eberhardt:2022wlc}, the large $k$ limit has also been implemented, where the approximation using the Weil-Petersson volumes of moduli space were used. There, they also found that higher genus partition functions get  \textbf{``oscillatory behavior''} in $k$, which its signature in island formalism and Hawking radiation could be detected. This behavior for instance could make the islands move from inside of the black hole toward slightly outside of it in the late times. The gravity in this regimes would not give the smooth contributions to the CFT partition function or the spectral form factor, where its effects can be injected in the lower dimensional JT gravity. The speed of the islands movement toward the horizon and the outward from it would then be directly related to this oscillation in $k$. In models of \cite{Hartman:2020swn}, it has also been found that since CGHS/RST model are defined on fixed topology where there is no replica wormholes, but by summing over ``topologies", as the bottom-up modification,  one can get the island rule formulation for the entropy of radiation, demonstrating again the effects of these higher genus topologies on the appearance and dynamics of the island.
 This is also related to the result of \cite{Hartman:2020swn} in the form of the relation $S_{\text{grav} } = \frac{c}{6} \rho= \Omega + \text{cosntant}$, where $\rho$ is the conformal factor, and $\Omega$ is the field in the solution of CGHS/RST, and the piece $\frac{c}{6} \rho$ comes from the Weyl transformation and from the $2d$ bath CFT, showing the theory knows the ``generalized entropy" gives the correct result. 
  
  The effects of these higher genus and various topologies can also be investigated by quantum error correction methods such as Petz map and modular flow, since in the JLMS and entanglement wedge reconstruction, the equality of relative entropy with and without the island as $S_{\text{rel}}(\rho_R | \sigma_R ) = S_{\text{rel}} (\rho_{R \cup I} ^{\text{semi-cl}} | \sigma_{R \cup I} ^{\text{semi-cl}})$ is a main basis as the states in the island $I$ are encoded in the radiation part $R$. The dynamics of the island would not change this structure and can be investigated using this conserved quantity.

  Another piece of evidence comes from the monotonicity of mutual information between the island and the region $R$ as  \cite{Hartman:2020swn}
  \begin{gather}
  \frac{d}{d\lambda_+} I_{\text{mat}} (I, R)  \ge 0,
  \end{gather}
  meaning the correlations between the island region and radiation increases, which is compatible with the picture that the island move from inside of the horizon toward outside increasing this mixed correlations. The above relation is part of a more general relation $\pm \frac{d}{d\lambda_{\pm} } S_{\text{gen} } (I)  \ge 0$
 
\begin{figure}[ht!]   
\begin{center}
\includegraphics[width=0.45\textwidth]{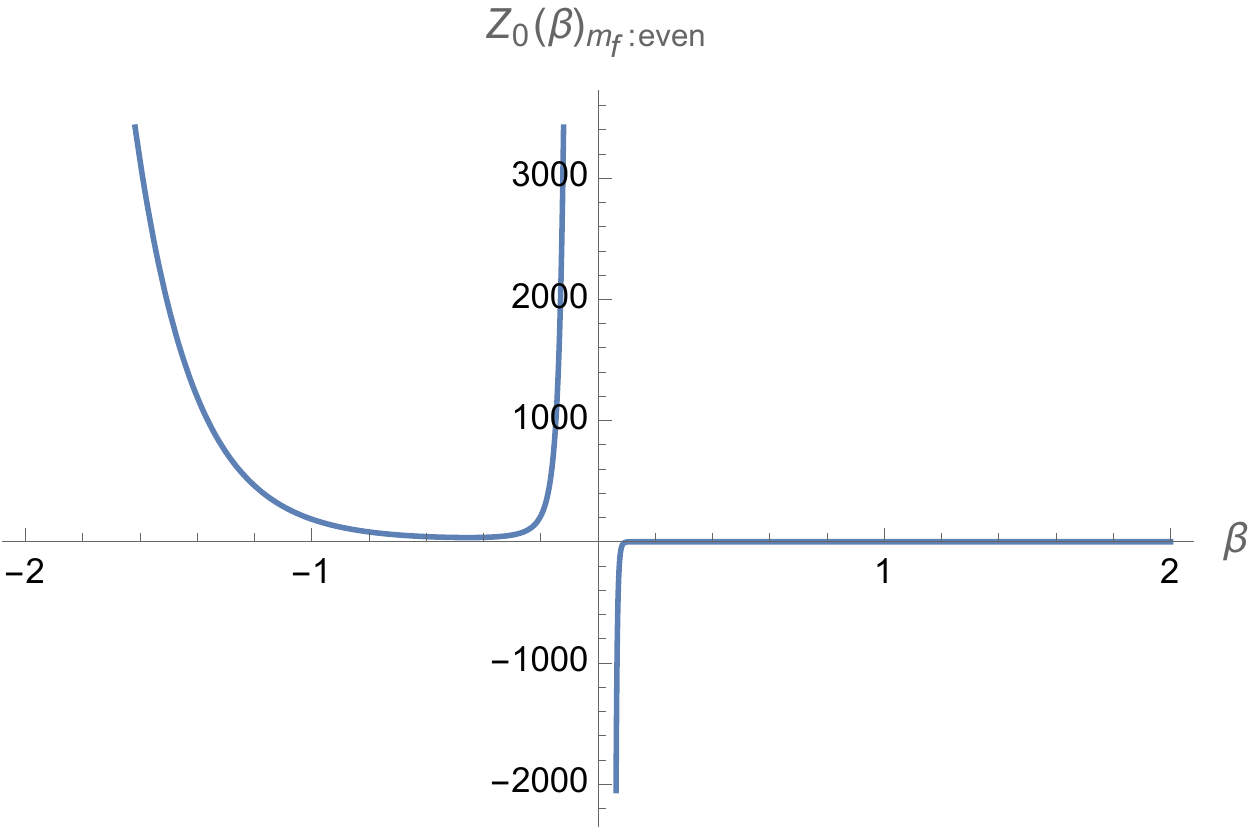}   \ \ \ \ \ \ \ \ \ \ \
\includegraphics[width=0.45\textwidth]{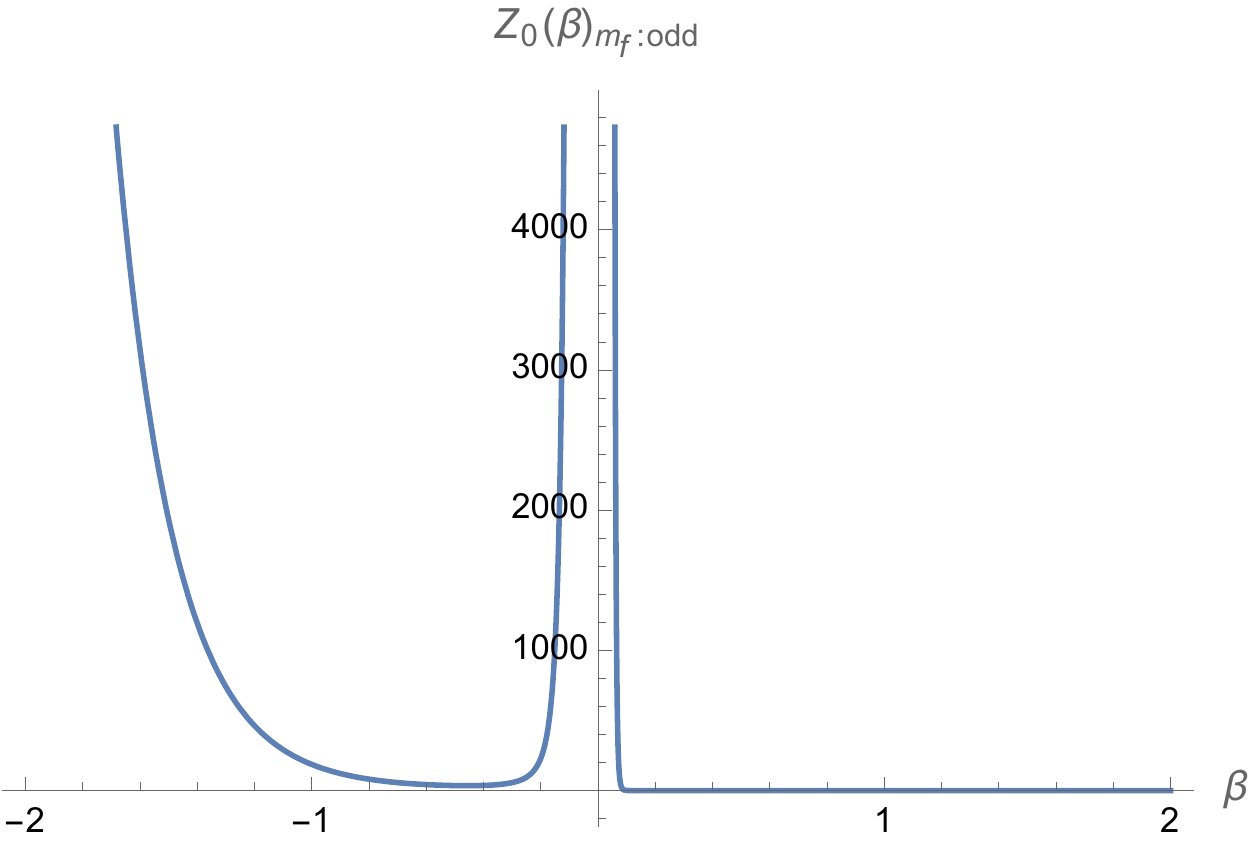}  
\caption{First order of the partition function, $Z_0 (\beta)$, versus $\beta$, for k=5, with the even or odd number of terms, $m$ \cite{Eberhardt:2022wlc}.}
\label{fig:z0}
\end{center}
\end{figure}

\begin{figure}[ht!]   
\begin{center}
\includegraphics[width=0.5\textwidth]{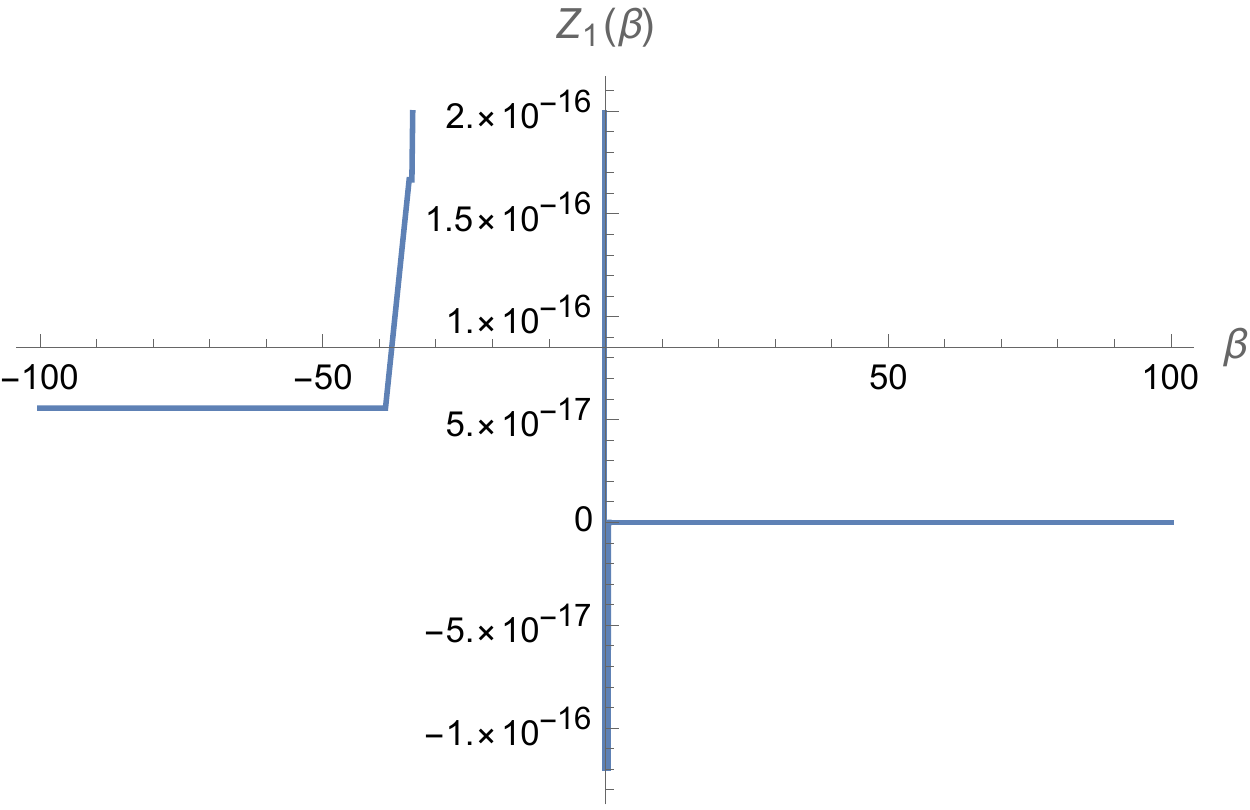}  
\caption{The next order of the partition function, $Z_1(\beta)$, versus $\beta$, for k=5, \cite{Eberhardt:2022wlc}.}
\label{fig:z1}
\end{center}
\end{figure}

 To get a better handle on this effect, we plot the behavior of $Z_0(\beta)$ and $Z_1(\beta)$ as shown in figures \ref{fig:z0} and \ref{fig:z1}.
 
 In the case of \cite{Eberhardt:2022wlc}, the limits where they could reach to JT gravity is $k \to \infty$ (related to small $\hbar$ limit) and $\beta \to 0$ (small thermal radius), while $\beta_{JT}= \frac{1}{k \beta}$ is being fixed.  The reason then that the oscillatory behavior in $k$ which depends on the whole properties of the topology and the background geometry, and the local dynamical behavior of the island depends on such a non-local parameter is due to the \textbf{Haar-invariance} and the delocalized quantum information. So, similar to the Gaussian Unitary Ensemble (GUE) model for Hamiltonian used in \cite{Cotler:2017jue}, the island dynamics which also depends on the long-time physics of a local system will similarly have become delocalized like the GUE models.  So the onset of the approximate Haar-invariance which can be characterized using the ``$k$-invariance" \cite{Cotler:2017jue} can determine if the correlation is local or completely non-local at each step, and when/if sharp phase transitions occurs.

Then, in \cite{Ashok:2021ffx}, in the coset manifold and in the $ ( \varphi, v, \bar{v} )$ coordinates, the action of the strings propagation has been written as
\begin{gather}
S=\frac{k_b}{\pi} \int d^2 z ( \partial \varphi \bar{\partial} \varphi + ( \partial \bar{v} + \bar{v} \partial \varphi) ( \bar{\partial} v + v \bar{\partial} \varphi)).
\end{gather}

The one loop string amplitude in the background of BTZ black hole has been derived in \cite{Ashok:2021ffx} as 
\begin{gather}
Z= \int_{\mathcal{F}_0} \frac{d^2 \tau}{4 \tau_2^2} Z_{\text{BTZ} } (r_\pm) Z_{gh} Z_{int}, 
\end{gather}
where the partition functions of the ghost and internal sectors are 
\begin{gather}
Z_{gh}= \tau_2  | \eta (\tau) |^4, \nonumber\\
Z_{int} = (q \bar{q} )^{- \frac{c_{int} }{24} } \sum_h d(h, \bar{h} ) q^h \bar{q}^{\bar{h}},  \nonumber\\
Z_{BTZ} = \frac{r_+ \sqrt{k_b-2} }{\sqrt{\tau_2} } \sum_{w,m} \frac{ e^{- \pi \frac{k_b}{\tau_2} r_+^2 | m - w \tau|^2 + \frac{2\pi}{\tau_2} \text{Im} (\bar{U}_{m,w}  )^2 } }{ | \theta_1 ( \bar{U}_{m, w} , \tau  ) |^2},
\end{gather}
where $d( h, \bar{h})$ denotes the degeneracy of the states of the internal CFT and $q=e^{2 \pi i \tau}$ is the elliptic norm. Also, here the holonomy $\bar{U}$ is
\begin{gather}
\bar{U}_{m,w} = (r_- - i r_+) ( m - w \tau).
\end{gather}

The $Z_{BTZ}$ part could be expanded as \cite{Ashok:2021ffx}
\begin{gather}
Z_{BTZ}= \frac{r_+ \sqrt{k_b-2} }{ \sqrt{\tau_2} } \frac{1}{ |\eta |^6} \sum_{w,m r, \bar{r}} S_r S_{\bar{r} } e^{2 \pi i m r - (r - \bar{r}) }  e^{2\pi m r_+ (1+ r + \bar{r} )}\nonumber\\
q^{- (r+ \frac{1}{2}) w ( r_- - i r_+ )} \bar{q}^ {- (\bar{r} + \frac{1}{2} ) w (r_- + i r_+) } e^{- \pi \frac{k_b}{\tau_2} r_+^2  | m-w\tau |^2+ \frac{2\pi}{\tau_2} \text{Im} (\bar{U}_{m,w}  )^2 }. 
\end{gather}

Then, this ``twisted partition function" of BTZ case can also be compared with the $2d$ JT case and then the Page curve can be derived, as the parameters $r_+$, $w$ and $\eta$ become time dependent. We leave the full calculations to the future works.

 It worths to mention here that, for studying the movement of the island, the conformal welding problem \cite{Almheiri:2019qdq} could also be considered, but instead one may use the average BCFT model of \cite{Kusuki:2022wns}, to study this problem. The more complete picture in fact can come from considering the time-dependence of bulk primaries $\phi_i$ and their two-point function, as they have ``finite" distance from each other. For doing that the higher orders of the following relations need to be considered.

\begin{gather} 
\phi_i(z) \sim \sum_I (2 \mathfrak{T}_z)^{h_I - h_i - \bar{h}_i} C_{i I}^a \phi_I ( \mathfrak{R}_z)+ . . . .,
\end{gather}

\begin{gather}
\langle \phi_i \phi_i \rangle_{\text{disk}}  \langle \phi_i \phi_i \rangle_{\text{disk}} = \sum_p (C_{iip})^2 (C_{p \mathbb{I}}^a)^2  \mathcal{F}_{ii}^{ii} ( h_p | z) \mathcal{F}_{ii}^{ii} ( h_p | z').
\end{gather}

 \section{Conclusion} \label{sec:conclusion}

The motivation of this work was to examine the saddles of Hawking radiation of black holes using mixed correlation measures and  geometric partial dimensional reduction first applied in \cite{Verheijden:2021yrb}.  This is done using mutual information, and in the setup of island formulation for deriving the Page curve, which we discussed for the case before and after the Page time separately.  Then, we applied quantum error correction code for each saddle and derived the inequalities for moving between them which give the consistent results for the possible phase transitions. Next, we extend the connections between $3d$ Einstein gravity and the dimensionally reduced $2d$ JT gravity used in \cite{Verheijden:2021yrb}, to other cases of $3d$ Chern-Simons gauged gravity, $2d$ boundary Liouville and $2d$ gauged WZW, and also $1d$ Schwarzian and $1d$ particles on group models. We commented on the connections between the parameters of partition functions of these models and their effects in the black hole evaporation. We also proposed that the island before the Page time would move inside the black hole toward the horizon, justifying it using different methods, and then commented on the links between its velocity and the specific parameters of the model, and particularly from $1d$ higher point of view.

In the appendix, we applied this idea, for the case of interconnections between negativity and island formulation, Kaluza-Klein and dilaton black holes, and also for the important case of extremal black holes. There, we utilized island formulation to get information about several other topics such as cosmological quantum fluctuations, Kondo effect in condensed matter systems, black hole secret sharing, quantum focusing and complexity, among others.

 \section*{Acknowledgments}
I would like to thank Junggi Yoon, Byoungjoon Ahn, Sang-Heon Yi, Jeongwon Ho, and Somyadip Thakur for useful  discussions.  This work indirectly has been supported by the science and technology promotion fund of the Korean government, by the Korean local governments - Gyeongsangbuk do province, and by the national research foundation of Korea (NRF) grant funded by the Korea government (MSIT) , (grant numbers 2021R1F1A1048531 and 2021R1A2C1010834).  The direct way of support is through an appointment to the JRG program of APCTP.

\appendix

\section{Mixed correlations and  Islands: the view from $1d$ higher}\label{sec:mixedmeasuresIsland}

In this section, we examine how to find the effects of island in various mixed correlation measures, models and examples from $1d$ higher point of view in the setup of VV \cite{Verheijden:2021yrb}. 

\subsection{The view of negativity and island from $1d$ higher}

It has been proposed in \cite{Kusuki:2019zsp}, that the holographic entanglement negativity is half of the R\'enyi reflected entropy of order $1/2$. In \cite{Shao:2022gpg, Basu:2022reu,Deng:2020ent, Afrasiar:2022ebi}, the entanglement negativity has been calculated for evaporating black holes in the island setups and in holographic models with a defect brane, demonstrating the island/$\text{BCFT}_2$ or island/defect models.
 The island formulation for entanglement negativity has also recently been proposed in \cite{KumarBasak:2020ams}.  Two proposals for the island contribution to negativity have been proposed there.
The first proposal involves extremizing the algebraic sum of the generalized R\'enyi entropies of order half, and the second one involves extremizing the sum of the area of a back-reacted brane on the entanglement wedge cross section (EWCS).

The first proposal could be written as
\begin{gather}\label{eq:negatavity1}
\mathcal{E} ^{\text{gen}} (A:B)= \frac{1}{2} \Bigg \lbrack S_{\text{gen}}^{(1/2)} (A \cup C) + S_{\text{gen}}^{(1/2)} (B \cup C) -S_{\text{gen}}^{(1/2)} (A \cup B \cup C) - S_{\text{gen}}^{(1/2)} ( C) \Bigg \rbrack \nonumber\\
\mathcal{E} (A:B) = \text{min} ( \text{ext}_{Q^{\prime \prime}} \{ \mathcal{E}^{\text{gen}} (A:B) \} ),
\end{gather}
where $C$ is the system which is between $A$ and $B$ and $Q^{\prime\prime} = \partial I_{s_{\mathcal{E}} } (A) \cap \partial I_{s_{\mathcal{E}}} (B)$, and  $S_{\text{gen}}^{(1/2)} $ which is the generalized R\'enyi entropy of order half has the relation
\begin{gather}
S^{(1/2)}_{\text{gen}} (A)= \frac{ \mathcal{A}^{(1/2)} \lbrack \partial I_s(A) \rbrack  }{4G_N} + S_{\text{eff}}^{(1/2)} ( A \cup I_s (A) ),
\end{gather}
and corresponds to the effective R\'enyi entropy of order half of the quantum matter fields coupled to semiclassical gravity.

The second proposal could be written in the most general form as
\begin{gather}
\mathcal{E}^{\text{gen}} (A:B) = \frac{\mathcal{A}^{(1/2)} (Q^{\prime \prime}= \partial I_{s_{\mathcal{E}}} (A) \cap \partial I_{s_{\mathcal{E}} } (B) ) }{4G_N} + \mathcal{E}^{\text{eff}} ( A \cup I_{s_{\mathcal{E}} } (A) : B \cup I_{s_{\mathcal{E}} } (B)) \nonumber\\
\mathcal{E} (A:B)= \text{min} ( \text{ext}_{Q^{\prime \prime}} \{ \mathcal{E}^{\text{gen}} (A:B) \} ).
\end{gather}

Then, using these two proposals, the contribution of island to entanglement of negativity in the setup of VV  \cite{Verheijden:2021yrb} can be studied. This can easily be done by replacing the each term of entropy in \ref{eq:negatavity1} with the one from $1d$ higher point of view of VV and then derive the Page curve in that setup.

\subsection{Islands in Kaluza-Klein black holes}

In \cite{Lu:2021gmv}, the behaviors of the islands in spherically symmetric Kaluza-Klein (KK) black holes have been studied. The KK charge denoted by $Q$ would deform the black holes from the Schwarzschild form. This charge slightly extends the boundary of the island in late times, increases the Page time by a factor of $(1+Q/r_h)$ and the scrambling time by a factor of $(1+Q/r_h)^{1/2}$.  The charge $Q$ also reduces the surface gravity and the Hawking temperature. This charge also depends on the $\omega$ in the higher genus partition functions, found for the off-shell action in \cite{Eberhardt:2022wlc}.

With the action of Kaluza-Klein theory with the Lagrangian
\begin{gather}
\mathcal{L}= \sqrt{-g} \left ( R- \frac{1}{2} (\partial \phi)^2 - \frac{1}{4} e^{\sqrt{2(D-1)/(D-2)} } F^2 \right ),
\end{gather}
after compactifying one of the spatial coordinates on a circle $\mathbb{S}^1$, a scalar and a vector field would be emerged.

The metric anasatz would also be
\begin{gather}
d\hat{s}^2_{D+1}= e^{2\alpha \phi} ds_D^2 + e^{2\beta \phi} (dz+ A_\mu dx^\mu)^2,
\end{gather}
where
\begin{gather}
\alpha^2=\frac{1}{2(D-1)(D-1)},  \ \ \ \ \ \ \ \beta= - (D-2) \alpha,
\end{gather}
which unlike the non-extremal Reissner–Nordström black hole has only one horizon.

The $4d$ Kaluza-Klein black hole which is a solution of the Lagrangian would be
\begin{gather}
ds^2= -W(r) dt^2 + \frac{dr^2}{W(r)} + H^{1/2} r^2 d\Omega^2,
\end{gather}
where
\begin{gather}
W(r)= f(r)/ \sqrt{H(r)}, \ \ \ \ f(r)=1- \frac{r_h}{r}, \ \ \ \ \ H(r)=1+\frac{Q}{r}.
\end{gather}

For the case without the island, the entanglement entropy of one interval $\lbrack b_-, b_+ \rbrack$ is
\begin{gather}
S_{\text{mat}} ( \lbrack b_- , b_+ \rbrack ) = \frac{c}{3} \log d ( b_-, b_+ ),
\end{gather}
where $d(x,y)$ can be found from
\begin{gather}
d^2 (x,y)=  | \Phi(x) \Phi(y) ( U(x) - U (y)) ( \bar{U} (x) - \bar{U} (y) ) |,
\end{gather}
and $\Phi$ are the elements of the Euclidean metric as $ds^2= \Phi^2 dU d \bar{U}$.

For the case with the island, the entanglement entropy for two intervals would be
\begin{gather}
S_{\text{mat}} (\mathcal{R} \cup I) = \frac{c}{3} \log \left \lbrack   \frac{d(a_+, a_-)  d(b_+, b_-)  d(a_+, b_+) d(a_-, b_-) }{d(a_+, b_-) d(a_-, b_+) } \right \rbrack.
\end{gather}

Then, using the KK metric above, the entropy $S$ for each case can be found and the Page curve be derived.  Note that another source of the movement of the island inside of the black hole toward the horizon is the tachyonic Kaluza Klein modes discussed in \cite{Malek:2020mlk} which makes the lower-dimensional supergravity masses to break the Breitenlohner-Freedman (BF) bound which then in the full 11 or 10 dimensions makes the vacuum perturbatively unstable.  Specifically, the ``brane-jet instability" of $SO(3) \times SO(3)$ vacuum \cite{Bena:2020xxb} would thrust the islands. This thrust is due to the net repulsive force of probe branes arising in certain areas of compactification manifold, which is the consequence of varying warp factors of the $11d$ solution, and this instability would be triggered by higher KK modes. These modes then change the structure of the islands.

\begin{figure}[ht!]   
\begin{center}
\includegraphics[width=0.5\textwidth]{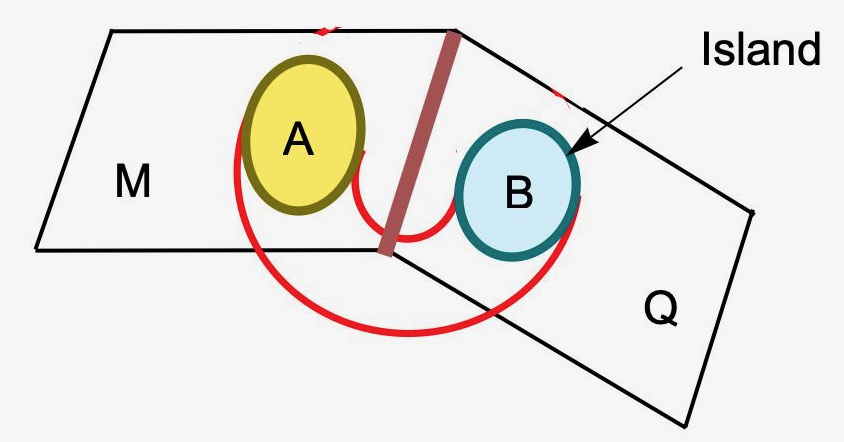}  
\caption{Island position on the end of the world brane relative to the bulk and boundary. The tension between the branes, as explained in the text, would push the island from inside of the black hole toward the outside.}
\label{fig:islandBCFT1}
\end{center}
\end{figure}

From the picture of AdS/BCFT and the island creation on the end of the world brane (EOW), \cite{Akal:2020twv}, as shown in figure \ref{fig:islandBCFT1}, it could understood better how the the tensions between the branes and these brane-jet instabilities can push the islands from inside of the black hole in the early stages to outside of the horizon in the later times. By looking from $1d$ higher, the various discontinuities in the entanglement contours observed in the presence of the island and the boundary, which can be explained by the localization-delocalization interpolations of entanglement, can be better observed. 

 From the behavior of the island and unitarity and resolving the information paradox it might be possible to explain why the tachyons can only appear in the symmetric representations $(\mathbf{k}, \mathbf{k})$ of $SO(3) \times SO(3) $, or other general theories.

Another evidence for the existence of an island inside of the black hole can come from the second solution for the set of differential equations, found in \cite{Pedraza:2021cvx} using the micro-canonical path integral, which could be written as 
\begin{gather}
V_K^{(2)} \approx V_K^B + \frac{(L^2 + 2 \mu U_K^B V_K^B)}{3 \mu U_K^B}, \ \ \ \ U_K^{(2)}\approx  \frac{5U_K^{B}}{3}+ \frac{1}{3\mu V_K^{(B)} } ,  
\end{gather}
where $(U_K, V_K)$ are elements of the Kruskal coordinate
\begin{gather}
ds^2= - e^{2 \rho (V_K, U_K) } dU_K dV_K, \ \ \ \ e^{2\rho} = \frac{4 \mu}{ (1+ \frac{\mu}{L^2} U_K V_K )^2 }.
\end{gather}

  \begin{figure}[ht!]   
\begin{center}
\includegraphics[width=0.6\textwidth]{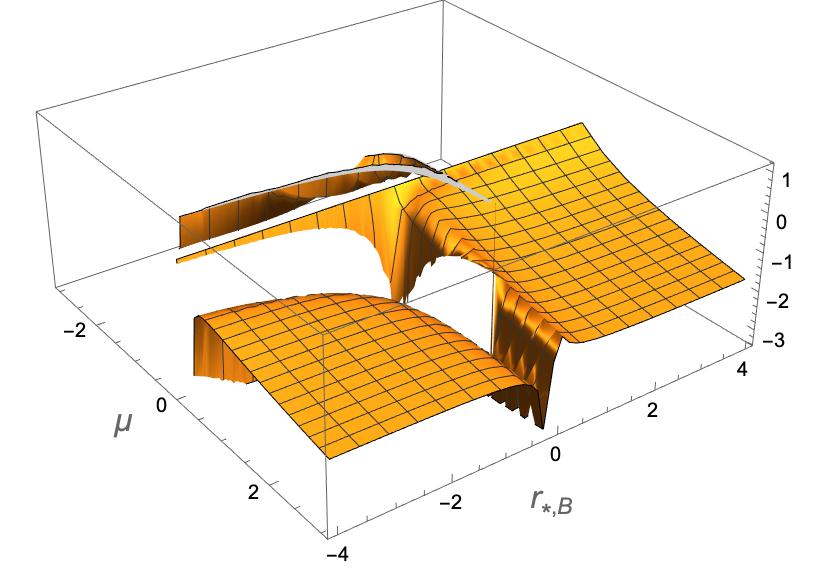}  
\caption{The structure of the second (``un-physical" from the point of view of an outer observer) solution of \cite{Pedraza:2021cvx} which can lie inside of the eternal black hole horizon.}
\label{fig:inQES}
\end{center}
\end{figure}

This solution from the point of view of an observer outside of the black hole horizon is unphysical. But from the point of view of an observer inside the horizon, it can be a physical solution of the set of equation of motion of quantum corrected version of JT gravity and it can be new saddle. This saddle, in specific cases, such as charged extremal case, can move toward the outside of the black hole horizon, leading again to the result that during the evaporation, island transfer from inside toward slightly outside of the horizon.

The radial position of the quantum extremal surface in the Schwarzchild coordinates is 
\begin{gather}
r_{\text{QES}} = L \sqrt{\mu} \left ( \frac{1- \frac{\mu}{L^2} U_K V_K}{1+ \frac{\mu}{L^2} U_K V_K} \right ) \approx r_H \left( -1+ \frac{18 \mu L^2 U_K^B V_K^B}{L^4+ 19 \mu L^2 U_K^B V_K^B + 25 \mu^2 (U_K^B V_K^B)^2} \right ).
\end{gather}

The structure of this solution is shown in figure \ref{fig:inQES}. Note that the thermodynamic for the static and nested wedge observer versus accelerating observers, would also be different, leading to the result that for another coordinate system and sets of observers, the above solution can indeed become physical.

\subsection{Islands for charged dilaton black holes and its view from higher $d$}

In this subsection, the effects of dilaton on island from higher $d$ point of view is examined. In \cite{Karananas:2020fwx}, the island prescription for $4d$ linear dilaton black holes with planar horizon has been studied. These black holes asymptotes to the linear dilaton background. In principle, these black holes are actually a two-dimensional Witten black hole with two additional free bosons. This picture in the string frame would help us more to understand how information from higher dimensions will be encoded in the lower dimensions. The picture is also that the entanglement entropy would have a ``running" behavior, as it changes along the RG flow with respect to the two-dimensional worldsheet length scale.
In fact, it has been shown that as the islands force the entropy to decrease, the dilaton runs toward the IR point.

In \cite{Karananas:2020fwx}, also, it has been found that without an island, the entropy behaves asymptotically as
\begin{gather}
S=S_{\text{matter}} \sim \frac{c}{3} \frac{t_b}{r_h},
\end{gather}
which grows linearly with time, while when an island is included in the computation one would get
\begin{gather}
S \approx \frac{k^{-2} e^{-\sigma} }{2 \alpha'} \Big |_{\text{horizon}} \approx \frac{r_h^2}{2 G_N},
\end{gather}
which is independent of time and is proportional to twice of the Bekenstein-Hawking entropy. Here $k$ is a constant with mass dimension $1$, and in each coordinate of the $x$ and $y$, the scale is $k^{-1}$. 

There, it has been shown that without the island the entropy behaves as $S \approx \frac{c}{3} \frac{t_b}{r_0} e^{-a \lambda}$, which blows up in the IR. However, when an island is included, the entropy scales as $S \approx \frac{r_h^2}{2G_N} = \frac{r_0^2}{2G_N} e^{2a\lambda}$, which has a decreasing behavior toward IR.

In \cite{Ahn:2021chg}, the island prescription for charged linear dilaton black holes has also been studied, both for the non-extremal and extremal cases. It has been found that the Page time is universal for all of the different models that they have been studied and it would be $t_{\text{Page}}= \frac{3}{\pi c} \frac{S_{\text{BH}}}{T_H}$. Now the question would be if this quantity is still universal looking from higher dimensions or it would change. 

The action considered in \cite{Ahn:2021chg} is four-dimensional dilaton action with a $U(1)$ gauge field in the Einstein frame as
\begin{gather}
I= \frac{1}{16 \pi G_N} \int d^4x \sqrt{g} \left ( R - \frac{1}{2} (\partial \sigma)^2 + 4 k^2 e^\sigma - \frac{1}{4} e^{\gamma \sigma} F_{\mu \nu} F^{\mu \nu} \right ),
\end{gather}
where $k$, $\gamma$ and $\sigma$ are constant and $\sigma$ is a scalar field.

\begin{figure}[ht!]   
\begin{center}
\includegraphics[width=0.32\textwidth]{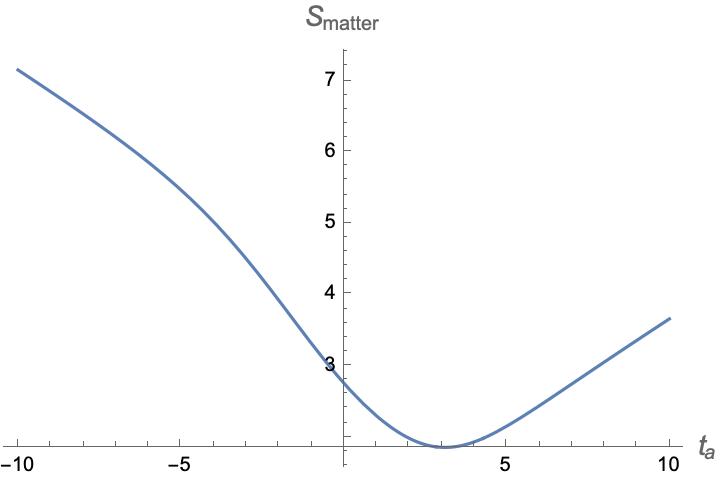} 
\includegraphics[width=0.32\textwidth]{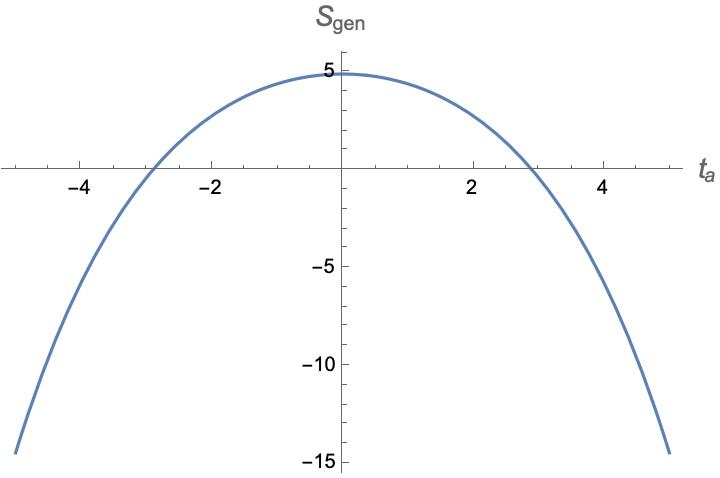}  
\includegraphics[width=0.32\textwidth]{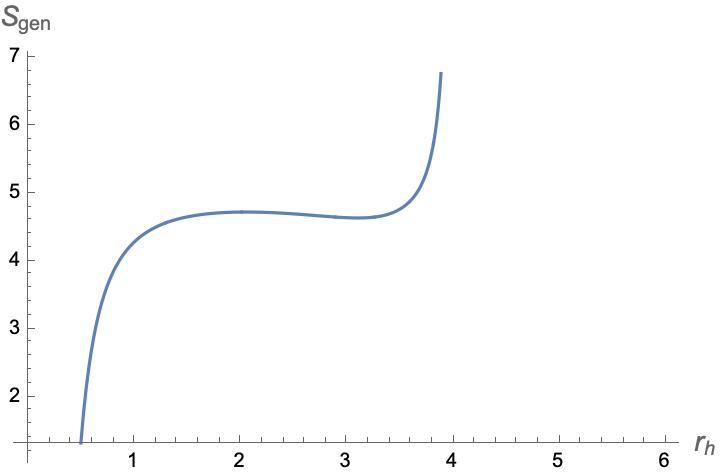} 
\caption{Diagrams coming from the results of relations found in \cite{Ahn:2021chg}. The left figure is $S_{\text{matter}}$ versus $t_a$, or $t_b$, for the non-extremal charged dilaton black hole, with the presence of an island, the middle is the plot of the generalized entropy, $S_{gen}$, versus $t_a$ or $t_b$, while one of them is being fixed for the same case, and finally the right one comes from generalized entropy $S_{\text{matter}}$, versus $r_h$ for the  ``extremal charged dilaton" black hole case.}
\label{fig:Ahnwork}
\end{center}
\end{figure}

For the case without an island, and by calculating the length of the geodesic, for the solution of the charged dilaton black hole, the entropy of the matter fields would be 
\begin{gather}
S_{\text{matter}} = \frac{c}{3} \log \left \lbrack 2 f(b) e^{\kappa_+ r^*(b)} \cosh k_+ t  \right \rbrack,
\end{gather}
 where for $r_+  \ll b$ would be
 \begin{gather}
 S_{\text{matter}}= \frac{c}{3} \log (2 \cosh k_+ t ) \simeq \frac{c}{3} \kappa_+ t,
 \end{gather}
which increases linearly with time. From the results of \cite{Ahn:2021chg}, the diagrams shown in figure \ref{fig:Ahnwork} can be found.

An interesting issue here is the claim in \cite{Ahn:2021chg}, which argued that the description for the extremal case cannot be obtained from the continuous limit of the non-extremal case, and without the island no correct result for the extremal case can be obtained. This is because the geodesics distances that need to be calculated without the island, i.e, eq. 5.1 of \cite{Ahn:2021chg} is ill-defined due to the divergences at $r=0$, or one could say since the origin is not part of the manifold, or the periodicity of the Euclidean time is not fixed.  In \cite{Yu:2021cgi}, the Page curve for Garfinkle-Horowitz-Strominger dilaton black holes has also been discussed. They also found that the charge $Q$ would have a significant impact on Page time. In addition, they  found that there is some peculiarity for the extremal case and there the Page time either diverges or would vanish, as the extremal black hole don't behave quite physically.

Note that as explained in \cite{Carroll:2009maa}, the entropy of an extremal black hole is zero from semi-classical method, and $S=\pi \rho^2/G$ from the string theory microstate counting, and the lack of the entropy from in the first approach could be explained by the creation of a separate dimension within the black hole. The island can be considered inside of this separated region going to other dimension or the whole region can be considered as the island of this case. So this way correct form of the Page curve for the extremal case would be related to the question of how information from higher $d$, for instance $4d$ Schwarzschild or Reissner–Nordstrom black holes would be encoded in the lower $2$  near horizon of black holes, leading to the pattern of entanglement entropy and island prescription for the extremal case.

\subsection{Extremality from higher $d$}

In \cite{DeWolfe:2013uba} the motivation also was to address the issue of finite entropy of extremal charged black hole (or black brane) at zero temperature, which can be addressed by the perspective we got here from looking from $1d$ higher to BH observables. For studying the problem of entropy of 3-charged extremal black holes, the authors studied the response to the probe Fermionic operators in different limits. Then, the fluctuations around the Fermi-surfaces in these different limits of charges and frequencies have been examined.

One interesting outcome of that work was  that the bottom-up approach to holography could give wrong or un-physical results and it would be necessary to address this problem in the top-down full string theory or supergravity backgrounds. For the case of $N=4$ SYM which has three independent chemical potentials (charges), when all the three charges are on, the horizon is non-singular and we have the issue of finite entropy in zero temperature. However, there are limits of charges (or approaching the extremality) in which this zero-temperature entropy is absent which gives insight into the nature of this entropy.

For $N=4$, when one charge is absent, (and for simplicity taking the other two charges equal) the entropy would vanish as the horizon have zero-area and is singular. Then, the Fermion responses were studied in this limit.
The interesting physical fact here is that this 2-charged case is the solution of $5D$ gauged supergravity background which comes from the reduction of type IIB supergravity background on $\text{AdS}_5 \times S^5$. This corresponds to a specific state in $N=4$ SYM where in this full theory, it was found that the masses depend on the running scalar field (dilaton) which diverges at the horizon/singularity. So for understanding the Fermionic responses in this limit of extremal black brane, one should consider these couplings and running dilatons from higher $d$ to lower $d$.

The other interesting finding of that work is that for the extremal two-charge black hole an energy scale  ``$\lambda$" exists that within this energy from the Fermi surface, fluctuations would not decay. In fact, there is a special sector with large density of states near the horizon that for generic three charge case is not gapped and extends down to Fermi surface leading to the non-zero entropy and unstable ground state but in the two-charge case, this sector is gapped which cannot get excited by energies less that  ``$\lambda$"  and so is stable. This gap removes the large entropy in zero temperature and changes the mechanism of Fermionic fluctuations and also the behavior of all the zero modes. As it has been studied further in \cite{Verheijden:2021yrb}, this idea is also related to the correct calculation of Euclidean saddles in gauge/gravity and the island prescriptions \cite{Penington:2019npb}, with correct boundary conditions needed for the in-falling particles.

In \cite{DeWolfe:2013uba}, they also noted that this stability might indicate that the fermionic fluctuations have no self-interaction which is intrinsic and that could be a large-N effect. Turning even a small charge $q$ other than the equal two charges would suddenly remove the gap. So in the regular three-charge case, the near-horizon and small-$\omega$ limits don't commute which indicates there are two regions, the inner and outer regions which are patched together. The inner region has the geometry of $\text{AdS}_2\times R^3$, where the outer region could help us solve our issue. It can be shown that in a two-charge case, this geometry can be lifted to a smooth $\text{AdS}_3\times R^3$ in six dimensions, leading to a well-behaved geometry, and the final stability. The Kaluza-Klein charge of the reduction is actually that $q_1$ that is being turned off in the two-charged case. The gap also can be understood in this picture as the minimum energy needed to turn a momentum vector which has a fixed amount of compact momentum time-like, which is also related to the amount of energy needed to push the island toward out of horizon. The angular component of the metric has the role of the dilaton which is the Kaluza-Klein charge of the reduction. The authors also suggest that the field theory explanations of this gap could also be constructed using the emergent non-chiral Virasoro algebra in the infrared which again points out to a universal feature. 

So this behavior is very similar to the Page phase transition of the black hole, where here the charge $q_1$ controls the phase transition. This gap is indeed like the  exchanges between the saddles of quantum extremal surfaces (QES) that jumps from the empty surface to a surface inside of the black hole horizon. Hence, this gap is like the island that appears after the Page time while here, when the charge $q_1$ is being turned on, the phase transition can be initiated. This issue can also be solved using a doubly holographic model for the gap inside the near horizon of the black brane.

Thus,  this two inner and outer regions that Gubser discussed is like the black hole and the bath in JT gravity where the $\text{AdS}_2$ part is like the black hole and the bath is in the higher dimensional uplifted case that is being controlled by the charge $q_1$. The entropy of $\text{AdS}_2$ would transfer to the higher dimensional part, making the whole geometry stable then. By turning on and off the $q_1$ charge like the case for black hole which connects the black hole interior to radiation, here too,  the higher dimension geometry can connect the $\text{AdS}_2$ with whatever needed to elevate it to the $\text{AdS}_3$. After turning on the charge $q_1$, the black hole interior becomes part of the entanglement wedge of (bath or radiation of BH for the case here) the geometry which is needed to be added to the $\text{AdS}_2$ part to elevate it to an $\text{AdS}_3$, so these islands decrease the large entropy seen for the $\text{AdS}_2$ case.  Note also that these islands are inside of the black hole horizon and so here the second geometry would  be inside of the black hole, while $\text{AdS}_2$ is the outer region.

In conclusion, the $\text{AdS}_2$ is like the black hole case and the bath is like the remainder of the geometry which we call here $G_{in}=\text{AdS}_3-\text{AdS}_2$. This is the remaining internal geometry which by turning on the charge $q_1$ would separate $G_{in}$ from $\text{AdS}_2$, making the near horizon unstable. When the charge $q_1$ becomes bigger, more of the entropy goes into the $\text{AdS}_2$, when $q_1=0$, the $G_{in}$ or the bath, appears inside of the black hole, and when $q_1$ becomes more negative, the size of the bath increases further.

\section{Measurable observables from $1d$ higher}\label{sec:measurableshighD}

In this section, several other topics that through dimensional reduction can be analyzed in the setup of island formulation will be discussed.

\subsection{Quantum Fluctuations}

So some of the results from the dimensional reductions can in fact be tested using  measurable quantities.

For instance, recently, in \cite{Gukov:2022oed}, using the dimensional reduction from $4d$ Einstein-Hilbert action to $2d$ JT gravity, the quantum fluctuations and the uncertainty in the time of a photon traveling from tip-to-tip of a causal diamond, in the $4d$ flat Minkowski case has been calculated. This is related to our study here connecting Hawking radiation in the higher $d$ case to the one dimensional lower.

Specifically, near the horizon, the causal development of a region in flat Minkowski spacetime can be described by the $2d$ dilaton theory which can be described by the JT action, and there the quantum fluctuations and uncertainty would lead to the metric fluctuations, where this uncertainty in the higher $4d$ case would lead to the uncertainty of the photon's  travel time from tip-to-tip of a causal diamond.

In principle, the dilaton field in the lower dimensional case is related to the area of the transverse directions. The modular fluctuations discussed in \cite{Gukov:2022oed} which has been proposed to be experimentally observable could actually be the reason that the island region moves inside the black hole toward the horizon. These fluctuations could be discussed from the higher dimensions point of view as in \cite{Gukov:2022oed}. The main point is to figure out how the island fluctuations would be Weyl rescaled between the dimensions. As explained in \cite{Gukov:2022oed}, the dynamics of the dilaton would control the size of $S^2$ part and also the relative position of the horizon (and therefore the island) with respect to the boundary. Since this dilaton has its own effective hydrodynamic description, it could effectively explain the dynamics of the island inside the horizon.

In \cite{Gukov:2022oed}, the near-horizon quantum dynamics of $4d$ Einstein gravity has been studied from $2d$ JT gravity model. The metric fluctuations in JT action, leads to the quantum uncertainty of the position of the horizon.
There, an important relation for our story had also been found, i,e, $\mathfrak{r}_s = L \frac{\Phi_h}{\Phi_b}$, noting that in the Schwarzchild coordinate, $\mathfrak{r}_s$ is the position of the Rindler horizon, $\Phi_h$ is the value of the dilaton at the horizon, and $\Phi_b$ is its value at the boundary. This relation then could be viewed from the higher dimension and also its relation to the behavior of the island as in \cite{Demulder:2022aij} and \cite{Verheijden:2021yrb}. From \cite{Demulder:2022aij}, the change of dilaton and metric components are connected by the relation $e^{2 \delta \phi} = \frac{\hat{\kappa} }{\kappa}$. So, in fact the dilaton parameter determines the ratio of the numbers of suspended D3 branes versus the number of (semi-infinite) D3 branes.

In addition, from the $2d$ JT gravity, the Hamiltonian of the reduced $1d$ action on its boundary can be written using the stress-energy tensor, as the ratio of the dilaton field at the horizon divided by its value at the boundary, i.e, $H_L=H_R= \frac{\Phi_h^2}{L \Phi_b}$. As suggested in \cite{Gukov:2022oed}, the time difference between the two sides of the boundary in the thermo-field double state which is a measurable quantity by an interferometer, can be found as $\delta = i \partial /\partial E$, where using the modular Hamiltonian, the time $\delta$ can be defined by $\delta= \frac{ \mathfrak{t}_L + \mathfrak{t}_R}{2}$. Then, $\delta$ can be written as $\delta=\frac{L \mathfrak{r}_s }{4 \mathfrak{r}_c^2 } t$, which subsequently can be written in terms of the values of the dilaton fields as $\mathfrak{r}_s= L \frac{\Phi_h}{\Phi_b}$ and $\mathfrak{r}_c^2= L^2 \gamma_{00} \big |_{\partial \tilde{M}_2} $, where $\gamma_{00}$ is the induced metric. These are parameters in the ``Schwarzchild coordinates". Therefore, the dimension reduction from $4d$ to $2d$ could be fully studied using these observables.

In \cite{Zurek:2022xzl}, also, among other methods, through dimensional reduction of the Einstein-Hilbert action to dilaton gravity, other observables such as quantum uncertainty at lightsheet horizons have been studied. One interesting point mentioned there is that, in the context of AdS/CFT, the stochastic energy or mass fluctuations which follows the relation $\Delta M \sim \sqrt{S} T = \frac{1}{\sqrt{2} \ell_p}, $ analogously could be written using the modular Hamiltonian $K \equiv \int  dB^\mu \zeta^\nu T_{\mu \nu}$. The fluctuations then could be found as $\langle \Delta K^2 \rangle = \frac{A_\Sigma}{4G_N}$.

These modular fluctuations can then source the fluctuations in the gravitational potential as
\begin{gather}
\langle \Phi^2 \rangle = \frac{\langle \Delta K^2 \rangle}{(d-2)^2} \left (\frac{4G_N}{A_\sigma}  \right )^2= \frac{1}{(d-2)^2} \frac{4 G_N}{A_\sigma}.
\end{gather}

So the subsequent diamonds created by the fluctuations of the modular Hamiltonian, would become statistically uncorrelated if the changes in the entanglement entropy increases as $\delta S_{\text{ent} } \sim \sqrt{S_{\text{ent} } } $. They also argued that, due to this property, the memory effects cannot exist in $d>4$ cases. Through the dimension reduction method of  \cite{Solodukhin:1998tc}. These effects and also black hole evaporation mechanisms, then can be looked at from $d=4$ to $d=3$, and $d=2$. By flowing between dimensions, $G_N$ would also flow affecting the dilaton fluctuations at each dimension.

\subsection{Kondo effect and island formulation}

Another observable which could be studied from $1d$ higher point of view is the Kondo effect which is related to AdS/BCFT \cite{Fujita:2011fp}. The island for BCFT has also recently been studied in \cite{Suzuki:2022xwv}.  The authors examined the connections between AdS/BCFT and the gravitational systems coupled to a $2d$ CFT  setup and showed that the one point function in AdS/BCFT can be reproduced using the gravity solution with scalar fields being turned on, and therefore they provided evidences for the island/BCFT correspondence.

As noted in \cite{Izumi:2022opi}, the brane dynamics and island formulation in holographic BCFTs have been studied. There, it has been shown that the scalar field perturbation in the AdS/BCFT would show complete reflection behavior which this fact also has been noted in the work of \cite{Gukov:2022oed}, which used this perturbation in lower $d$ to calculate the uncertainty in travel time of photons moving from tip-to-tip of causal diamonds in higher $4d$ case. Therefore, the results of these two works indicate that many of the Island/BCFT results could lead to measurable quantities in the higher $d$ spaces, like what has been predicted in \cite{Izumi:2022opi}, that the dynamics of the EOW brane would have significant effects on the evaporation of the black holes in higher dimensions.  In \cite{Kusuki:2022wns}, it was shown that the dual BCFT of the island can be found by the average of the boundary conditions, as this fact can help to better understand the evolution of the black hole and the flow of information during the Hawking radiation. This can be another reason why looking from the one dimension higher can indeed give correct insight about the evolution of the black hole.

An interesting observation regarding the connections of black hole physics and condensed matter systems, is also that the diagram seen in the time evolution of screening process due to the formation of the Kondo cloud, and also the behavior of the electric flux (modeled by the dimensionless quantity $D= \frac{a'(v,1) }{a'(0,1)}-1$, where $a'_v (v,y)$ is proportional to the electric flux), in the condensed phase, is very similar to the exchange of the saddle points and the resulting Hawking-Page phase diagram. For instance, check Figure. 10 of \cite{Erdmenger:2016msd}, which studied the evolution of the screening of the impurity during a Gaussian quench using the flux through the horizon in the dual holographic black hole model. So the decreasing of the degrees of freedom which are being screened can also be viewed by the creation of the island inside of the black hole which hides away those degrees of freedom. Therefore, there is a direct connection between the critical temperature $T_c$ of the Kondo effect and Page time of black holes.

As for further connections between condensed matter systems and Hawking-Page phase transitions, note that in \cite{Banerjee:2022jnv},  it has been shown that the Berry phase can distinguish between different characterizations of Hilbert space and the authors could also specify the topological phase transitions of entanglement entropy when the black hole forms. In addition, in \cite{Anous:2022wqh}, the island formulation in interface field theories (ICFTs) has been checked where the connections to Janus solution has been investigated. We will back to the role of Berry phase in island formulation in future sections.

\subsection{Black hole hair from higher $d$}

As for further results for the effects of higher $d$ on black hole evolution, one can look the connections between the quantum hair and black hole information as in the case of \cite{Calmet:2021cip}, where it has been argued that the quantum state of a graviton field which is outside of a black hole horizon can carry information about the internal structure of black holes, \cite{Raju:2020smc}.  This story can be further extended using our picture of dimensional reduction and tracing information from higher dimensions to lower dimensions, as it has also been shown that the final state would be a complex superposition of the initial black hole state, as the gravity would prevent the local storing of quantum information. So the information in radiation turns out to be highly mixed and non-local.

After the radiation of the first quantum $r_1$, the exterior state of the black hole could be written as
\begin{gather}
\Psi_i \to \sum_n \sum_{r_1} c_n \alpha(E_n, r_1) | g(E_n - \Delta_1) , r_1 \rangle,
\end{gather}
where $r_1$ is the radiation here and $g$ is the exterior geometry, which in principe can have one dimension higher than the background of the black hole where the quanta can hide in that part.

If the quanta remains in the same dimension, the next emission then leads to
\begin{gather}
\sum_n \sum_{r_1, r_2} c_n \alpha(E_n, r_1) \alpha(E_n - \Delta_1 , r_2)  | g(E_n - \Delta_1 - \Delta_2), r_1, r_2 \rangle.
\end{gather}

If no quanta runs to the higher $d$, then the final radiation state could be written as
\begin{gather}
\sum_n \sum_{r_1, r_2, ..., r_N} c_n \  \alpha(E_n, r_1) \  \alpha(E_n - \Delta_1, r_2)  \ \alpha(E_n- \Delta_1 - \Delta_2, r_3)  \ . . .  \ |r_1 r_2 ... r_N \rangle.
\end{gather}

Note that as mentioned in \cite{Calmet:2021cip}, the quanta of black hole radiation are very localized in space and time rather than being in the form of a plane wave state, therefore, the chance of these quanta being penetrated in higher dimensions would be very high leading to our perspective of looking to black hole observables from one dimension higher very plausible, however this would also point out to the fact that the argument of \cite{Calmet:2021cip} cannot be complete, as the total information of black hole cannot be retained from the coefficients $c_n$ in the lower $d$ cases. So  in the lower dimensions the picture would not be unitary and in order to retain all the information, access to higher $d$ would be necessary.

\subsection{Connectivity of spacetimes}

There are interesting connections between the strength of the mixed correlations and connectivity of spacetimes. 

In the recent work of \cite{Engelhardt:2022qts}, the canonical purification of the black hole radiation after the Page time has been investigated and the connections with $\text{ER}=\text{EPR}$ has been studied, where it has been demonstrated that the multipartite entanglement gives a more complete picture of the evaporation. In that work also a question has been raised that if at a time before the page time, $t_p$, and another one after the Page time, the R\'enyi entropies are equal, but the first one before $t_p$ has disconnected wedges while the one after $t_p$ is connected, therefore entanglement entropy (and even computational complexity) would not be enough to determine the connectivity of spacetime and they proposed that other measures defined based on, for instance quantum error correction, could distinguish the \textbf{connectivity} of spacetime. However, the answer is simply the \textbf{Berry phase} or modular Berry phase. Specifically, in \cite{Banerjee:2022jnv}, it has been shown that the Berry phase can determine the non-factorization quality of the Hilbert space and detect the presence of the wormholes.

When there is a wormhole present, the symplectic form it creates would give rise to the Berry phase. There, three different categories of Berry phases have been introduced, namely the Virasoro, the gauge and the modular Berry phase, where each has their own distinguished spacetime wormhole geometry. It has also been shown that the Berry curvature which can be written in terms of the Crofton form can characterize the topological transitions of the entanglement entropy, for instance during the formation of a black hole in spacetime. This study is also similar to the result of our work \cite{Ghodrati:2021ozc}, which we showed the connections between the Crofron form and mixed correlation measures such as critical distance between two subsystems where mutual information drops to zero,  as both can distinguish the phase structure of entanglement entropy in any confining geometry.

Note that the fact that island exists and various saddles can dominate during the evaporation of the black hole, indicates that like outside of the event horizon, inside also contain various structures with different characteristics even before the Page time, where each subregion can be characterized by a specific Berry Phase and modular Berry curvature. This picture is very similar to the process of melting of snow on the ground, which the speed of the process would depend on the material type on the background in each region. Similarly, the evaporation of the information and its flow toward outside would depend on the fine-grained structure of spacetime at each specific region inside of the black hole, where each can be told-apart by the modular Berry phase. Therefore, inside of the black hole is not just a singularity and empty or a homogenous spacetime structure, whereas it has still structure and various texture with differrent Berry curvature at each point, depending on the surrounding of the black hole which can lead to a more complex model of black hole evaporation, which would lead one to propose the need for a  ``tomography" way of imaging the black hole, where we speculate further.

As another evidence for this proposal, we could mention the work \cite{Caceres:2022smh}, where the ``trans-IR" flows for describing physics inside the black holes have been introduced, and they argued that the a-function that they constructed would vanish flowing toward the singularity, so most of the degrees of the freedom ``inside" of the black hole are accumulated close to the horizon and in fact around the singularity there are no degrees of freedom. This also indicates that at early stages of black hole radiation, the islands are within the horizon and very close to it. Due to the radiation the island moves closer and closer to the horizon and at the Page time slides outside of the horizon and then at later times most of the degrees of freedom are in fact gathered in the island. So during the radiation island moves along the a-function and the trans-IR flows constructed in \cite{Caceres:2022smh}. In fact, the bulk gravitational dynamics are the dynamics of the RG flow \cite{deBoer:1999tgo, deBoer:2000cz} and therefore indeed the dynamics of this flow can depict the dynamics of bulk and in this case the island inside of the black hole. As  \cite{Caceres:2022smh} argued that action complexity and 2-point correlation can probe this trans-IR, they could also probe the dynamics of the island inside of the black hole.

In \cite{Lee:2022efh}, the specific  ``brane tomography" for the black hole microstates has been proposed, where the propagation of the information in the CFT has been modeled by adding dilaton gravity to the end of the world brane, where the properties of the brane or local gravitational dynamics on it could be probed using the von-Neumann entropy and the quantum extremal island in the double holography and brane descriptions, which again shows how viewing from one dimension higher can give rich information about the brane structures and information properties of the quantum gravity. 
The use of such quantum information measures to check supergravity models, in a similar way, has been studied in our previous works \cite{Ghodrati:2022kuk,Ghodrati:2021ozc}.

So a JT coupling at each subregion can be defined as
\begin{gather}
\alpha \equiv \frac{G_N \varphi_1}{G_N^{\text{brane}} L}.
\end{gather}

For the metric
\begin{gather}
ds^2= \frac{1}{\cos^2 y} \left ( - L^2 d\tau^2 + L^2 dy^2 + \frac{r_+^2}{L^2} \cos^2 (\tau) dx^2 \right ),
\end{gather}
and for the case without JT gravity on the boundary a conserved quantity, $\mathcal{Q}_E$, which is written as 
\begin{gather}
\mathcal{Q}_E=\frac{1}{\cos y} \frac{\tau' (y)}{\sqrt{1- \tau'(y)^2}},
\end{gather}
would vanish, but for the case with a JT gravity on the boundary, the conserved charge would depend on the brane time $\tau_{\text{brane}}$, as $\mathcal{Q}_E= - \alpha \cos(\tau_{\text{brane}} )$. This boundary time $\tau$, and therefore the conserved charge $\mathcal{Q}_E$, would be related to the modular Berry phase where at each time and each saddle/phase would behave differently.

The complicated distribution of information inside the black hole with different dilaton or Berry phase or couplings can also be modeled differently, such as using the replica wormholes, as seen in previous works.  In \cite{Almheiri:2019qdq}, the form of the gravitational action used with replica manifolds and twist operators for the matter sectors which are inserted at various singularities, have been written as 
\begin{gather}
- \frac{1}{n} I_{\text{grav}} = \frac{S_0}{4\pi} \left \lbrack \int_{\Sigma_2} R + \int_{\partial \Sigma_2} 2K \right \rbrack + \int_{\Sigma_2} \frac{\phi}{4\pi} (R+2) + \frac{\phi_b}{4\pi} \int_{\partial \Sigma_2} 2K - (1- \frac{1}{n} ) \sum_i \lbrack S_0 + \phi(w_i) \rbrack,
\end{gather}
leading to the equation of motion
\begin{gather}
-4 \partial_w \partial_{\bar{w}} \rho+ e^{2 \rho} = 2 \pi ( 1- \frac{1}{n} ) \sum_i \delta^2( w - w_i). 
\end{gather}

Here, $w_i$ are the positions of the conical singularities or cosmic branes, and so they affect the form of the evolution of the black hole and the evaporation of the information. The tomography there is instead modeled by different copies of the black hole which are connected by different complexified wormholes, where these complex saddles could also be imagined as instantons, where this is also related to the conformal welding problem. In fact, the imprints of these wormholes in the limit of $n \to 1$ are the same as the tomography model of \cite{Lee:2022efh}. The distance between the twist fields $w_i$, which determines the size of the mouth of the wormhole and then the size of the island, would be related to the JT coupling $\alpha$, and modular Berry phase in the tomographic model and also to the dilaton $\varphi$, and the partial reduction parameter of \cite{Verheijden:2021yrb}.

Also, a metric ansatz for this form of distribution of matter field and singularities could be written as \cite{Almheiri:2019qdq},
\begin{gather}
ds^2 = e^{2\rho}  dw d\bar{w}, \ \ \ \ \ \ \ e^{2\rho} = \frac{4}{ (1-  | w  |^2 )^2 } e^{2 \delta \rho},
\end{gather}
where $\delta \rho$ is 
\begin{gather}
\delta \rho \sim - \frac{ (1- | w| )^2 }{3} U (\theta), \ \ \ \ \text{as} \ \ |w| \to 1.
\end{gather}

Another piece of evidence for the fact that the full and correct holographic picture should be time-dependent with the consideration of all the quantum fluctuations, came from the model of the work \cite{Geng:2022slq}, which showed that considering the  fluctuations of RS branes and the associated radion/dilaton mode could resolve the issue of the degeneracy of RT surfaces for the defects in wedge holography as an example, and therefore can resolve several issues in this direction.  Orbifolding the radion which plays the role of the dilaton in the JT gravity, has actually the same meaning as the size of the wormhole mouth determined by the distance between the conical singularities or defect/cosmic branes, $w_i$ in the replica wormholes \cite{Almheiri:2019qdq}, and is related to the JT couplings in the tomographic model of \cite{Lee:2022efh}. The fluctuations of these branes then can induce various phase transitions. In that work, it was also shown that the JT gravity can naturally be realized in the Karch-Randall braneworld, again indicating that in quantum gravity, looking from higher dimensions to lower dimensions can catch the gist of the dynamics and specifically the black hole evolution.

\subsection{Black hole secret sharing and Kaluza-Klein reduction}

As another way to depict the effects of the wormholes on the black hole evaporation, the ``inception black hole" model  in the setup of $\text{ER}=\text{EPR}$ and double holography has been used in \cite{Balasubramanian:2020hfs,Balasubramanian:2021xcm}.

In \cite{Balasubramanian:2020hfs}, the authors discussed geometric secret sharing in a mixed model of Hawking radiation, where they found that certain subregions of the black hole interior might not be reconstructable with specific subsystems of radiation and for reconstructing those subregions, ``all" of the radiation might be needed. 
However, we claim that the inception geometry, in addition to having a different Newton's constant $G'_N$ and horizon, $r'_h$ could also be in other dimensional spacetimes, $d'$, glued to the real spacetime, which then can give a better picture of the story, specifically for tracking the information of cases like the extremal black holes.

So these secret sharing models could be investigated in the setup of \cite{Verheijden:2021yrb} to check how the information would be distributed in Kaluza-Klein dimension reduction models. These study would actually help to understand better how holography and AdS/CFT works as it could help to depict how information layer-by-layer sit in different dimensions and become entangled to create the spacetime and also make the evaporation of the black hole unitary leading to the Page curve.  Specifically, this idea is compatible with novel Euclidean wormholes found between replicas in \cite{Almheiri:2019qdq, Penington:2019kki}, using replica trick for computing the radiation entropy. Also, the idea of \text{``inception geometry"} inside another incepted geometry of \cite{Balasubramanian:2020hfs} could be extended to higher dimensions, and check how in various dimensions they interact. The quantum secret sharing scheme of Hawking radiation then could be traced over to lower dimensions during the Kaluza-Klein dimension reduction in the scheme of \cite{Verheijden:2021yrb}. In fact, the form that the information are layered in different dimensions, could be imagined in their model as pieces of paper which are folded in a convex format. In \cite{Balasubramanian:2020hfs}, the real and inception geometries based on ER$=$EPR proposal and black hole complementarity have been considered to be ``on top of each other''. The multi-boundary setup of \cite{Balasubramanian:2020hfs} can also have different phases and configurations.

\begin{figure}[htbp]   
\begin{center}
\includegraphics[width=0.3\textwidth]{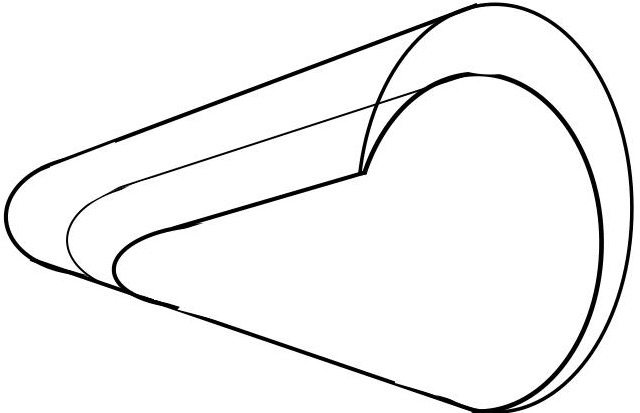}  
\caption{The real and inception geometries are folded in the covex-to-covex configuration. This then could be dimensionally reduced to $1d$ lower case. This inception model \cite{Balasubramanian:2020hfs}  could be connected to the replica wormhole picture \cite{Dong:2021oad}, tomographic brane model \cite{Lee:2022efh}, and also flows across dimensions \cite{GonzalezLezcano:2022mcd}.}
\label{fig:folded}
\end{center}
\end{figure}

The way the real physical geometry and the inception geometry are folded together would be as shown in figure \ref{fig:folded}, which is a convex surface on top of another convex surface. However, in other models, this can also be in the form of convex to concave gluing as well.

In principle, since the radiation and the CFT on the EOW brane are maximally entangled, any measurement on the radiation would result in the projection on the brane and vice versa.

The case below the Page transition would correspond to microstates of brane (or the number of Hawking quanta of radiation) less than exponential of  black hole microstates, as $k < e^{S_{\text{BH}} }$ and after Page transition the other way around. After the Page time, the islands, which are regions disconnected from the boundary of space and are reconstructable from the radiation, would form between the real black hole horizon and EOW as explained in \cite{Balasubramanian:2020hfs}. This then should be projected in the lower dimension after the Kaluza-Klein dimensional reduction. 

In our previous works \cite{Ghodrati:2021ozc, Ghodrati:2015rta}, the position of the cutoff brane $r_\Lambda$ have been changed in various ten-dimensional supergravity backgrounds to probe the phase transitions. In the model of ``inception geometry'' of \cite{Balasubramanian:2020hfs}, this could be written in terms of the constants of real space and the inception space as
\begin{gather}
r_t= \sqrt{\frac { \ell^2 G_N^2 r_h'^2 - \ell'^2 G_N'^2 r_h^2}{\ell^2 G_N^2 - \ell'^2 G_N'^2} },
\end{gather}
where here the prime corresponds to the inception case. This relation can also be extended for the case where one has flows between various dimensions as well, such as \cite{GonzalezLezcano:2022mcd}, where the flows between real and inception geometries could also be studied.  In addition, the various phases in confining models could also be parameterized by these constants as well. Specially, the Page transition happens at $r_h/G_N = r'_h/ G_N'$, or $r_h/d \ G_N = r'_h/  d' \ G_N'$ which then corresponds to confinement/deconfinement in those $10-d$ supergravity models. Note that during these evolutions, generally we take the central charge of the inception geometry to be constant, i.e, $c'=3\ell' /2 G_N'$. In fact, in \cite{Balasubramanian:2020hfs} it has been shown that the central charge of inception CFT would be higher than the real CFT, which would make sense for the confining models studied in \cite{Ghodrati:2021ozc} as well, since the inception geometry would correspond to the deconfined case and the real geometry to the confined case. So moving the UV cutoff in those confining models would actually correspond to moving the splicing locus of the inception and real geometries, leading to the change in the amount of entanglement between the microstates and radiation in the model of \cite{Balasubramanian:2020hfs}, which is also related to the running of the parameters $G_N'$, $\ell'$, and even $d'$.

Then, the EOW branes are added to the setup to model the microstates of the black hole. In principle, the splitting of the auxiliary systems or radiation (such as in the case of \cite{Dong:2021clv}) into multiple distinct parts could be modeled by purifying the additional black holes on the EOW brane with a multi-boundary wormhole where each leg would correspond to different parts of radiation. Interestingly, in \cite{Balasubramanian:2020hfs}, in their model, the part of the RT surface that is in the real geometry corresponds to effective field theory contribution to the generalized entropy and the part that is in the inception part corresponds to the brane segment entropy. Therefore, if we want to connect this model to other models, such as those of \cite{Fallows:2021sge}, the term in the form of $\frac{1}{4G} \tanh^{-1} T$ would correspond to the real part leading to the effective field theory contribution and the term in the form of $\frac{1}{4G} \ln \frac{L}{\epsilon}$ comes from terms that are in the inception geometry and are related to the boundary of islands. In addition, partial islands as in \cite{Almheiri:2019hni, Balasubramanian:2020hfs} could also be formed in the inception geometry, again related to the picture of tomography brane model, and higher corrections to JT couplings.

\subsection{Quantum focussing conjecture in $2d$ from $3d$ point of view}

An important criteria needed to be considered when connecting spacetimes with various parameters through replica wormholes, inception geometry or tomographic brane models, is the Quantum Null Energy Condition discussing the quantum expansion along a null congruence $N$. 

The Quantum Focusing Conjecture (QFC) \cite{Bousso:2015mna} claims that the quantum expansion $\Theta$ which is a functional derivative of the generalized entropy along the null congruences, which are orthogonal to a surface in the bulk $\sigma$, and which need not to lie on a horizon, cannot increase along any congruence. In other words, 
\begin{gather}
\frac{d\Theta}{d\lambda} \le 0.
\end{gather}

In most of the models islands lie behind the horizon, however, in \cite{Almheiri:2019yqk}, using QFC, the authors showed that islands can be extended to outside of the black hole horizon without the violation of causality, as QFC could save it even in that case.

As a practical application, in \cite{Almheiri:2019yqk}, it has been shown that the physics of the near-horizon region can be encoded in the state of the bath which is far away from the black hole, but the QFC would guarantee that the signals from islands cannot reach the physical AdS boundary.

It would then be interesting to check how QFC resolves this paradox from the $3d$ point of view looking to the $2d$ case, and how looking from the $3d$, the information in the near-horizon of the nearly-$\text{AdS}_2$ black hole in JT gravity is encoded.

The two dimensional version of QFC could be written as \cite{Bousso:2015mna,Almheiri:2019yqk}
\begin{gather}\label{eq:QFC2d}
\nabla^2_+ (\phi+S) \le 0,
\end{gather}
where $S$ is the entropy of the bulk fields, or it could be considered as the entropy of an interval along the light ray reaching to the end point of the interval in the bulk. This relation then could be viewed from one $d$ higher case.

The interesting point is that \ref{eq:QFC2d} together with the condition of having extremality at the starting point would imply that the generalized entropy will decrease along the light ray which starts at the quantum extremal surface and ends on the boundary. This conjecture would be still true in the three-dimensional case as well.

Note that in general, as one approaches the boundary, the dilaton grows while the entanglement entropy saturates. In the one dimension higher, the dilaton field  $\Phi$ actually controls the size of the circle as
\begin{gather}
\int ds_{S^1} = \int_0^{2\pi R} dt e^{\phi(x^\mu)} = 2\pi R e^{\phi(x^\mu)},
 \end{gather}
where $R e^{\phi(x^\mu)}$ is the effective radius of the circle at $x^\mu$. In addition, we have the relation $\sqrt{g_{3d}}= e^\phi \sqrt{g_{2d}}$. So from the size of the circle $S^1$ at any fixed point, one can see that $x^\mu$ would increase in the higher $d$ case. However, based on QFC, it would reach a singularity or goes beyond the physical boundary of $\text{AdS}_2$.

The condition for the island to be outside of the horizon is that the interval that one considers, namely $\lbrack b, b' \rbrack$ should be very large so that its entanglement wedge can contain the island,  the central charge $c$ should be relatively large, i.e,  $c \gg 1$,  and $\phi_0/c$  should not be too large. Therefore, the degrees of freedom of the island can be encoded in a very large distance correlation far from the horizon and in the exterior. Looking at this situation from the higher dimensional point of view would then be more interesting as the correlations, or quanta degrees of freedom could run away to the upper dimension with a higher chance.

For the JT gravity, using the equations of motion, the QFC could be written as
\begin{gather}\label{eq:WZWeq}
-2\pi T_{++} + \nabla_+^2 S \le 0.
\end{gather}
However, in \cite{Almheiri:2019yqk}, it has been shown actually one should look for a stronger inequality for general $2d$ holographic theories, as
\begin{gather}
-2\pi  T_{++} + \nabla_+^2 S + \frac{6}{c} ( \nabla_+ S)^2 \le 0.
\end{gather}

This then could also be written for $3d$ BTZ and also for WZW action. For the WZW model we have the relation
\begin{gather}
T_{\ell \ell'}= e^{2 \pi i \left (\frac{\ell(\ell+1) }{k+2} - \frac{k}{8 (k+2) }    \right ) } \delta_{\ell, \ell'}.
\end{gather}

The energy momentum tensor of the coset theory can also be written as
\begin{gather}
T= \frac{(J^+ J^-)- \partial J^3}{k+2} - \frac{2 ( J^3 J^3)}{k(k+2)},
\end{gather}

Putting this relation into \ref{eq:WZWeq} would lead to a Bousso bound for the WZW model case. The same can also be done for the Chern-Simons, BF, or $1d$ Schwarzian models.

\subsection{Connections with complexity}

As we saw in section \ref{sec:phasesmixed}, there would be four distinct behaviors in the correlation structures of the two mixed systems, and these should show their effects on the complexity, or rather complexity of purification \cite{Ghodrati:2019hnn} or complexity spreading \cite{Ghodrati:2018hss,Zhou:2019jlh,Ghodrati:2019bzz,Ghodrati:2017roz} as well. Interestingly these four distinct features have also been recently observed in \cite{Balasubramanian:2022tpr}. So, in order to investigate the saddles of Hawking radiation, complexity for JT models of gravity and island formulation needs to be further studied, specifically from our setup of looking from $1d$ higher point of view. Note that in this regard, the JT gravity itself actually arises from the dimensional reduction of charged black holes. In fact, in recent work of \cite{Brown:2018bms} and also in \cite{Alishahiha:2018swh}, the complexity of JT gravity has been studied which now needs to be embedded in the island formulation.

The complexity can specifically help to understand extremal limit of black holes. The $\text{AdS}_2$ factor in the near extremal limit of charged black holes significantly affects the complexity, as without considering the UV cutoff, it makes it constant at the later times. However, in \cite{Akhavan:2018wla}, it was shown that the UV cutoff at the boundary induces a cutoff behind the horizon which removes some part of the space time inside the horizon, which this is actually the island part, or the gap in the case of \cite{DeWolfe:2013uba}. After removing those parts of the spacetime, the rest shows a linear growth for the complexity which would be the expected behavior for the $\text{AdS}_2$ solution with a constant dilaton. So the island would decrease the complexity too and force it to become constant, instead of the linear growth, similar to the case of entropy.  Removing this part behind the horizon, as done in  \cite{Akhavan:2018wla} though, makes the complexity to increase linearly.

For a two dimensional gravity model which is obtained from a four dimensional Maxwell-Einstein gravity, by a dimensional reduction, in \cite{Alishahiha:2018swh}, the author found the connections between complexity growth rate, $dI/d\tau$, and entropy as
\begin{gather}\label{eq:extremalC}
\frac{dI}{d\tau}= S_0 T + \frac{\pi \ell}{G} T^2,
\end{gather}
where $S_0$ is the entropy of the extremal black hole, and therefore \ref{eq:extremalC} leads to the complexity of extremal black hole. This is for the case with cutoff which actually removes the island behind the horizon. Remember that keeping the patch behind the horizon where the island is, makes the entropy and complexity to become saturated at late times.

In fact, in \cite{Bhattacharya:2021jrn}, the complexity for the island part has been calculated and at the Page time, a phase transition and a jump in the volume has been detected, where because of the appearance of the island, the entanglement entropy becomes saturated, which correspondingly affects the complexity as well.

Then, the complexity of island creation itself, after the Page time can be computed. Note that always when we look from the higher dimensions to lower dimensions, the angular metric components in the higher $d$ would become dilaton in the lower $d$ case. So as found in \cite{Bhattacharya:2021jrn}, when in higher $d$, the Hartman-Maldacena (HM) surface and HM volume depend upon the angle between the gravitational brane and the conformal brane, and the complexity in lower $d$ would depend on the ``dilaton" coming from the dimension reduction of the space between HM surface and the conformal boundary.
The resulting dilaton in the lower dimensional space then would depend on the position of the brane in the higher $d$ case. Since, there is a critical angle in higher $d$, there is a critical value for the dilaton in lower $d$ as well, where we denote by $\phi_c$. When the area of HM surface becomes bigger than the island surface, the island surface would saturate the Page curve. This corresponds to the change of the two corresponding dilaton fields with respect to each other in the lower dimension space.

As the physical angle increased, which means the strength of the dilaton in lower $d$ increased, the volume of the island would decrease. This is exactly what we have found in section \ref{sec:calcb}, which demonstrated that increasing the angle $\mu$ or $\nu$ would decrease the critical bath size which corresponds to the island. Consequently, our results in section \ref{sec:calcb} is compatible with the results of \cite{Bhattacharya:2021jrn}, and therefore it would be compatible with the covariant prescription of subregion complexity.

After the Page time, the degrees of freedom in the right brane get access to the volume bounded by the HM surface, the left brane and the island surface. In the lower dimensional space, the two corresponding dilatons get mixed. This mixing at the Page time would cause the jump in the complexity, critical distance \cite{Ghodrati:2021ozc}, and other mixed correlation measures such as logarithmic negativity or entanglement of purification, as one would expect and also observed in \cite{Ghodrati:2019hnn}. The mixing of the dilaton in lower $d$ also affects the mutual complexity \cite{Hernandez:2020nem} or complexity of purification \cite{Ghodrati:2019hnn}, as in the higher $d$, the entanglement wedge of radiation gets access to the island degrees of freedom. This mixing then lowers the CoP of black hole degrees of freedom and increases the CoP of radiation, with a first order phase transition jump. 

Also, varying the parameters which controls the dimensional reductions by time, would introduce dynamics in the lower dimensional case, again leading to a model for the ``geometrical" evaporation of the black hole. So the nature of the jumps observed in higher $d$, can be explained by the mixing of the dilaton fields in the partially dimensionally reduced, lower $d$ case.

Note that generally the Einstein-Hilbert action changes after the partial dimensional reduction from
\begin{gather}
S= \frac{1}{16 \pi G^{(3)} } \int d^3 x \sqrt{-g} ( R^{(3)} - 2 \Lambda ),
\end{gather}
to the following action
\begin{gather}
S= \frac{2 \pi \alpha \ell_3}{16 \pi G^{(3)} } \int d^2 x \sqrt{-h} \phi ( R^{(2)} - 2 \Lambda ),
\end{gather}
where $\alpha \in (0, 1 \rbrack $ is a parameter which controls the partial reduction. Then, one can see how the parameter $\alpha$ will enter the rate of growth of complexity in a linear way, unlike the entropy which behaves as $S \propto \log \left \lbrack \sinh (c_0 \alpha) \right \rbrack$.

In \cite{Hernandez:2020nem}, the generalized CV (subregion complexity$=$volume) has also been introduced, where by varying the island profile $\tilde{\mathcal{B}}$, in the relation
\begin{gather}
\mathcal{C}_V^{\text{sub} (\mathbf{R})} = \text{max} \left \lbrack \frac{ \tilde{W}_{\text{gen}} (\tilde{\mathcal{B}}) + \tilde{W}_K(\tilde{\mathcal{B}}) }{G_{\text{eff}} \ell'} + \mathcal{C} ( R \cup \tilde{\mathcal{B}}) \right \rbrack,
\end{gather}
the CV complexity could be extremized. From this relation also one can see that the partial dimensional reduction would linearly affects the generalized volume $W_{\text{gen}}$, and also $W_K$ and therefore the complexity can give the corresponding Page curve. So the geometric evaporation can work for complexity as well.

\subsection{Connections with other works}

The four saddles we studied here and the connections with the geometric partial reduction and wormholes can be studied along other measures, which here we comment on several of them.

These four solutions are related to the number of saddles of partition function in the calculation of negativity as in \cite{Dong:2021oad}. Indeed for the partition function of negativity, there are four solutions for each case, before and after the Page time, which are directly connected to the four saddles that always could be found in any mixed correlation measure. In \cite{KumarBasak:2021rrx}, additional measures of entanglement negativity have been found for the setup of partial geometric reduction of  \cite{Verheijden:2021yrb}, and the Page curve for this mixed measure through the geometric evaporation has been constructed. Further measures such as partial entanglement entropy or subregion complexity  \cite{Bhattacharya:2021dnd}, or quantum error correction similar to \cite{Kibe:2021gtw}, would lead to the same results. Also R\'enyi negativity has been studied in the setup of quench dynamics \cite{Murciano:2021zvs}, and the entanglement bound for the thermalized states has been found in \cite{Vardhan:2021npf}, which is related to the quantum null energy condition and QFC.

In \cite{Shapourian:2020mkc},  by coupling the system to a bath and using another measure of random mixed states, namely the partial transpose and logarithmic negativity, and then using the diagrammatic method, the entanglement properties have been studied. In that work, similar to our result here, the authors found a critical size for the bath where when the bath is smaller than the system size, the logarithmic negativity shows an initial increase and then a final decrease, similar to the Page curve.

In \cite{Geng:2021mic}, the entanglement structure of a holographic BCFT in the black hole setup has been further studied. Their system is considered to be  doubly holographic which is dual to an eternal black string with an embedded Karch-Randall (KR) brane parameterized by its angle. The emergence of islands have been shown to depend on the angle for such branes where a critical angle $\theta_{\text{crit}}$ exists where below it, at zero temperature, the islands cannot be forged. Generally in such setups, from the $d+1$-dimensional bulk perspective, at the early time, the entropy would be controlled by the extremal \textit{Hartman-Maldacena (HM) surface}  which crosses the black string horizon and its surface increases by the growth of the Einstein-Rosen bridge. However, in the late-time the ``island surface" would control the entropy. The island surface which are the standard RT surfaces, lies between KR brane and the conformal boundary and is constant in time. The angle there would follow the Israel junction conditions which connects the tension of the brane with this angle.

In the finite temperature, however, it has been shown in \cite{Geng:2021mic} that with a black string, islands could be present, even with branes where the angles between them is below the critical angle $\theta_{\text{crit}}$. These islands however, only would be present in the finite connected region on the brane where they have been dubbed ``atoll''. Similar to our studies, in their case too, the size of the subregion and the brane angle would determine the behavior of entanglement entropy, so that it remains constant in time, or to follow the Page curve. So these parameters essentially determine the size and behavior of the  replica wormholes, inception geometry or tomographic brane picture, as we have observed in other models as well. Also, in \cite{Geng:2021mic}, it has been pointed out that the nested surfaces could taken shape by decreasing the brane angle $\theta_b$ or by increasing the anchor point $\Sigma$. So, the entropy depends on both the subregion size and brane angle. In general, it has been observed that decreasing the size of the radiation region or the brane angle would decrease the island size. 

In works such as \cite{Krishnan:2020oun,Krishnan:2020fer, Ghosh:2021axl}, the issues of island formations in the setups beyond AdS, in higher dimensions, and in singly and doubly holographic scenarios have been discussed where even the back-reactions from the bath has been considered.

Recently, also, in \cite{Akal:2021dqt}, the behavior of the Page curve under final state projection has been studied.
In that context, a final state boundary condition is imposed on the spacelike singularity. Then, the effect of this final state projection using the behavior of pseudo-entropy has been investigated. The real part of the pseudo-entropy in fact could estimate the amount of quantum entanglement or the number of Bell pairs averaged over histories between the initial state and the post-selected final state. Then, the extended version of pseudo-entropy for mixed system which can capture the historical behavior of mixed correlations would be constructed and then its island contribution could be investigated. As in  \cite{Akal:2021dqt}, the decreasing behavior of the Page curve can be arisen due to the past evolution of the post-selected final state. Both of these explanations actually use the modifications of Hilbert space structure inside the black hole.

In \cite{Akal:2021dqt}, the global $\text{AdS}_3$ has been written as
\begin{gather}
ds^2=- \frac{T^2}{\pi^2} \cosh^2 \rho dt^2 + d\rho^2 + \frac{T^2}{\pi^2} \sinh^2 \rho dx^2,
\end{gather}
and the EOW brane has been written in the form of
\begin{gather}
\cosh \rho \sin \frac{\pi t}{T} = \cosh \eta_0,
\end{gather}
which describes a two-dimensional de Sitter spacetime. Then, the tension of this brane $\mathcal{T}$ is related to $\eta_0$ as $\mathcal{T}= - \coth \eta_0$. The tension which is dual to the boundary of the BCFT has the range $\mathcal{T} < -1$. The complex-valued boundary entropy then could be written as
\begin{gather}
S_{\text{bdy}} = \frac{c}{6} \log \sqrt{\frac{| \mathcal{T} | -1 }{ | \mathcal{T}| +1} } - i \frac{\pi c}{12},
\end{gather}
which again the angle which determines the tension of the brane determines the entropy, and again using this relation one can see that a critical value for the angle would exist. The further connections between the models studied here such as inception geometry, replica wormholes and tomography could be further studied in this way.

In \cite{Dong:2021clv}, the connections between holographic entanglement negativity and replica symmetry breaking has been studied. There it has been shown that the R\'{e}nyi negativities are often dominated by bulk solutions that break the replica symmetry. It would be important to check how such solutions which break the replica symmetry contribute to the islands, compared to those that preserve the symmetry.

In \cite{Wang:2021aog}, the connections between the reflected entropies of multipartite mixed states in $\text{CFT}_2$ and hyperbolic string vertices of closed string field theory (CSFT) \cite{Moosavian:2017qsp, Moosavian:2017sev} have been established as both have the same Riemann surfaces. It would be interesting to analyze the critical bath size we have found in terms of these hyperbolic string vertices. The connections between the Page curve and the dynamical equation, i.e, the Batalin-Vilkovisky (BV) master equation then could be analyzed.

Other corrections such as corner term could also play a role in the evaporation and in the analysis of black hole phases. In \cite{Arias:2021ilh}, the Hayward term for the corners of the geometry in JT gravity has been looked into. These corners arise in the computation of the Hartle-Hawking wave functions and reduce the density matrices. These terms with extra Nambu-Goto term would be compatible with the cosmic brane prescription.
This Hayward term for the JT gravity would be
\begin{gather}
I_H^{d=2} \equiv \frac{1}{8\pi G} \text{cos}^{-1} ( n . \bar{n} ) \Phi_\Gamma,
\end{gather}
where $\Gamma$ is a codim-2 corner where split the boundary region $\Sigma= B \cup \bar{B}$ as $\Gamma= B \cap \bar{B}$, and $\Phi_\Gamma$ is the dilaton field at the point $\Gamma$ which follows from an implicit standard dimensional reduction scheme. So for the setup of geometric picture of black hole evaporation of \cite{Verheijden:2021yrb}, this term could also be considered. Therefore, the effective action with the Hayward term could be written as \cite{Arias:2021ilh}
\begin{gather}\label{eq:JTH}
I_{JTH}=-\frac{\phi_b}{16 \pi G_N} \frac{\theta^2}{\beta} + \frac{1}{8 \pi G_N} (\Phi_0 + \Phi_\Gamma) (2\pi-\theta),
\end{gather}
where the constant value  on the boundary is $\phi_\partial(u)=\phi_b$, and $\Phi_0 \gg 1$ is the constant in the JT gravity corresponding to the extremal entropy. The action \ref{eq:JTH}, can be written in terms of the deficit angle $\alpha=2\pi - \theta$ as well. In addition, in that work, by integrating the identity $\hat{S}_n= n^2\partial_n \left ( \frac{n-1}{n} S_n \right )$, 
 the R\'enyi entropy for the case with the boundary and Hayward term, can be found as
 \begin{gather}
 S_n= \frac{\phi}{4G} \frac{\pi}{\beta} \frac{n+1}{n},
 \end{gather}
where $n$ is the Rényi index. So adding these Hayward term would correspond to conical defects in the AdS spacetimes, which then affects the partition functions, the states and the phases of the mixed correlations in the radiation. These Hayward terms would also definitely affect the replica wormholes, as the bulk dilaton field would get shifted by the dilaton in the corners, and also the positions of the twist fields, the JT couplings in the tomographic brane model, and the real part of the inception geometry, i.e, $\frac{1}{4G} \tanh^{-1} T $.

In \cite{Dvali:2021ofp},  the vortex structure in the black hole has been studied, which was based on the graviton-condensate description of black holes and also the correspondence between the black holes and ``saturons". There, it has been shown that both the black holes and also Q-ball-type saturon of a renormalizable theory would obey the same extremality bound on the spin. This correspondence would also be interesting from the bath structure points of view and also the structures of mixed correlations. For instance, for these saturons, the structure of negativity and its island contribution, could be studied.
Specifically, the effects of interactions between the randomly oriented and scattered pairs of vortices/anti-vortices on the mixed correlations and island of negativity could be probed.

In \cite{Gautam:2022akq}, it has been proposed that instead of considering the entanglement between the spatial degrees of freedom, one could consider the gauge degrees of freedom and construct the so called ``matrix entanglement". Using this setup, the evaporation of black hole can be modeled using the entanglement between the confined and deconfined sectors. One could imagine that the presence of the confined degrees of freedom in the partially deconfined states could be modeled again by the replica wormholes or inception geometry, or in the setup of void formations of \cite{Liu:2019svk}. The presence of these partially deconfined states in the confined degrees of freedom could also be understood from the effects of the dilaton from the higher dimensions in the lower dimensions, and also wormholes in the setup of $\text{ER}=\text{EPR}$.

 In \cite{Hashimoto:2020cas}, the island formulation in $4d$ has been studied where it has been shown that the entanglement entropy has an area-like divergence, and to resolve this issue the authors replaced the Newton's constant $G_N$ by its renormalized version as
 \begin{gather}
 \frac{1}{4G_N^{(r)}}  = \frac{1}{4G_N} + \frac{1}{\epsilon^2}, 
 \end{gather}
 and rederived the Page curve. In the higher dimension case then, their proposal for the island would be \cite{Hashimoto:2020cas}
\begin{gather}
S(R)=\text{min} \left \{ \text{ext} \left\lbrack \frac{\text{Area}(\partial I)  }{4 G_N^{(r)} } + S_{\text{matter}}^{( \text{finite} ) } ( R \cup I ) \right \rbrack  \right \}.
\end{gather} 
The exact $2d$ version of that work has been also done in \cite{Anegawa:2020ezn}.

 So effectively it is the Newton's  constant that is being renormalized which reaches to the bare Newton's constant in the lower dimensional case. Also, note that in $4d$, the Newton's constant has dimension $(\text{length})^2$ which then lead to a dilaton field $\Phi$ which also has the dimension of $(\text{length})^2$ being interpreted as the area. However, in $2d$, $G_N$ and therefore $\Phi_r$ in \cite{Verheijden:2021yrb} would be dimensionless, as just a number. This also matches with the result of \cite{Eberhardt:2022wlc} as for the case of  $k \to \infty$ (related to small $\hbar$ limit) the partition functions of higher dimensional case would turned out to be the lower dimensional case.
 
 So the effects of the ``areas" in higher dimension case would lead to ``points" or dimensionless quantities in the lower dimensional case. This is again consistent with the result of  \cite{Hashimoto:2020cas} where noted that the $4d$ matter fields in $2d$ subspace would become massless fields which are the lowest mode of the KK towers. However, in \cite{Demulder:2022aij}, it was shown that this would not be exactly massless, and from the island formulation, a lower bound can be found for the mass of the graviton field. Also, the distance between the two boundaries would be always associated with the ``complexity" or volume.

It would also be interesting to check the duality between the apparent horizon of pure state black hole, i.e, $\frac{1}{4} \text{Area}(\gamma)$, and $S_{vN} ( \bar {\rho} = - \sum_n (\rho)_{nn} \log(\rho)_{nn}$ called ``diagonal entropy", \cite{Chandra:2022fwi}, from higher dimension point of view, and dimensional reduction as well. This can specifically be done by the large central charge $c$ expansion in the $2d$ CFT.

Furthermore, as shown in \cite{Baiguera:2022sao},  the charged R\'enyi entropy could also help to classify the phase structures in entanglement, specifically using the global symmetry in different charged sectors.  Its form can be written as
\begin{gather}
S_n(\mu) \equiv \frac{1}{1-n} \log \text{Tr} \Big \lbrack  \rho_A \frac{e^{\mu Q_A} }{n_A(\mu)}  \Big \rbrack^n,
\end{gather}
where $Q_A$ is the charge operator for each region, $\mu$ is the chemical potential which is the conjugate for the charge $Q_A$, $\rho_A= \text{Tr}_{\bar{A}} \rho$ is the reduced density matrix over a region $A$, and $n_A(\mu) \equiv \text{Tr}  \lbrack \rho_A e^{\mu Q_A} \rbrack$ is the normalization constant. This quantity then could be observed from a higher dimension which could have interesting results for further understanding of the symmetry-protected topological (SPT) states.

 \medskip
 
\bibliography{verlinde.bib}

\providecommand{\href}[2]{#2}\begingroup\raggedright\begin{thebibliography}{100}

\bibitem{Verheijden:2021yrb}
E.~Verheijden and E.~Verlinde, {\it {From the BTZ black hole to JT gravity:
  geometrizing the island}},  {\em JHEP} {\bf 11} (2021) 092,
  [\href{http://xxx.lanl.gov/abs/2102.00922}{{\tt arXiv:2102.00922}}].

\bibitem{Almheiri:2019hni}
A.~Almheiri, R.~Mahajan, J.~Maldacena, and Y.~Zhao, {\it {The Page curve of
  Hawking radiation from semiclassical geometry}},  {\em JHEP} {\bf 03} (2020)
  149, [\href{http://xxx.lanl.gov/abs/1908.10996}{{\tt arXiv:1908.10996}}].

\bibitem{Almheiri:2019psf}
A.~Almheiri, N.~Engelhardt, D.~Marolf, and H.~Maxfield, {\it {The entropy of
  bulk quantum fields and the entanglement wedge of an evaporating black
  hole}},  {\em JHEP} {\bf 12} (2019) 063,
  [\href{http://xxx.lanl.gov/abs/1905.08762}{{\tt arXiv:1905.08762}}].

\bibitem{Ghodrati:2022kuk}
M.~Ghodrati, {\it {Critical distance and Crofton form in confining
  geometries}},  {\em J. Korean Phys. Soc.} {\bf 81} (2022), no.~2 77--90,
  [\href{http://xxx.lanl.gov/abs/2205.03565}{{\tt arXiv:2205.03565}}].

\bibitem{Ghodrati:2020vzm}
M.~Ghodrati, {\it {Entanglement wedge reconstruction and correlation measures
  in mixed states: Modular flows versus quantum recovery channels}},  {\em
  Phys. Rev. D} {\bf 104} (2021), no.~4 046004,
  [\href{http://xxx.lanl.gov/abs/2012.04386}{{\tt arXiv:2012.04386}}].

\bibitem{Dong:2021oad}
X.~Dong, S.~McBride, and W.~W. Weng, {\it {Replica Wormholes and Holographic
  Entanglement Negativity}},  \href{http://xxx.lanl.gov/abs/2110.11947}{{\tt
  arXiv:2110.11947}}.

\bibitem{Anderson:2020vwi}
L.~Anderson, O.~Parrikar, and R.~M. Soni, {\it {Islands with gravitating baths:
  towards ER = EPR}},  {\em JHEP} {\bf 21} (2020) 226,
  [\href{http://xxx.lanl.gov/abs/2103.14746}{{\tt arXiv:2103.14746}}].

\bibitem{Agrawal:2021nkw}
S.~Agrawal, O.~DeWolfe, J.~Levin, and G.~Smith, {\it {Phase Transitions of
  Correlations in Black Hole Geometries}},
  \href{http://xxx.lanl.gov/abs/2112.09704}{{\tt arXiv:2112.09704}}.

\bibitem{Jain:2020rbb}
P.~Jain and S.~Mahapatra, {\it {Mixed state entanglement measures as probe for
  confinement}},  {\em Phys. Rev. D} {\bf 102} (2020) 126022,
  [\href{http://xxx.lanl.gov/abs/2010.07702}{{\tt arXiv:2010.07702}}].

\bibitem{Ghodrati:2021ozc}
M.~Ghodrati, {\it {Correlations of mixed systems in confining backgrounds}},
  {\em Eur. Phys. J. C} {\bf 82} (2022), no.~6 531,
  [\href{http://xxx.lanl.gov/abs/2110.12970}{{\tt arXiv:2110.12970}}].

\bibitem{Dong:2021clv}
X.~Dong, X.-L. Qi, and M.~Walter, {\it {Holographic entanglement negativity and
  replica symmetry breaking}},  {\em JHEP} {\bf 06} (2021) 024,
  [\href{http://xxx.lanl.gov/abs/2101.11029}{{\tt arXiv:2101.11029}}].

\bibitem{Akers:2021pvd}
C.~Akers, T.~Faulkner, S.~Lin, and P.~Rath, {\it {Reflected entropy in random
  tensor networks}},  \href{http://xxx.lanl.gov/abs/2112.09122}{{\tt
  arXiv:2112.09122}}.

\bibitem{Vardhan:2021npf}
S.~Vardhan, J.~Kudler-Flam, H.~Shapourian, and H.~Liu, {\it {Bound entanglement
  in thermalized states and black hole radiation}},
  \href{http://xxx.lanl.gov/abs/2110.02959}{{\tt arXiv:2110.02959}}.

\bibitem{Shapourian:2020mkc}
H.~Shapourian, S.~Liu, J.~Kudler-Flam, and A.~Vishwanath, {\it {Entanglement
  Negativity Spectrum of Random Mixed States: A Diagrammatic Approach}},  {\em
  PRX Quantum} {\bf 2} (2021), no.~3 030347,
  [\href{http://xxx.lanl.gov/abs/2011.01277}{{\tt arXiv:2011.01277}}].

\bibitem{Jeong:2022zea}
H.-S. Jeong, K.-Y. Kim, and Y.-W. Sun, {\it {Holographic entanglement density
  for spontaneous symmetry breaking}},
  \href{http://xxx.lanl.gov/abs/2203.07612}{{\tt arXiv:2203.07612}}.

\bibitem{Akal:2022qei}
I.~Akal, T.~Kawamoto, S.-M. Ruan, T.~Takayanagi, and Z.~Wei, {\it {Zoo of
  holographic moving mirrors}},  \href{http://xxx.lanl.gov/abs/2205.02663}{{\tt
  arXiv:2205.02663}}.

\bibitem{Haehl:2022uop}
F.~M. Haehl and Y.~Zhao, {\it {Collisions of localized shocks and quantum
  circuits}},  \href{http://xxx.lanl.gov/abs/2202.04661}{{\tt
  arXiv:2202.04661}}.

\bibitem{Penington:2019kki}
G.~Penington, S.~H. Shenker, D.~Stanford, and Z.~Yang, {\it {Replica wormholes
  and the black hole interior}},
  \href{http://xxx.lanl.gov/abs/1911.11977}{{\tt arXiv:1911.11977}}.

\bibitem{Demulder:2022aij}
S.~Demulder, A.~Gnecchi, I.~Lavdas, and D.~Lust, {\it {Islands and Light
  Gravitons in type IIB String Theory}},
  \href{http://xxx.lanl.gov/abs/2204.03669}{{\tt arXiv:2204.03669}}.

\bibitem{Ghodrati:2019hnn}
M.~Ghodrati, X.-M. Kuang, B.~Wang, C.-Y. Zhang, and Y.-T. Zhou, {\it {The
  connection between holographic entanglement and complexity of purification}},
   {\em JHEP} {\bf 09} (2019) 009,
  [\href{http://xxx.lanl.gov/abs/1902.02475}{{\tt arXiv:1902.02475}}].

\bibitem{Akal:2021foz}
I.~Akal, Y.~Kusuki, N.~Shiba, T.~Takayanagi, and Z.~Wei, {\it {Holographic
  moving mirrors}},  {\em Class. Quant. Grav.} {\bf 38} (2021), no.~22 224001,
  [\href{http://xxx.lanl.gov/abs/2106.11179}{{\tt arXiv:2106.11179}}].

\bibitem{Grimaldi:2022suv}
G.~Grimaldi, J.~Hernandez, and R.~C. Myers, {\it {Quantum Extremal Islands Made
  Easy, Part IV: Massive Black Holes on the Brane}},
  \href{http://xxx.lanl.gov/abs/2202.00679}{{\tt arXiv:2202.00679}}.

\bibitem{Alishahiha:2022kzc}
M.~Alishahiha, S.~Banerjee, and J.~Kames-King, {\it {Complexity via Replica
  Trick}},  \href{http://xxx.lanl.gov/abs/2205.01150}{{\tt arXiv:2205.01150}}.

\bibitem{Balasubramanian:2022tpr}
V.~Balasubramanian, P.~Caputa, J.~Magan, and Q.~Wu, {\it {A new measure of
  quantum state complexity}},  \href{http://xxx.lanl.gov/abs/2202.06957}{{\tt
  arXiv:2202.06957}}.

\bibitem{Balasubramanian:2022fiy}
V.~Balasubramanian, A.~Kar, Yue, Li, and O.~Parrikar, {\it {Quantum Error
  Correction in the Black Hole Interior}},
  \href{http://xxx.lanl.gov/abs/2203.01961}{{\tt arXiv:2203.01961}}.

\bibitem{Mertens:2018fds}
T.~G. Mertens, {\it {The Schwarzian theory \textemdash{} origins}},  {\em JHEP}
  {\bf 05} (2018) 036, [\href{http://xxx.lanl.gov/abs/1801.09605}{{\tt
  arXiv:1801.09605}}].

\bibitem{Suzuki:2021zbe}
K.~Suzuki and T.~Takayanagi, {\it {JT gravity limit of Liouville CFT and matrix
  model}},  {\em JHEP} {\bf 11} (2021) 137,
  [\href{http://xxx.lanl.gov/abs/2108.12096}{{\tt arXiv:2108.12096}}].

\bibitem{Blommaert:2020yeo}
A.~Blommaert, T.~G. Mertens, and H.~Verschelde, {\it {Unruh detectors and
  quantum chaos in JT gravity}},  {\em JHEP} {\bf 03} (2021) 086,
  [\href{http://xxx.lanl.gov/abs/2005.13058}{{\tt arXiv:2005.13058}}].

\bibitem{Kawamoto:2022etl}
T.~Kawamoto, T.~Mori, Y.-k. Suzuki, T.~Takayanagi, and T.~Ugajin, {\it
  {Holographic local operator quenches in BCFTs}},  {\em JHEP} {\bf 05} (2022)
  060, [\href{http://xxx.lanl.gov/abs/2203.03851}{{\tt arXiv:2203.03851}}].

\bibitem{PhysRevD.44.314}
E.~Witten, {\it String theory and black holes},  {\em Phys. Rev. D} {\bf 44}
  (Jul, 1991) 314--324.

\bibitem{Ashok:2021ffx}
S.~K. Ashok and J.~Troost, {\it {Twisted Strings in Three-dimensional Black
  Holes}},  \href{http://xxx.lanl.gov/abs/2112.08784}{{\tt arXiv:2112.08784}}.

\bibitem{GonzalezLezcano:2022mcd}
A.~Gonz\'alez~Lezcano, J.~Hong, J.~T. Liu, L.~A. Pando~Zayas, and C.~F.
  Uhlemann, {\it {$c$-Functions in Flows Across Dimensions}},
  \href{http://xxx.lanl.gov/abs/2207.09360}{{\tt arXiv:2207.09360}}.

\bibitem{Akal:2020twv}
I.~Akal, Y.~Kusuki, N.~Shiba, T.~Takayanagi, and Z.~Wei, {\it {Entanglement
  Entropy in a Holographic Moving Mirror and the Page Curve}},  {\em Phys. Rev.
  Lett.} {\bf 126} (2021), no.~6 061604,
  [\href{http://xxx.lanl.gov/abs/2011.12005}{{\tt arXiv:2011.12005}}].

\bibitem{Miyaji:2021ktr}
M.~Miyaji, T.~Takayanagi, and T.~Ugajin, {\it {Spectrum of End of the World
  Branes in Holographic BCFTs}},  {\em JHEP} {\bf 06} (2021) 023,
  [\href{http://xxx.lanl.gov/abs/2103.06893}{{\tt arXiv:2103.06893}}].

\bibitem{Mertens:2020hbs}
T.~G. Mertens and G.~J. Turiaci, {\it {Liouville quantum gravity -- holography,
  JT and matrices}},  {\em JHEP} {\bf 01} (2021) 073,
  [\href{http://xxx.lanl.gov/abs/2006.07072}{{\tt arXiv:2006.07072}}].

\bibitem{Tian:2022pso}
J.~Tian, {\it {Islands in Generalized Dilaton Theories}},
  \href{http://xxx.lanl.gov/abs/2204.08751}{{\tt arXiv:2204.08751}}.

\bibitem{Ahn:2021chg}
B.~Ahn, S.-E. Bak, H.-S. Jeong, K.-Y. Kim, and Y.-W. Sun, {\it {Islands in
  charged linear dilaton black holes}},  {\em Phys. Rev. D} {\bf 105} (2022),
  no.~4 046012, [\href{http://xxx.lanl.gov/abs/2107.07444}{{\tt
  arXiv:2107.07444}}].

\bibitem{Seiberg:2003nm}
N.~Seiberg and D.~Shih, {\it {Branes, rings and matrix models in minimal
  (super)string theory}},  {\em JHEP} {\bf 02} (2004) 021,
  [\href{http://xxx.lanl.gov/abs/hep-th/0312170}{{\tt hep-th/0312170}}].

\bibitem{Mertens:2019tcm}
T.~G. Mertens and G.~J. Turiaci, {\it {Defects in Jackiw-Teitelboim Quantum
  Gravity}},  {\em JHEP} {\bf 08} (2019) 127,
  [\href{http://xxx.lanl.gov/abs/1904.05228}{{\tt arXiv:1904.05228}}].

\bibitem{Israel:2003ry}
D.~Israel, C.~Kounnas, and M.~P. Petropoulos, {\it {Superstrings on NS5
  backgrounds, deformed AdS(3) and holography}},  {\em JHEP} {\bf 10} (2003)
  028, [\href{http://xxx.lanl.gov/abs/hep-th/0306053}{{\tt hep-th/0306053}}].

\bibitem{Nippanikar:2021skr}
O.~V. Nippanikar, A.~Sharma, and K.~P. Yogendran, {\it {The BTZ black hole
  spectrum and partition function}},
  \href{http://xxx.lanl.gov/abs/2112.11253}{{\tt arXiv:2112.11253}}.

\bibitem{Eberhardt:2022wlc}
L.~Eberhardt, {\it {Off-shell Partition Functions in 3d Gravity}},
  \href{http://xxx.lanl.gov/abs/2204.09789}{{\tt arXiv:2204.09789}}.

\bibitem{Hartman:2020swn}
T.~Hartman, E.~Shaghoulian, and A.~Strominger, {\it {Islands in Asymptotically
  Flat 2D Gravity}},  {\em JHEP} {\bf 07} (2020) 022,
  [\href{http://xxx.lanl.gov/abs/2004.13857}{{\tt arXiv:2004.13857}}].

\bibitem{Cotler:2017jue}
J.~Cotler, N.~Hunter-Jones, J.~Liu, and B.~Yoshida, {\it {Chaos, Complexity,
  and Random Matrices}},  {\em JHEP} {\bf 11} (2017) 048,
  [\href{http://xxx.lanl.gov/abs/1706.05400}{{\tt arXiv:1706.05400}}].

\bibitem{Almheiri:2019qdq}
A.~Almheiri, T.~Hartman, J.~Maldacena, E.~Shaghoulian, and A.~Tajdini, {\it
  {Replica Wormholes and the Entropy of Hawking Radiation}},  {\em JHEP} {\bf
  05} (2020) 013, [\href{http://xxx.lanl.gov/abs/1911.12333}{{\tt
  arXiv:1911.12333}}].

\bibitem{Kusuki:2022wns}
Y.~Kusuki, {\it {Semiclassical Gravity from Averaged Boundaries in
  two-dimensional BCFTs}},  \href{http://xxx.lanl.gov/abs/2206.03035}{{\tt
  arXiv:2206.03035}}.

\bibitem{Kusuki:2019zsp}
Y.~Kusuki, J.~Kudler-Flam, and S.~Ryu, {\it {Derivation of Holographic
  Negativity in AdS$_3$/CFT$_2$}},  {\em Phys. Rev. Lett.} {\bf 123} (2019),
  no.~13 131603, [\href{http://xxx.lanl.gov/abs/1907.07824}{{\tt
  arXiv:1907.07824}}].

\bibitem{Shao:2022gpg}
Y.~Shao, M.-K. Yuan, and Y.~Zhou, {\it {Entanglement Negativity and Defect
  Extremal Surface}},  \href{http://xxx.lanl.gov/abs/2206.05951}{{\tt
  arXiv:2206.05951}}.

\bibitem{Basu:2022reu}
D.~Basu, H.~Parihar, V.~Raj, and G.~Sengupta, {\it {Defect extremal surfaces
  for entanglement negativity}},
  \href{http://xxx.lanl.gov/abs/2205.07905}{{\tt arXiv:2205.07905}}.

\bibitem{Deng:2020ent}
F.~Deng, J.~Chu, and Y.~Zhou, {\it {Defect extremal surface as the holographic
  counterpart of Island formula}},  {\em JHEP} {\bf 03} (2021) 008,
  [\href{http://xxx.lanl.gov/abs/2012.07612}{{\tt arXiv:2012.07612}}].

\bibitem{Afrasiar:2022ebi}
M.~Afrasiar, J.~Kumar~Basak, A.~Chandra, and G.~Sengupta, {\it {Islands for
  Entanglement Negativity in Communicating Black Holes}},
  \href{http://xxx.lanl.gov/abs/2205.07903}{{\tt arXiv:2205.07903}}.

\bibitem{KumarBasak:2020ams}
J.~Kumar~Basak, D.~Basu, V.~Malvimat, H.~Parihar, and G.~Sengupta, {\it
  {Islands for Entanglement Negativity}},
  \href{http://xxx.lanl.gov/abs/2012.03983}{{\tt arXiv:2012.03983}}.

\bibitem{Lu:2021gmv}
Y.~Lu and J.~Lin, {\it {Islands in Kaluza\textendash{}Klein black holes}},
  {\em Eur. Phys. J. C} {\bf 82} (2022), no.~2 132,
  [\href{http://xxx.lanl.gov/abs/2106.07845}{{\tt arXiv:2106.07845}}].

\bibitem{Malek:2020mlk}
E.~Malek, H.~Nicolai, and H.~Samtleben, {\it {Tachyonic Kaluza-Klein modes and
  the AdS swampland conjecture}},  {\em JHEP} {\bf 08} (2020) 159,
  [\href{http://xxx.lanl.gov/abs/2005.07713}{{\tt arXiv:2005.07713}}].

\bibitem{Bena:2020xxb}
I.~Bena, K.~Pilch, and N.~P. Warner, {\it {Brane-Jet Instabilities}},  {\em
  JHEP} {\bf 10} (2020) 091, [\href{http://xxx.lanl.gov/abs/2003.02851}{{\tt
  arXiv:2003.02851}}].

\bibitem{Pedraza:2021cvx}
J.~F. Pedraza, A.~Svesko, W.~Sybesma, and M.~R. Visser, {\it {Semi-classical
  thermodynamics of quantum extremal surfaces in Jackiw-Teitelboim gravity}},
  {\em JHEP} {\bf 12} (2021) 134,
  [\href{http://xxx.lanl.gov/abs/2107.10358}{{\tt arXiv:2107.10358}}].

\bibitem{Karananas:2020fwx}
G.~K. Karananas, A.~Kehagias, and J.~Taskas, {\it {Islands in linear dilaton
  black holes}},  {\em JHEP} {\bf 03} (2021) 253,
  [\href{http://xxx.lanl.gov/abs/2101.00024}{{\tt arXiv:2101.00024}}].

\bibitem{Yu:2021cgi}
M.-H. Yu and X.-H. Ge, {\it {Islands and Page curves in charged dilaton black
  holes}},  {\em Eur. Phys. J. C} {\bf 82} (2022), no.~1 14,
  [\href{http://xxx.lanl.gov/abs/2107.03031}{{\tt arXiv:2107.03031}}].

\bibitem{Carroll:2009maa}
S.~M. Carroll, M.~C. Johnson, and L.~Randall, {\it {Extremal limits and black
  hole entropy}},  {\em JHEP} {\bf 11} (2009) 109,
  [\href{http://xxx.lanl.gov/abs/0901.0931}{{\tt arXiv:0901.0931}}].

\bibitem{DeWolfe:2013uba}
O.~DeWolfe, S.~S. Gubser, and C.~Rosen, {\it {Fermionic response in a zero
  entropy state of $\mathcal N=$ 4 super-Yang-Mills}},  {\em Phys. Rev. D} {\bf
  91} (2015), no.~4 046011, [\href{http://xxx.lanl.gov/abs/1312.7347}{{\tt
  arXiv:1312.7347}}].

\bibitem{Penington:2019npb}
G.~Penington, {\it {Entanglement Wedge Reconstruction and the Information
  Paradox}},  {\em JHEP} {\bf 09} (2020) 002,
  [\href{http://xxx.lanl.gov/abs/1905.08255}{{\tt arXiv:1905.08255}}].

\bibitem{Gukov:2022oed}
S.~Gukov, V.~S.~H. Lee, and K.~M. Zurek, {\it {Near-Horizon Quantum Dynamics of
  4-d Einstein Gravity from 2-d JT Gravity}},
  \href{http://xxx.lanl.gov/abs/2205.02233}{{\tt arXiv:2205.02233}}.

\bibitem{Zurek:2022xzl}
K.~M. Zurek, {\it {Snowmass 2021 White Paper: Observational Signatures of
  Quantum Gravity}},  \href{http://xxx.lanl.gov/abs/2205.01799}{{\tt
  arXiv:2205.01799}}.

\bibitem{Solodukhin:1998tc}
S.~N. Solodukhin, {\it {Conformal description of horizon's states}},  {\em
  Phys. Lett. B} {\bf 454} (1999) 213--222,
  [\href{http://xxx.lanl.gov/abs/hep-th/9812056}{{\tt hep-th/9812056}}].

\bibitem{Fujita:2011fp}
M.~Fujita, T.~Takayanagi, and E.~Tonni, {\it {Aspects of AdS/BCFT}},  {\em
  JHEP} {\bf 11} (2011) 043, [\href{http://xxx.lanl.gov/abs/1108.5152}{{\tt
  arXiv:1108.5152}}].

\bibitem{Suzuki:2022xwv}
K.~Suzuki and T.~Takayanagi, {\it {BCFT and Islands in Two Dimensions}},
  \href{http://xxx.lanl.gov/abs/2202.08462}{{\tt arXiv:2202.08462}}.

\bibitem{Izumi:2022opi}
K.~Izumi, T.~Shiromizu, K.~Suzuki, T.~Takayanagi, and N.~Tanahashi, {\it {Brane
  Dynamics of Holographic BCFTs}},
  \href{http://xxx.lanl.gov/abs/2205.15500}{{\tt arXiv:2205.15500}}.

\bibitem{Erdmenger:2016msd}
J.~Erdmenger, M.~Flory, M.-N. Newrzella, M.~Strydom, and J.~M.~S. Wu, {\it
  {Quantum Quenches in a Holographic Kondo Model}},  {\em JHEP} {\bf 04} (2017)
  045, [\href{http://xxx.lanl.gov/abs/1612.06860}{{\tt arXiv:1612.06860}}].

\bibitem{Banerjee:2022jnv}
S.~Banerjee, M.~Dorband, J.~Erdmenger, R.~Meyer, and A.-L. Weigel, {\it {Berry
  phases, wormholes and factorization in AdS/CFT}},
  \href{http://xxx.lanl.gov/abs/2202.11717}{{\tt arXiv:2202.11717}}.

\bibitem{Anous:2022wqh}
T.~Anous, M.~Meineri, P.~Pelliconi, and J.~Sonner, {\it {Sailing past the End
  of the World and discovering the Island}},
  \href{http://xxx.lanl.gov/abs/2202.11718}{{\tt arXiv:2202.11718}}.

\bibitem{Calmet:2021cip}
X.~Calmet and S.~D.~H. Hsu, {\it {Quantum hair and black hole information}},
  {\em Phys. Lett. B} {\bf 827} (2022) 136995,
  [\href{http://xxx.lanl.gov/abs/2112.05171}{{\tt arXiv:2112.05171}}].

\bibitem{Raju:2020smc}
S.~Raju, {\it {Lessons from the information paradox}},  {\em Phys. Rept.} {\bf
  943} (2022) 2187, [\href{http://xxx.lanl.gov/abs/2012.05770}{{\tt
  arXiv:2012.05770}}].

\bibitem{Engelhardt:2022qts}
N.~Engelhardt and r.~Folkestad, {\it {Canonical Purification of Evaporating
  Black Holes}},  \href{http://xxx.lanl.gov/abs/2201.08395}{{\tt
  arXiv:2201.08395}}.

\bibitem{Caceres:2022smh}
E.~Caceres, A.~Kundu, A.~K. Patra, and S.~Shashi, {\it {Trans-IR Flows to Black
  Hole Singularities}},  \href{http://xxx.lanl.gov/abs/2201.06579}{{\tt
  arXiv:2201.06579}}.

\bibitem{deBoer:1999tgo}
J.~de~Boer, E.~P. Verlinde, and H.~L. Verlinde, {\it {On the holographic
  renormalization group}},  {\em JHEP} {\bf 08} (2000) 003,
  [\href{http://xxx.lanl.gov/abs/hep-th/9912012}{{\tt hep-th/9912012}}].

\bibitem{deBoer:2000cz}
J.~de~Boer, {\it {The Holographic renormalization group}},  {\em Fortsch.
  Phys.} {\bf 49} (2001) 339--358,
  [\href{http://xxx.lanl.gov/abs/hep-th/0101026}{{\tt hep-th/0101026}}].

\bibitem{Lee:2022efh}
J.~H. Lee, D.~Neuenfeld, and A.~Shukla, {\it {Bounds on gravitational brane
  couplings and tomography in AdS3 black hole microstates}},
  \href{http://xxx.lanl.gov/abs/2206.06511}{{\tt arXiv:2206.06511}}.

\bibitem{Geng:2022slq}
H.~Geng, A.~Karch, C.~Perez-Pardavila, S.~Raju, L.~Randall, M.~Riojas, and
  S.~Shashi, {\it {Jackiw-Teitelboim Gravity from the Karch-Randall
  Braneworld}},  \href{http://xxx.lanl.gov/abs/2206.04695}{{\tt
  arXiv:2206.04695}}.

\bibitem{Balasubramanian:2020hfs}
V.~Balasubramanian, A.~Kar, O.~Parrikar, G.~S\'arosi, and T.~Ugajin, {\it
  {Geometric secret sharing in a model of Hawking radiation}},  {\em JHEP} {\bf
  01} (2021) 177, [\href{http://xxx.lanl.gov/abs/2003.05448}{{\tt
  arXiv:2003.05448}}].

\bibitem{Balasubramanian:2021xcm}
V.~Balasubramanian, B.~Craps, M.~Khramtsov, and E.~Shaghoulian, {\it
  {Submerging islands through thermalization}},  {\em JHEP} {\bf 10} (2021)
  048, [\href{http://xxx.lanl.gov/abs/2107.14746}{{\tt arXiv:2107.14746}}].

\bibitem{Ghodrati:2015rta}
M.~Ghodrati, {\it {Schwinger Effect and Entanglement Entropy in Confining
  Geometries}},  {\em Phys. Rev. D} {\bf 92} (2015), no.~6 065015,
  [\href{http://xxx.lanl.gov/abs/1506.08557}{{\tt arXiv:1506.08557}}].

\bibitem{Fallows:2021sge}
S.~Fallows and S.~F. Ross, {\it {Islands and mixed states in closed
  universes}},  {\em JHEP} {\bf 07} (2021) 022,
  [\href{http://xxx.lanl.gov/abs/2103.14364}{{\tt arXiv:2103.14364}}].

\bibitem{Bousso:2015mna}
R.~Bousso, Z.~Fisher, S.~Leichenauer, and A.~C. Wall, {\it {Quantum focusing
  conjecture}},  {\em Phys. Rev. D} {\bf 93} (2016), no.~6 064044,
  [\href{http://xxx.lanl.gov/abs/1506.02669}{{\tt arXiv:1506.02669}}].

\bibitem{Almheiri:2019yqk}
A.~Almheiri, R.~Mahajan, and J.~Maldacena, {\it {Islands outside the horizon}},
   \href{http://xxx.lanl.gov/abs/1910.11077}{{\tt arXiv:1910.11077}}.

\bibitem{Ghodrati:2018hss}
M.~Ghodrati, {\it {Complexity growth rate during phase transitions}},  {\em
  Phys. Rev. D} {\bf 98} (2018), no.~10 106011,
  [\href{http://xxx.lanl.gov/abs/1808.08164}{{\tt arXiv:1808.08164}}].

\bibitem{Zhou:2019jlh}
Y.-T. Zhou, M.~Ghodrati, X.-M. Kuang, and J.-P. Wu, {\it {Evolutions of
  entanglement and complexity after a thermal quench in massive gravity
  theory}},  {\em Phys. Rev. D} {\bf 100} (2019), no.~6 066003,
  [\href{http://xxx.lanl.gov/abs/1907.08453}{{\tt arXiv:1907.08453}}].

\bibitem{Ghodrati:2019bzz}
M.~Ghodrati, {\it {Complexity and emergence of warped AdS$_{3}$ space-time from
  chiral Liouville action}},  {\em JHEP} {\bf 02} (2020) 052,
  [\href{http://xxx.lanl.gov/abs/1911.03819}{{\tt arXiv:1911.03819}}].

\bibitem{Ghodrati:2017roz}
M.~Ghodrati, {\it {Complexity growth in massive gravity theories, the effects
  of chirality, and more}},  {\em Phys. Rev. D} {\bf 96} (2017), no.~10 106020,
  [\href{http://xxx.lanl.gov/abs/1708.07981}{{\tt arXiv:1708.07981}}].

\bibitem{Brown:2018bms}
A.~R. Brown, H.~Gharibyan, H.~W. Lin, L.~Susskind, L.~Thorlacius, and Y.~Zhao,
  {\it {Complexity of Jackiw-Teitelboim gravity}},  {\em Phys. Rev. D} {\bf 99}
  (2019), no.~4 046016, [\href{http://xxx.lanl.gov/abs/1810.08741}{{\tt
  arXiv:1810.08741}}].

\bibitem{Alishahiha:2018swh}
M.~Alishahiha, {\it {On complexity of Jackiw\textendash{}Teitelboim gravity}},
  {\em Eur. Phys. J. C} {\bf 79} (2019), no.~4 365,
  [\href{http://xxx.lanl.gov/abs/1811.09028}{{\tt arXiv:1811.09028}}].

\bibitem{Akhavan:2018wla}
A.~Akhavan, M.~Alishahiha, A.~Naseh, and H.~Zolfi, {\it {Complexity and Behind
  the Horizon Cut Off}},  {\em JHEP} {\bf 12} (2018) 090,
  [\href{http://xxx.lanl.gov/abs/1810.12015}{{\tt arXiv:1810.12015}}].

\bibitem{Bhattacharya:2021jrn}
A.~Bhattacharya, A.~Bhattacharyya, P.~Nandy, and A.~K. Patra, {\it {Islands and
  complexity of eternal black hole and radiation subsystems for a doubly
  holographic model}},  {\em JHEP} {\bf 05} (2021) 135,
  [\href{http://xxx.lanl.gov/abs/2103.15852}{{\tt arXiv:2103.15852}}].

\bibitem{Hernandez:2020nem}
J.~Hernandez, R.~C. Myers, and S.-M. Ruan, {\it {Quantum extremal islands made
  easy. Part III. Complexity on the brane}},  {\em JHEP} {\bf 02} (2021) 173,
  [\href{http://xxx.lanl.gov/abs/2010.16398}{{\tt arXiv:2010.16398}}].

\bibitem{KumarBasak:2021rrx}
J.~Kumar~Basak, D.~Basu, V.~Malvimat, H.~Parihar, and G.~Sengupta, {\it {Page
  Curve for Entanglement Negativity through Geometric Evaporation}},
  \href{http://xxx.lanl.gov/abs/2106.12593}{{\tt arXiv:2106.12593}}.

\bibitem{Bhattacharya:2021dnd}
A.~Bhattacharya, A.~Bhattacharyya, P.~Nandy, and A.~K. Patra, {\it {Partial
  islands and subregion complexity in geometric secret-sharing model}},
  \href{http://xxx.lanl.gov/abs/2109.07842}{{\tt arXiv:2109.07842}}.

\bibitem{Kibe:2021gtw}
T.~Kibe, P.~Mandayam, and A.~Mukhopadhyay, {\it {Holographic spacetime, black
  holes and quantum error correcting codes: A review}},
  \href{http://xxx.lanl.gov/abs/2110.14669}{{\tt arXiv:2110.14669}}.

\bibitem{Murciano:2021zvs}
S.~Murciano, V.~Alba, and P.~Calabrese, {\it {Quench dynamics of R\'enyi
  negativities and the quasiparticle picture}},
  \href{http://xxx.lanl.gov/abs/2110.14589}{{\tt arXiv:2110.14589}}.

\bibitem{Geng:2021mic}
H.~Geng, A.~Karch, C.~Perez-Pardavila, S.~Raju, L.~Randall, M.~Riojas, and
  S.~Shashi, {\it {Entanglement Phase Structure of a Holographic BCFT in a
  Black Hole Background}},  \href{http://xxx.lanl.gov/abs/2112.09132}{{\tt
  arXiv:2112.09132}}.

\bibitem{Krishnan:2020oun}
C.~Krishnan, V.~Patil, and J.~Pereira, {\it {Page Curve and the Information
  Paradox in Flat Space}},  \href{http://xxx.lanl.gov/abs/2005.02993}{{\tt
  arXiv:2005.02993}}.

\bibitem{Krishnan:2020fer}
C.~Krishnan, {\it {Critical Islands}},  {\em JHEP} {\bf 01} (2021) 179,
  [\href{http://xxx.lanl.gov/abs/2007.06551}{{\tt arXiv:2007.06551}}].

\bibitem{Ghosh:2021axl}
K.~Ghosh and C.~Krishnan, {\it {Dirichlet baths and the not-so-fine-grained
  Page curve}},  {\em JHEP} {\bf 08} (2021) 119,
  [\href{http://xxx.lanl.gov/abs/2103.17253}{{\tt arXiv:2103.17253}}].

\bibitem{Akal:2021dqt}
I.~Akal, T.~Kawamoto, S.-M. Ruan, T.~Takayanagi, and Z.~Wei, {\it {On the Page
  curve under final state projection}},
  \href{http://xxx.lanl.gov/abs/2112.08433}{{\tt arXiv:2112.08433}}.

\bibitem{Wang:2021aog}
P.~Wang, H.~Wu, and H.~Yang, {\it {Connections between reflected entropies and
  hyperbolic string vertices}},  \href{http://xxx.lanl.gov/abs/2112.09503}{{\tt
  arXiv:2112.09503}}.

\bibitem{Moosavian:2017qsp}
S.~F. Moosavian and R.~Pius, {\it {Hyperbolic geometry and closed bosonic
  string field theory. Part I. The string vertices via hyperbolic Riemann
  surfaces}},  {\em JHEP} {\bf 08} (2019) 157,
  [\href{http://xxx.lanl.gov/abs/1706.07366}{{\tt arXiv:1706.07366}}].

\bibitem{Moosavian:2017sev}
S.~F. Moosavian and R.~Pius, {\it {Hyperbolic geometry and closed bosonic
  string field theory. Part II. The rules for evaluating the quantum BV master
  action}},  {\em JHEP} {\bf 08} (2019) 177,
  [\href{http://xxx.lanl.gov/abs/1708.04977}{{\tt arXiv:1708.04977}}].

\bibitem{Arias:2021ilh}
R.~Arias, M.~Botta-Cantcheff, and P.~J. Martinez, {\it {Pacman geometries and
  the Hayward term in JT gravity}},
  \href{http://xxx.lanl.gov/abs/2112.10799}{{\tt arXiv:2112.10799}}.

\bibitem{Dvali:2021ofp}
G.~Dvali, F.~Kuhnel, and M.~Zantedeschi, {\it {Vortexes in Black Holes}},
  \href{http://xxx.lanl.gov/abs/2112.08354}{{\tt arXiv:2112.08354}}.

\bibitem{Gautam:2022akq}
V.~Gautam, M.~Hanada, A.~Jevicki, and C.~Peng, {\it {Matrix Entanglement}},
  \href{http://xxx.lanl.gov/abs/2204.06472}{{\tt arXiv:2204.06472}}.

\bibitem{Liu:2019svk}
H.~Liu and S.~Vardhan, {\it {Void Formation in Operator Growth, Entanglement,
  and Unitarity}},  {\em JHEP} {\bf 03} (2021) 159,
  [\href{http://xxx.lanl.gov/abs/1912.08918}{{\tt arXiv:1912.08918}}].

\bibitem{Hashimoto:2020cas}
K.~Hashimoto, N.~Iizuka, and Y.~Matsuo, {\it {Islands in Schwarzschild black
  holes}},  {\em JHEP} {\bf 06} (2020) 085,
  [\href{http://xxx.lanl.gov/abs/2004.05863}{{\tt arXiv:2004.05863}}].

\bibitem{Anegawa:2020ezn}
T.~Anegawa and N.~Iizuka, {\it {Notes on islands in asymptotically flat 2d
  dilaton black holes}},  {\em JHEP} {\bf 07} (2020) 036,
  [\href{http://xxx.lanl.gov/abs/2004.01601}{{\tt arXiv:2004.01601}}].

\bibitem{Chandra:2022fwi}
J.~Chandra and T.~Hartman, {\it {Coarse graining pure states in AdS/CFT}},
  \href{http://xxx.lanl.gov/abs/2206.03414}{{\tt arXiv:2206.03414}}.

\bibitem{Baiguera:2022sao}
S.~Baiguera, L.~Bianchi, S.~Chapman, and D.~A. Galante, {\it {Shape
  Deformations of Charged R\'enyi Entropies from Holography}},
  \href{http://xxx.lanl.gov/abs/2203.15028}{{\tt arXiv:2203.15028}}.

\end{thebibliography}\endgroup
\bibliographystyle{JHEP}
\end{document}